\documentclass[a4paper,11pt]{article}
\pdfoutput=1

\usepackage[utf8]{inputenc}
\usepackage{jheppub}
\hypersetup{pdfencoding=unicode, bookmarksopen=true, bookmarksnumbered}
\usepackage[T1]{fontenc}
\usepackage{microtype}
\usepackage{mathrsfs}
\usepackage{subcaption}
\usepackage[all]{xy}
\usepackage{tikz}
\usepackage{diagbox}
\usepackage{physics}
\usepackage{longtable}
\usepackage{afterpage}
\usepackage{lscape}

\usetikzlibrary{decorations.markings}
\usetikzlibrary{arrows.meta}
\tikzset{->-/.style={decoration={
  markings,
  mark=at position #1 with {\arrow{>}}},postaction={decorate}}}

\newcolumntype{L}[1]{>{\raggedright\let\newline\\\arraybackslash\hspace{0pt}}m{#1}}
\newcolumntype{C}[1]{>{\centering\let\newline\\\arraybackslash\hspace{0pt}}m{#1}}
\newcolumntype{R}[1]{>{\raggedleft\let\newline\\\arraybackslash\hspace{0pt}}m{#1}}

\pgfarrowsdeclare{:}{:}{}{}

\let\a=\alpha  \let\g=\gamma \let\d=\delta 
    
 \let\m=\mu \let\n=\nu   
\let\s=\sigma     
   \let\G=\Gamma \let\D=\Delta  
 \let\P=\Pi \let\S=\Sigma

\DeclareMathOperator{\Gal}{Gal}
\DeclareMathOperator{\Li}{Li}
\newcommand\sqprod{\mathbin{\text{\scalebox{.84}{$\square$}}}}

\def\a{\alpha}

\def\CF{{\cal F}}

\def\CH{{\cal H}}
\def\CI{{\cal I}}

\def\CL{{\cal L}}
\def\CM{{\cal M}}
\def\CN{{\cal N}}

\def\CS{{\cal S}}
\def\CT{{\cal T}}

\def\CW{{\cal W}}

\def\CZ{{\cal Z}}

\def\beq#1\eeq{\begin{align}#1\end{align}}

\makeatletter
\newcommand*{\rom}[1]{\expandafter\romannumeral #1}

\makeatother

\title{3D TFTs and boundary VOAs from BPS spectra of $(G,G')$ Argyres-Douglas theories}

\abstract{We explore 3d $ \mathcal{N}=4 $ theories arising from twisted compactification of 4d $ \mathcal{N}=2 $ $ (G, G') $ Argyres-Douglas superconformal field theories (SCFTs), together with the 2d vertex operator algebras (VOAs) supported on the holomorphic boundary of their topologically twisted sector. Starting from the Coulomb branch BPS spectra of the $ (G,G') $ Argyres-Douglas theories, we develop a systematic and efficient method to obtain the ellipsoid partition functions of associated 3d theories using quiver mutations and wall-crossing invariants. This allows us to extract the modular data of the boundary VOAs, which are related to the Schur sectors of the 4d theories through the 4d SCFT/2d VOA correspondence. Our results provide a useful computational bridge between 4d SCFTs and 2d VOAs through interpolating 3d topological field theories.}

\author[a]{Minsung Kim,}
\author[b]{Sungjoon Kim}

\affiliation[a]{Quantum Universe Center, Korea Institute for Advanced Study, Seoul 02455, Korea}
\affiliation[b]{Korea Institute for Advanced Study, 85 Hoegiro, Dongdaemun-Gu, Seoul 02455, Korea}

\emailAdd{minsung1@kias.re.kr}
\emailAdd{sungjoon@kias.re.kr}

\begin{document}
\preprint{KIAS-Q25021}
\maketitle



\section{Introduction}

Supersymmetric quantum field theory has revealed deep structural connections between different types of quantum field theories. One fascinating example is the correspondence between 4d $\mathcal{N}=2$ superconformal field theories (SCFTs) and 2d vertex operator algebras (VOAs) \cite{Beem:2013sza}, commonly referred to as the 4d SCFT/2d VOA correspondence, or simply the SCFT/VOA correspondence. More precisely, this correspondence asserts that a protected subsector of a 4d $\CN=2$ SCFT, called the {\it Schur sector}, is mapped to a VOA which constitutes the chiral algebra of a 2d conformal field theory. This leads to a non-trivial relation
\begin{align}
    \CI_{\text{Schur}}(q) = \chi_0(q) \, ,
    \label{eq: index-character}
\end{align}
where $\CI_{\text{Schur}}(q)$ is the Schur index which counts the operators in the Schur sector of the 4d theory, and $\chi_0(q)$ is the vaccum character of the associated 2d VOA. The central charges of the 2d VOA and the 4d SCFT satisfy
\begin{align}
    c_{2d} = - 12 c_{4d}\,.
\end{align}
It is notable that an infinite chiral symmetry algebra appears in a subsector of any 4d $\CN=2$ SCFT, and this structure has since been studied extensively \cite{Xie:2016evu,Song:2016yfd,Song:2017oew,Beem:2017ooy,Fluder:2017oxm,Bonetti:2018fqz,Creutzig:2018lbc,Oh:2019bgz,Jeong:2019pzg,Beem:2019snk,Auger:2019gts,Xie:2019zlb,Dedushenko:2019yiw,Xie:2019vzr,Dedushenko:2019mzv,Adamovic:2020lvj,Dedushenko:2023cvd}.

A remarkable formula has been proposed that computes the Schur index from the BPS particle spectrum on the Coulomb branch \cite{Cordova:2015nma}:
\begin{align}
    \CI_{\text{Schur}}(q) = (q)_\infty^{2r} \Tr M(q)^{-1} \, ,
    \label{eq: cordova-shao}
\end{align}
where $(q)_\infty = \prod_{i=1}^\infty (1-q^i)$ is the $q$-Pochhammer symbol, $r$ is the dimension of the Coulomb branch, and $M(q)$ is the monodromy operator of the quantum torus algebra \cite{Cecotti:2010fi}
\begin{align}
    M(q) = \prod_{\g}^{\curvearrowright} \Psi_q(X_\g) \, .
\end{align}
The operaotor $ M(q) $ is defined as an ordered product over all BPS particles with electromagnetic charge $\g$, arranged according to the phases of their central charges, $\arg(\CZ_\g)$, and involves the function $\Psi_q(X)$ given by
\begin{align}
    \Psi_q(X) = \prod_{n\geq 0}(1+q^{n+\frac{1}{2}}X) = \sum_{n\geq 0} \frac{q^{\frac{n^2}{2}}}{(q)_n} X^n \,.
\end{align}
The formula \eqref{eq: cordova-shao} is surprising in that it uses {\it massive} BPS particle data to compute the spectrum of {\it conformal} primaries. Another important feature is that the monodromy operator $M(q)$ is invariant under wall-crossing on the Coulomb branch. Thus, the various powers of the monodromy operator $M(q)^n$ are also wall-crossing invariants, which give rise to a natural extension of the Schur index formula \cite{Cecotti:2015lab}. Indeed, recent studies verified that such extension produces a families of VOAs that lie in the same Galois orbit \cite{Kim:2024dxu,Go:2025ixu}.

In spite of the usefulness of the Schur index formula \eqref{eq: cordova-shao}, there are two obstacles that one has to address. First, the formula requires detailed knowledge of the BPS particle spectrum as input data. In particular, we have to determine how many BPS particles are present at a given Coulomb vacuum, together with their electromagnetic charges and the ordering of their central charge phases. For a generic 4d $ \mathcal{N}=2 $ SCFT, this is a highly non-trivial task. Another issue is more practical in nature. Even when the formula can be written down explicitly, comparing it with the vacuum character of a specific 2d VOA typically requires performing a series expansion. However, this expansion involves infinite sums over integers and the number of sums grows with the number of BPS particles, making the series expansion extremely challenging for theories with complicated BPS spectrum. Due to this technical subtlety, explicit computations have so far been limited to relatively simple cases.

In this work, we provide an improved resolution to both obstacles for a family of 4d $ \mathcal{N}=2 $ Argyres-Douglas theories \cite{Argyres:1995jj, Argyres:1995xn} labeled by two simply-laced Lie algebras $ (G, G') $, which admit a geometric realization via type IIB string theory on isolated hypersurface singularities \cite{Eguchi:1996vu, Eguchi:1996ds, Cecotti:2010fi}. These theories admit BPS quivers \cite{Cecotti:2011rv} which encode the dynamics of the BPS particles on their Coulomb vacua. Using the quiver mutation method, we determine the BPS particle spectrum at a Coulomb vacuum of the $(G,G')$ theory, thereby enabling the application of the formula.\footnote{The same mutations are discussed in \cite{Keller:2010bq,Cecotti:2010fi,Xie:2012gd} as well.} To address the second technical issue, we extensively use the functional identities that encapsulate the wall-crossing phenomena: for $X_{\g_1} X_{\g_2} =q^{-1}\, X_{\g_2} X_{\g_1}$, we have
\begin{align}
    \begin{aligned}\label{eq: Psi id}
        \Psi_q(X_{\g_1}) \Psi_q(X_{\g_2})
        &= \Psi_q(X_{\g_2}) \Psi_q(X_{\g_1+\g_2}) \Psi_q(X_{\g_1}) \, , \\
        \Theta(X_{\g_1}) \Psi_q(X_{\g_2})
        &= \Psi_q(X_{\g_1+\g_2}) \Theta(X_{\g_1}) \, , \\
        \Theta(X)
        = \Psi_q(X)\Psi_q(X^{-1})
        &= \frac{1}{(q)_\infty} \sum_{n\in\mathbb{Z}} q^{\frac{n^2}{2}} X^n \, .
    \end{aligned}
\end{align}
These identities allow us to reduce the number of $ \Psi_q(X) $ functions appearing in $ M(q) $ whenever possible, which in turn decreases the number of infinite summations. This would instantly improve the difficulty; however, no systematic procedure for achieving such simplification is currently known. Nevertheless, we have developed an efficient machinery that performs this task by formulating it as a discrete optimization problem. The corresponding \texttt{Mathematica} implementation is provided \cite{code}, and detailed instructions are given in Appendix~\ref{app: code}. Through case-by-case checks, we confirmed that our code almost always produces the most simplified expressions. For instance, in the $(A_2,A_6)$ Argyres-Douglas theory, the fifth power of the monodromy operator given by
\begin{align}
    \CI_{\text{Schur}}^{(A_2,A_6)^5}(q)
    = (q)_{\infty}^{12} \Tr M(q)^5\,,
\end{align}
has 180 factors of $\Psi_q(X)$ inside the trace in the minimal chamber. Our {\tt Mathematica} code simplifies this into 126 $\Theta(X)$ functions with no remaining $\Psi_q(X)$ factors, demonstrating that further simplification using the identities \eqref{eq: Psi id} is not possible.

However, this progress still does not fully resolve the difficulties arising from the complexity of the BPS spectrum, and the problem is inevitably reappears in theories with sufficiently large $G$ and $G'$. Furthermore, potential convergence issues may appear in the series expansion of the Schur index formula. 

To overcome them, we adopt an alternative approach based on a recently suggested wall-crossing invariant formula \cite{Gaiotto:2024ioj}
\begin{align}
    Z_{S_b^3} = \Tr_\CH \bigg( \prod_{\g}^{\curvearrowleft} \Phi_b(x_\g)
    \bigg) \, ,
    \label{eq: intro tr formula}
\end{align}
where the trace is over an auxiliary Hilbert space $\CH$ associated with the Weyl algebra variables $x_\g$ determined by the Dirac paring of charges $\g$. The formula involves ordered products of the Faddeev's quantum dilogarithm given by
\begin{align}
    \Phi_b(z) = \exp\left( \frac{1}{4} \int_{\mathbb{R}+i 0^+} \frac{dt}{t}\frac{e^{-2izt}}{\sinh(bt)\sinh(b^{-1}t)} \right)\, ,
\end{align}
whose properties analogous to \eqref{eq: Psi id} are reviewed in Appendix~\ref{app: QDL}. 

The formula \eqref{eq: intro tr formula}, which we will simply refer to as the {\it trace formula}, is also a wall-crossing invariant that computes the 3d ellipsoid partition function $Z_{S_b^3}$ of a topologically twisted $ \mathcal{N}=4 $ SCFT. This 3d $ \mathcal{N}=4 $ SCFT arises from the $ U(1)_r $ twisted circle compactification of a 4d $ \mathcal{N}=2 $ SCFT, and its topological twist yields a topological field theory (TFT) of cohomological type \cite{Witten:1990bs} that supports a 2d boundary VOA on its holomorphic boundary \cite{Costello:2018fnz,Dedushenko:2018bpp,Gang:2023rei,Ferrari:2023fez,Dedushenko:2023cvd}.\footnote{See \cite{ArabiArdehali:2024ysy,ArabiArdehali:2024vli} for an alternative construction of the 3d TFTs using high-temperature effective field theory from the 4d $\CN=1$ Lagrangian description of Argyres-Douglas theories \cite{Maruyoshi:2016tqk,Maruyoshi:2016aim,Agarwal:2016pjo}.} In this paper, we refer to such cohomological TFTs simply as TFTs. They are topologically twisted subsectors of the 3d SCFTs, and the metric dependence of the partition function \eqref{eq: intro tr formula} appears only through an overall phase. Surprisingly, the VOA supported on the boundary of this TFT is precisely the one appearing in the 4d SCFT/2d VOA correspondence \cite{Gaiotto:2024ioj,Go:2025ixu,Kim:2025klh}. Thus, we can study the 4d SCFT/2d VOA correspondence by constructing an interpolating 3d theory whose modular data agree with those of the 2d boundary VOA. The modular data can be extracted from the 3d TFT perspective by utilizing the localization techniques developed in \cite{Closset:2018ghr}.

In this paper, we focus on general $(G,G')$ Argyres-Douglas theories and identify the corresponding 3d TFTs arising from their $U(1)_r$ twisted circle compactifications. We extract the partial modular data of these 3d TFTs and confirm that they are compatible with the modular data expected from the associated 2d VOAs. These results provide strong evidence for the existence of consistently interpolating 3d TFTs for the 4d SCFT/2d VOA correspondence. Compared to other approaches, the topological nature of the 3d theory enables access to aspects of the SCFT/VOA correspondence that would otherwise be prohibitively difficult to analyze. For instance, the partial modular data of the VOA corresponding to the $(A_3,D_4)$ theory, which is also dual to the $(A_2,E_6)$, can be calculated from the 3d TFT perspective, which is the first result as far as we aware of. Taken together, these observations show that our 3d TFT approach not only broadens the range of Argyres–Douglas theories amenable to detailed analysis, but also offers a powerful method for exploring the SCFT/VOA correspondence.\medskip

The paper is organized as follows. In section~\ref{sec: tr}, we introduce the trace formula that presents a 3d TFT arising from the $U(1)_r$ twisted circle compactification of a 4d $\CN=2$ SCFT. We also review the BPS quivers and mutations which read the BPS particle spectrum required for the trace formula. In~section \ref{sec: GG}, we provide several examples of $(G,G')$ Argyres-Douglas theories and confirm that the modular data of the resulting 3d TFTs are compatible with those of the desired 2d VOAs. We then conclude by outlining future directions in section~\ref{sec: discussion}. In Appendix~\ref{app:voa}, we briefly explain the construction of the modular data for the VOAs appearing in the main text. Appendix~\ref{app: 3d review} contains a survey of a particular class of 3d SCFTs called the {\it rank-0 theories} along with relevant computational tools. In Appendix~\ref{app: embed}, we supply additional examples of trace formula computations, including checks of non-trivial isomorphisms of $(G,G')$ theories at the level of the trace formula. Instructions for the accompanying {\tt Mathematica} code that simplifies wall-crossing invariants are provided in Appendix~\ref{app: code}.

\section{3D TFTs from 4D \texorpdfstring{$\CN=2$}{N=2} SCFTs} \label{sec: tr}

In this section, we describe a novel method for constructing 3d topological field theories from 4d $\CN=2$ superconformal field theories. More precisely, for a given Coulomb branch data of a 4d $\CN=2$ SCFT, we read a 3d TFT from a wall-crossing invariant which computes the 3d ellipsoid partition function of the TFT \cite{Gaiotto:2024ioj}. It turns out that this 3d TFT provides a natural bridge connecting the two sides of the 4d SCFT/2d VOA correspondence \cite{Beem:2013sza}, offering an mediating perspective for understanding the correspondence. Namely, this 3d TFT is equivalent to the recently suggested one \cite{Dedushenko:2023cvd} that arises from the $U(1)_r$ twisted compactification of the 4d SCFT along the cigar circle after performing the holomorphic-topological twist \cite{Kapustin:2006hi} combined with an $\Omega$-deformation along the topological direction \cite{Oh:2019bgz,Jeong:2019pzg}. The resulting 3d TFT admits a holomorphic 2d boundary on which the desired VOA is supported.

One way to obtain such 3d TFTs is to consider a Janus-like loop on the Coulomb branch of the 4d SCFT, along which an effective 3d $\CN=2$ Abelian Chern-Simons matter theory arises with a proper superpotential \cite{Gaiotto:2024ioj}. As noted above, the ellipsoid partition function of the resulting 3d theory is obtained from a wall-crossing invariant, allowing the detailed properties of the 3d theory to be inferred directly from this invariant. In the following subsections, we first introduce the wall-crossing invariant formula that computes the 3d ellipsoid partition function from the Coulomb branch data of a given 4d $\CN=2$ SCFT. Throughout the paper, we refer to this as the \emph{trace formula}. We then review the {\it quiver mutation method} for characterizing the Coulomb branch data of a 4d $\CN=2$ SCFT, which serves as the input for the trace formula. Although the trace formula can in principle be applied to general 4d $\CN=2$ SCFTs, we focus on a particular class of 4d SCFTs, the $(G,G')$ Argyres-Douglas theories, for which we provide a complete prescription for the trace formula.

\subsection{Trace formula}

Consider a 4d $ \mathcal{N}=2 $ SCFT with Coulomb branch moduli space $ \mathcal{M}_{\text{Coulomb}} $. The effective theory on the Coulomb branch is described by an Abelian gauge theory. Suppose we have a $ U(1)^r $ gauge symmetry at a generic point on the Coulomb branch, where $ r $ is called the rank of the Coulomb branch, together with a flavor symmetry of rank $ f $. Then the BPS particles carry $ r $ electric, $ r $ magnetic and $ f $ flavor charges, so the charge lattice $ \Gamma $ of the BPS particles have
\begin{align}
    \rank(\G) = 2r+f  \, .
\end{align}
The electromagnetic charges obey the Dirac quantization condition, which implies the existence of an integral antisymmetric pairing
\begin{align}
    \langle \gamma_1, \gamma_2 \rangle = e_1 \cdot m_2 - m_1 \cdot e_2 \in \mathbb{Z} \, , \quad
    \gamma_i = (e_i, m_i, f_i) \in \Gamma \, ,
\end{align}
called the Dirac pairing. The central charge $ \mathcal{Z}_\gamma : \Gamma \to \mathbb{C} $ of 4d $ \mathcal{N}=2 $ supersymmetry is a complex valued linear function of charge $\gamma\in \Gamma$, and it provides a lower bound on the mass $ M_\gamma $ of states in a supermultiplet as $ M_\gamma \geq |\mathcal{Z}_\gamma| $. States that saturate this bound are BPS states, which can be divided into {\it particles} and {\it anti-particles}: the former lie in the upper half of the complex $\CZ$-plane, while the latter lie in the lower half. If there exists a BPS particle of charge $\g$, then CPT invariance ensures the presence of its CPT conjugate, an anti-particle of charge $-\g$.

We now consider the 3d TFT obtained from the $U(1)_r$ twisted circle compactification of the 4d $\CN=2$ SCFT. The UV description of the 3d theory can be read from the trace formula \cite{Gaiotto:2024ioj}
\begin{align}
    Z_{S_b^3} = \Tr_{\CH} \bigg( \prod_{\g \in \G_{\text{BPS}}}^{\curvearrowleft} \Phi_b(x_\g)
    \bigg) \, ,
    \label{eq: trace formula}
\end{align}
which computes the ellipsoid partition function of the resulting 3d theory. Here, $\G_{\text{BPS}}$ is an ordered set of stable BPS (and anti-BPS) particles of charge $\g$, arranged such that their central charge phases, $\arg(\CZ_\g)$, increase monotonically. Thus, for a given Coulomb branch data, the formula \eqref{eq: trace formula} is a trace over an auxiliary Hilbert space $\CH = L^2(\mathbb{R}^{\rank(\G/\G_f)/2})$ of an ordered product of the Faddeev's quantum dilogarithms $\Phi_b(x_\g)$ defined in \eqref{eq: QDL def}, where $\Gamma_f \subset \Gamma$ is the sublattice of flavor charges. The operators $x_\g$ obey the Weyl algebra relations
\begin{align}
    [x_\g , x_{\g'}] = \frac{1}{2\pi i} \langle \g,\g' \rangle \, .
\end{align}
Note that the formula \eqref{eq: trace formula} may appear to depend on the location of the Coulomb branch. In general, the number of BPS particles changes when crossing a {\it wall} on the Coulomb branch which divides two distinct regions called chambers. Consequently, the form of the expression \eqref{eq: trace formula} will look different in distinct chambers. However, the pentagon identity of the Faddeev's quantum dilogarithm
\begin{align}
    \Phi_b(x_\g) \Phi_b(x_{\g'})
    =
    \Phi_b(x_{\g'})
    \Phi_b(x_\g+x_{\g'})
    \Phi_b(x_\g)
    \;\;
    \text{if}\;\;
    \langle \g,\g' \rangle =1 \, ,
    \label{eq: pentagon id}
\end{align}
ensures that the formula \eqref{eq: trace formula} remains invariant under such wall-crossings. In other words, we may evaluate the formula at any convenient point on the Coulomb branch, and the result will be identical in all chambers. Thus, the formula \eqref{eq: trace formula} is a wall-crossing invariant.

\begin{figure}[tbp]
    \centering
    \begin{tabular}{C{35ex}cC{35ex}}
        \includegraphics{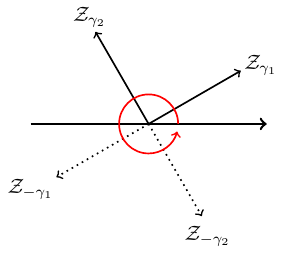} & 
        $ \overset{\text{wall-crossing}}{\longleftrightarrow} $ &
        \includegraphics{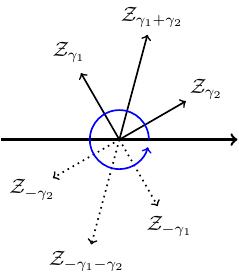}
    \end{tabular}
\caption{\label{fig: A2 chamber} Central charges in two distinct chambers on the Coulomb branch of the $(A_1,A_2)$ Argyres-Douglas theory. In the LHS chamber, there are two BPS (anti-)particles, represented by solid(dashed) vectors on the complex plane, whereas in the RHS chamber there are three. The trace formula \eqref{eq: trace formula} reads the Faddeev's quantum dilogarithms in the counterclockwise order of the central charge phases, indicated by the red and blue circular arrows in the LHS/RHS chambers, respectively.}
\end{figure}

\paragraph{Example: \texorpdfstring{$(A_1,A_2)$}{(A1,A2)}} Consider the 4d $\CN=2$ $(A_1,A_2)$ Argyres-Douglas theory which has two distinct chambers as depicted in Figure~\ref{fig: A2 chamber}. In the left chamber, there are two stable BPS particles having charges $\g_1$ and $\g_2$ with the Dirac paring $\langle \g_1,\g_2 \rangle=1$, and their central charges satisfy $\arg(\CZ_{\g_1}) < \arg(\CZ_{\g_2})$. The central charges of the CPT conjugates(anti-particles) are indicated by dashed lines. We therefore obtain an ordered set of electromagnetic charges $\G_{\text{BPS}}$ by reading the phases of the central charges in a counterclockwise direction, as illustrated by the red circular arrow:
\begin{align}
	&\G_{\text{BPS}} = \{ \g_1 , \g_2 , - \g_1 , -\g_2 \} \, .
    \label{eq: A2 2ptcl}
\end{align}
The trace formula \eqref{eq: trace formula} reads
\begin{align}
	Z_{S_b^3}^{(\textcolor{red}{\text{L}})}
	=
	\Tr\Big(
	\Phi_b(x_{\g_1})\Phi_b(x_{\g_2})\Phi_b(-x_{\g_1})\Phi_b(-x_{\g_2})
	\Big) \, ,
\end{align}
where the superscript (\textcolor{red}{L}) denotes the expression obtained in the left chamber. Note that we used $x_{-\g} = - x_{\g}$, since the Weyl algebra variables $x_\g$ depend linearly on the charge $\g$. 

Now, suppose we move along the Coulomb branch $u\in\CM_{\text{Coulomb}}$ within the left chamber, so that the central charges $\CZ_{\g_i}(u)$ vary smoothly without any intersecting. However, once $u$ is located exactly on the wall separating the left and right chambers, the two central charges will be aligned, $\arg(\CZ_{\g_1}) = \arg(\CZ_{\g_2}) $. Upon crossing into the right chamber, the ordering of the two central charges are reversed, and another BPS particle of charge $\g_1 + \g_2$ is generated, as drawn in Figure~\ref{fig: A2 chamber}. Its central charges satisfy $\arg(\CZ_{\g_1}) > \arg(\CZ_{\g_1+\g_2}) > \arg(\CZ_{\g_2})$. Consequently, we obtain the ordered set $\G_{\text{BPS}}$ in the right chamber by following the blue circular arrow as
\begin{align}
    \G_{\text{BPS}} &= \{ \g_2 ,\g_1+\g_2 , \g_1, -\g_2 ,-\g_1-\g_2 , -\g_1 \} \, ,
    \label{eq: A2 3ptcl}
\end{align}
and the formula \eqref{eq: trace formula} now yields
\begin{align}
    Z_{S_b^3}^{(\textcolor{blue}{\text{R}})} = \Tr \Big(
    \Phi_b(x_{\g_2})
    \Phi_b(x_{\g_1}+x_{\g_2})
    \Phi_b(x_{\g_1})
    \Phi_b(-x_{\g_2})
    \Phi_b(-x_{\g_1}-x_{\g_2})
    \Phi_b(-x_{\g_1})
    \Big) \, ,
\end{align}
where (\textcolor{blue}{R}) denotes the expression evaluated in the right chamber. Thanks to the pentagon identity \eqref{eq: pentagon id}, one can easily check that
\begin{align}
    Z_{S_b^3}^{(\textcolor{red}{\text{L}})}
    =
    Z_{S_b^3}^{(\textcolor{blue}{\text{R}})}
    \equiv
    Z_{S_b^3}^{(A_1,A_2)}
\end{align}
which illustrates that the trace formula is a wall-crossing invariant.

Note that the Weyl algebra variables satisfying $[x_{\g_1},x_{\g_2}]=(2\pi i)^{-1}$ can be regarded as canonical momentum and position operators by identifying $ x_{\g_1}=\hat{p} $ and $ x_{\g_2}=\hat{q} $, which obey
\begin{align}
    [\hat{p},\hat{q}] = \frac{1}{2\pi i} \, .
\end{align}
With this identification, we can use the completeness relations and inner products of their eigenstates $\hat{p}\ket{p}=p\ket{p}$ and $\hat{q}\ket{q}=q\ket{q}$ as
\begin{align}
    {\bf 1} = \int dp \dyad{p}{p} \, , \quad
    {\bf 1} = \int dq \dyad{q}{q} \, , \quad
    \braket{p}{q} = e^{2\pi i pq}\,.
\end{align}
Thus, we can manipulate the operators in the trace using the pentagon identity \eqref{eq: pentagon id} and the fusion relation $\Phi_b(x)\Phi_b(-x) = \Phi_b(0)^2 e^{\pi i x^2}$, where $\Phi_b(0)=e^{\pi i(b^2+b^{-2})/24}$, as
\begin{align}
    Z_{S_b^3}^{(A_1,A_2)}
    &= \Tr \Big( \Phi_b(\hat{p}) \Phi_b(\hat{q}) \Phi_b(-\hat{p}) \Phi_b(-\hat{q}) \Big) 
    = \Tr \Big( \Phi_b(\hat{q}) \Phi_b(\hat{p}+\hat{q}) \Phi_b(\hat{p}) \Phi_b(-\hat{p}) \Phi_b(-\hat{q}) \Big) \nonumber\\
    &= \Phi_b(0)^4 \Tr \Big( e^{\pi i \hat{q}^2} \Phi_b(\hat{p}+\hat{q}) e^{\pi i \hat{p}^2} \Big) 
    = \Phi_b(0)^4 \Tr \Big( e^{\pi i \hat{q}^2} e^{\pi i \hat{p}^2} \Phi_b(\hat{q}) \Big) \nonumber\\
    &=\Phi_b(0)^4\int dp\, dq\, \Phi_b(q) e^{\pi i q^2} e^{\pi i p^2} 
    =i^{\frac{1}{2}}\Phi_b(0)^4 \int dq\,e^{\pi i q^2}\Phi_b(q)\,.
    \label{eq: gang-yamazaki}
\end{align}
Here, we use $f(\hat p + \hat q) e^{\pi i \hat p^2} = e^{\pi i \hat p^2} f(\hat q)$ in the fourth equality. Up to a $b$-dependent overall phase arising from the SUSY gravitational CS term and background CS term of the $U(1)_R$ symmetry of 3d $\CN=2$ supersymmetry, the result resembles that of the ellipsoid partition function of the 3d $\CN=2$ Gang-Yamazaki theory \cite{Gang:2018huc} which has $\CN=4$ SUSY enhancement in the infrared. Indeed, the result \eqref{eq: gang-yamazaki} coincides with the partition function of the topologically A-twisted Gang-Yamazaki theory, up to an overall phase. It has been checked that the A-twisted Gang-Yamazaki theory is a 3d TFT that supports the $M(2,5)$ Virasoro minimal model on its holomorphic 2d boundary \cite{Gang:2021hrd,Gang:2023rei,Ferrari:2023fez,ArabiArdehali:2024ysy,Go:2025ixu}. Surprisingly, this is precisely the VOA to which the Schur sector of the $(A_1,A_2)$ Argyres-Douglas theory is mapped. This example illustrates that the 3d TFT obtained from the trace formula naturally bridges the 4d SCFT and the 2d VOA in the SCFT/VOA correspondence.

Before proceeding, let us make a few remarks. The Argyres-Douglas theories have Coulomb branch operators with fractional conformal dimensions. Under the $ U(1)_r $ twisted compactification, a Coulomb branch operator $ \mathcal{O} $ of conformal dimension $ \Delta $ acquires a phase $ \mathcal{O} \to e^{2\pi i \Delta} \mathcal{O} $. Hence, if all Coulomb branch operators of an Argyres-Douglas theory without a Higgs branch have non-integer conformal dimensions, the resulting 3d theory after the $U(1)_r$ twisted reduction will have empty Coulomb and Higgs branches. Such theories are called the 3d {\it rank-zero} theory \cite{Gang:2018huc,Gang:2021hrd} which are observed to have $\CN = 4$ or $5$ supersymmetry in the infrared. These rather simple but still interacting SCFTs have been studied in various contexts. The Gang-Yamazaki theory is the first example of the rank-0 SCFT and additional examples arising from the general $(G,G')$ Argyres-Douglas theories will appear frequently in section~\ref{sec: ex}. For interested readers, we summarize recent developments of the 3d rank-0 theories in Appendix~\ref{app: rank-0}, along with relevant computational techniques.

Another remark is that the simplification using the pentagon and fusion identities of $\Phi_b(x)$'s in \eqref{eq: gang-yamazaki} is technically crucial when we extract the 3d TFT data. However, there is no algorithmic procedure or general pattern that guarantees the reduction. Nevertheless, we have developed an efficient {\tt Mathematica} code that drastically simplifies the trace formula. 

Lastly, in contrast to the simple case of the $(A_1,A_2)$ theory, it is not that straightforward to embed the Weyl algebra into a canonical quantum mechanical framework for general $(G,G')$ theories or more generic 4d $\CN=2$ SCFTs. For computational completeness, we provide a systematic procedure for performing such an embedding in Appendix~\ref{app: embed}.

\subsection{BPS spectra from mutation of BPS quiver \label{sec: mutation}}

The trace formula requires the Coulomb branch data $\G_{\text{BPS}}(u) = \{\g\}$, which is an ordered set of electromagnetic charges of stable BPS particles and anti-particles in a given chamber. The charges $\g$ are arranged in increasing order of the phases of their central charges $\arg(\CZ_\g(u))$. Characterizing these data is in general a highly nontrivial problem. However, it can be achieved for a large class of $\mathcal{N}=2$ theories that admit a \emph{BPS quiver}. Such theories admit a basis $ \{\gamma_i\} $ of the charge lattice $ \Gamma $ such that any charge vector $ \gamma \in \Gamma $ lies in either $ \gamma \in \Gamma_+ $ or $ \gamma \in -\Gamma_+ $, where $ \Gamma_+ = \bigoplus_{i=1}^{\rank \Gamma} \mathbb{Z}_{\geq 0} \gamma_i $. Consequently, the central charge $ \mathcal{Z}_\gamma $ can be also represented as a positive (or negative) linear combination of the central charges $ \mathcal{Z}_{\gamma_i} $ of the basis elements. Since the central charges of all BPS particles lie in the upper half of the complex $\CZ$-plane, the set of occupied BPS charges form a cone in this upper half-plane, bounded by the leftmost and rightmost central charges. This is called the {\it cone of BPS particles}. Not all 4d $ \mathcal{N}=2 $ theories admit such a cone, but whenever do, the corresponding BPS quiver can be constructed using the following rules:
\begin{enumerate}
    \item Draw $2r+f$ nodes and assign basis charges $\g_i$ to them, one to each node.
    \item For each pair of $(\gamma_i,\gamma_j)$, draw $\langle \g_i,\g_j \rangle$ arrows directed from $i$ to $j$ whenever $\langle \g_i,\g_j \rangle > 0$.
\end{enumerate}

\begin{figure}[tbp]
    \centering
    $ \displaystyle \vcenter{\hbox{\includegraphics{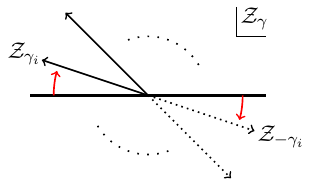}}}
    \quad \overset{\textcolor{red}{\mathfrak{M}_i}}{\longrightarrow} \quad
    \vcenter{\hbox{\includegraphics{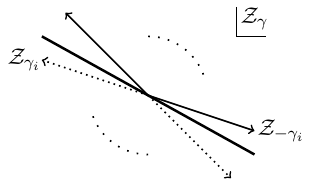}}} $
    \caption{\label{fig: cone mutation} A mutation $\mathfrak{M}_i$ at the $i$-th node, where the central charge $\CZ_{\g_i}$ of the associated basis charge $\g_i$ lies on the left boundary of the cone of BPS particles, corresponds to a clockwise rotation of the upper-half $\CZ$-plane. Under this rotation, $\CZ_{\g_i}$ exits the half-plane and becomes an anti-particle, while its CPT conjugate $\CZ_{-\g_i}$ enters the half-plane and becomes a particle. In the figure, solid vectors represent BPS particles and dotted vectors represent anti-particles.}
\end{figure}

As one may observe, the shape of the BPS quiver depends on the choice of basis charges. There is a special change of basis charges called {\it quiver mutation}, or simply a {\it mutation}. If we perform a mutation at the $i$-th node, whose assigned charge is $\g_i$, the new basis charges $\mathfrak{M}_i(\g_j)$ are given by
\begin{align} \label{eq: mutation}
    \mathfrak{M}_i(\g_j) = 
    \left\{
    \begin{array}{cc}
       \g_j + \langle \g_i,\g_j \rangle \, \g_i  & \;\;\text{if}\;\; \langle \g_i,\g_j \rangle > 0 \\
       (-1)^{\d_{ij}}\, \g_j  & \;\; \text{if}\;\; \langle \g_i,\g_j \rangle \leq 0
    \end{array} \, ,
    \right.
\end{align}
and a new BPS quiver is constructed from the Dirac pairing. It turns out that this mutation captures not only the ordering of the phases of the central charges but also the charges of the occupied BPS particles \cite{Alim:2011kw}. Suppose we perform a mutation at $\gamma_i$. This produces a new basis in which the original basis $\gamma_i$ is replaced by $-\gamma_i$. The situation corresponds to rotating the upper half of the $\mathcal{Z}$-plane clockwise, so that $\CZ_{\g_i}$ exits the upper half-plane while $-\CZ_{\g_i}$ enters it, as illustrated in Figure~\ref{fig: cone mutation}. More precisely, the {\it left} boundary of the cone of BPS particles moves out of the half-plane, while the central charge of $-\gamma_i$ simultaneously enters and becomes the new {\it right} boundary of the cone. In this way, a mutation can be understood as the change of basis charges that occurs precisely when $\gamma_i$ is about to leave the upper half of the $\mathcal{Z}$-plane under clockwise rotation, while $-\gamma_i$ is about to enter.

Now, suppose we rotate the upper half of the $\CZ$-plane by $180^\circ$ by performing a particular sequence of mutations that sweeps all central charges of the occupied BPS particles. After this rotation, every original BPS particle becomes an anti-particle, while every original anti-particles becomes a particle. This implies that the effect of this mutation sequence is to flip the signs of all basis charges simultaneously:
\begin{align}
    \left\{ \mathfrak{M}_{\s_\CL} \circ \cdots \circ \mathfrak{M}_{\s_2} \circ \mathfrak{M}_{\s_1} (\g_j) \,\middle|\, j = 1,\cdots, \rank(\G) \right\}
    = \left\{ -\g_1,-\g_2,\cdots,-\g_{\rank(\G)} \right\} \, .
\end{align}
where the sequence of mutations is applied along the nodes $\s_1\to \s_2 \to \cdots\to \s_\CL$. Since each mutation step corresponds to sweeping the central charge of the BPS particle that is about to leave the upper half $\CZ$-plane, one can determine the electromagnetic charges of all occupied BPS particles by reading off the charges $\tilde{\gamma}_i$ assigned to the node that is about to be mutated:
\begin{gather}
    \begin{gathered} \label{eq: mutated charges}
        \tilde{\g}_1 \equiv  \g_{\s_1} \, , \quad
        \tilde{\g}_2 \equiv \mathfrak{M}_{\s_1}(\g_{\s_2}) \, , \quad
        \tilde{\g}_3 \equiv \mathfrak{M}_{\s_2} \circ \mathfrak{M}_{\s_1}(\g_{\s_3}) \, , \cdots ,\\
        \tilde{\g}_k \equiv \mathfrak{M}_{\s_{k-1}} \circ \cdots \circ \mathfrak{M}_{\s_1}(\g_{\s_k}) \, , \quad \cdots \, , \quad
        \tilde{\g}_{\CL} \equiv \mathfrak{M}_{\s_{\CL-1}}\circ \cdots\circ \mathfrak{M}_{\s_1}(\g_{\s_\CL}) \, .
    \end{gathered}
\end{gather}
Thus, the number of BPS particles is $\CL$, and their central charges satisfy
\begin{align}
    \arg(\CZ_{\tilde{\g}_1})
    >
    \arg(\CZ_{\tilde{\g}_2})
    >
    \cdots
    >
    \arg(\CZ_{\tilde{\g}_\CL}) \, .
\end{align}
Consequently, the ordered set of BPS particle charges $\G_{\text{BPS}}$ for the trace formula is
\begin{align}
    \G_{\text{BPS}} = \{
    \tilde{\g}_\CL, \cdots, \tilde{\g}_2 , \tilde{\g}_1 ,\; -\tilde{\g}_\CL, \cdots, -\tilde{\g}_2 , -\tilde{\g}_1 
    \}\,.
\end{align}

To summarize, for a given 4d $\mathcal{N}=2$ theory that admits a BPS quiver, if one can find a sequence of mutations that flips the signs of all basis charges, then the inverse order of the charges $\tilde{\g}_i$ that are about to be mutated \eqref{eq: mutated charges} determines the charges of the occupied BPS particles and the ordering of their central charge phases, thereby characterizing the set $\G_{\text{BPS}}$. For clarity, let us consider an example.

\begin{figure}[tbp]
    \centering
    \includegraphics{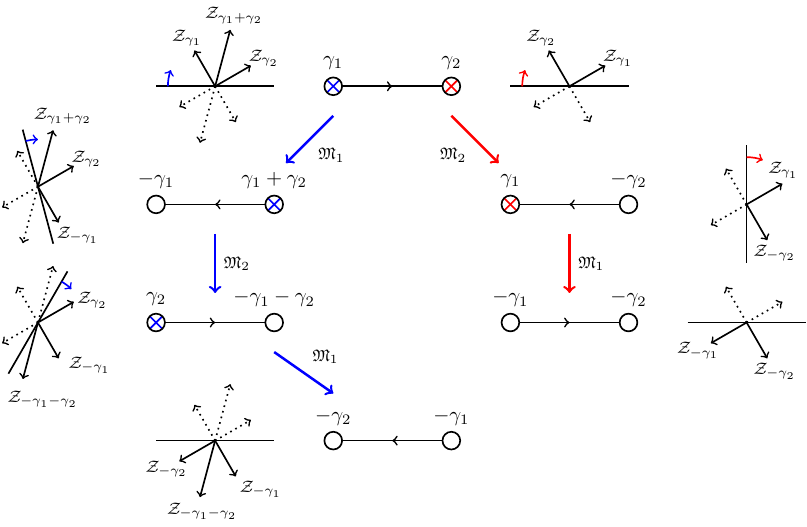}
    \caption{Two sequences of mutations that flip the signs of all basis charges of $(A_1,A_2)$ Argyres-Douglas theory. The sequence shown in blue corresponds to the chamber with three BPS particles, while the sequence shown in red corresponds to chamber with two BPS particles.}\label{fig: A1A2 mutation}
\end{figure}

\paragraph{Example: mutation of the $(A_1,A_2)$ theory}

Consider the BPS quiver of the $(A_1,A_2)$ Argyres-Douglas theory, which is the $A_2$ Dynkin diagram as depicted at the top of Figure~\ref{fig: A1A2 mutation}. As discussed in the previous subsection, this theory has two chambers containing two and three BPS particles as illustrated in Figure~\ref{fig: A2 chamber}. If we mutate the quiver with a sequence $\mathfrak{M}_1 \circ \mathfrak{M}_2 \circ \mathfrak{M}_1$, indicated in blue on the nodes, all basis charges flip their signs. This corresponds to a $180^\circ$ rotation of the upper half of the $\CZ$-plane that sweeps through the three occupied BPS particles whose central charge phases are ordered as
\begin{align}
    \arg(\CZ_{\g_2})
    <
    \arg(\CZ_{\g_1+\g_2})
    <
    \arg(\CZ_{\g_1})\,.
\end{align}
Thus, by simply reading off the assigned charges that are about to be mutated, we can determine $\G_{\text{BPS}}$ as in \eqref{eq: A2 3ptcl}. Similarly, the mutation sequence $\mathfrak{M}_1\circ \mathfrak{M}_2$ indicated in red on the nodes, also flips the signs of all basis charges which captures the ordering of central charge in the chamber with two BPS particles:
\begin{align}
    \arg(\CZ_{\g_1})
    <
    \arg(\CZ_{\g_2}) \, .
\end{align}
By reading the charges assigned to the nodes that are about to be mutated, we obtain $\G_{\text{BPS}}$ as in \eqref{eq: A2 2ptcl}.

To emphasize once more, if we identify a sequence of mutations that flips the signs of all basis charges, then the increasing order of the central charge phases is determined by reading off the charges assigned to the nodes that are about to be mutated, but in the reverse order of the mutation sequence. This data is sufficient for constructing the trace formula. Although we presented two such mutation sequences in this example, it is enough to find just one, since the trace formula does not depend on the choices of chambers.

\subsection{BPS quivers of \texorpdfstring{$(G,G')$}{(G,G')} Argyres-Douglas theories}

The main objective of this paper is to apply the trace formula to a family of 4d $ \mathcal{N}=2 $ Argyres-Douglas SCFTs labeled by two simply-laced Lie algebras $ \mathfrak{g} $ and $ \mathfrak{g}' $, or by their Dynkin diagrams $ G $ and $ G' $. These theories can be geometrically engineered by considering type IIB string theory on an isolated hypersurface singularity \cite{Cecotti:2010fi}
\begin{align}\label{eq: singularity}
    \big\{ (z_1,z_2,z_3,z_4) \mid W(z_1,z_2,z_3,z_4) = 0 \big\} \subset \mathbb{C}^4 \, ,
\end{align}
where $ W = W_{\mathfrak{g}}(z_1,z_2) + W_{\mathfrak{g}'}(z_3,z_4) $ with
\begin{gather}
    \begin{gathered}
        W_{A_n}(z,w) = z^2 + w^{n+1} \, , \quad
        W_{D_n}(z,w) = z^{n-1} + z w^{2} \, , \\
        W_{E_6}(z,w) = z^3 + w^4 \, , \quad
        W_{E_7}(z,w) = z^3 + zw^3 \, , \quad
        W_{E_8}(z,w) = z^3 + w^5\,.
    \end{gathered}
\end{gather}
Let us briefly review the quantities that characterize the $(G,G')$ theory. Due to the superconformal symmetry in 4d, there exists a $ \mathbb{C}^* $-action $ W(\lambda^{q_i} z_i) = \lambda W(z_i) $ with scaling weights $ q_i>0 $ satisfying $ \sum_i q_i>1 $ \cite{Gukov:1999ya, Shapere:1999xr}. A deformation of the hypersurface singularity is described by the smooth hypersurface
\begin{align}
    \widehat{W}(z_i) = W(z_i) + \sum_{a=1}^\mu t_a x_a \, ,
\end{align}
where $ \{x_a\} $ is a monomial basis of the Milnor ring defined as
\begin{align}\label{eq: Milnor}
    \mathcal{M}(W) = \mathbb{C}[z_1,z_2,z_3,z_4] / \mathcal{J} \, , \quad
    \mathcal{J} = \left\langle \frac{\partial W}{\partial z_1}, \cdots, \frac{\partial W}{\partial z_4} \right\rangle \, ,
\end{align}
and $ \mu = \dim \mathcal{M}(W) $ is called the Milnor number. We note that two hypersurface singularities $ \{ W_1=0\} $ and $ \{W_2=0\} $ are biholomorphically equivalent if and only if their Milnor rings $ \mathcal{M}(W_1) $ and $ \mathcal{M}(W_2) $ are isomorphic. This may lead to a non-trivial isomorphism between two Argyres-Douglas theories. For each monomial $ x_a $, we define the scaling dimension $ \Delta_a $ as
\begin{align}
    \Delta_a = \frac{1-q(x_a)}{\sum_{i=1}^4 q_i - 1} \, ,
\end{align}
where $ q(x_a) $ is the $ \mathbb{C}^* $-charge of the monomial $ x_a $ and $ 1-q(x_a) $ is the scaling dimension of the deformation parameter $ t_a $. Deformations with $ \Delta_a>1 $ correspond to Coulomb branch operators with scaling dimensions $ \Delta_a $. Deformations with $ \Delta_a < 1 $ are paired with deformations $ \Delta_b>1 $ satisfying $ \Delta_a+\Delta_b=2 $, and are interpreted as chiral deformations. In the case of $ \Delta_a=1 $, the corresponding deformations are associated with conserved currents of the flavor symmetry. From the Coulomb branch spectrum, the central charges $ a_{4d} $ and $ c_{4d} $ are given by \cite{Shapere:2008zf, Cecotti:2013lda}
\begin{align}
    \begin{aligned}
        a_{4d} &= \frac{1}{4}\sum_{\Delta_a>1} (\Delta_a-1) + \frac{r_{\mathfrak{g}} r_{\mathfrak{g}'} h_{\mathfrak{g}} h_{\mathfrak{g}'}}{24(h_{\mathfrak{g}}+h_{\mathfrak{g}'})} + \frac{5}{24}r  \, , \\
        c_{4d} &= \frac{r_{\mathfrak{g}} r_{\mathfrak{g}'} h_{\mathfrak{g}} h_{\mathfrak{g}'}}{12(h_{\mathfrak{g}}+h_{\mathfrak{g}'})} + \frac{1}{6}r \, ,
    \end{aligned}
\end{align}
where $ r_{\mathfrak{g}} $ and $ h_{\mathfrak{g}} $ are rank and Coxeter number of the simply-laced Lie algebra $ \mathfrak{g} $, respectively, and $ r $ is the dimension of the Coulomb branch, i.e., the number of deformations with $ \Delta_a>1 $.

Now, let us discuss on the BPS quiver of the $(G,G')$ theory that encodes the spectrum of the BPS particles on the Coulomb branch. The BPS quiver of $ (A_1, G) $ theory in the \emph{canonical chamber} is defined by the quiver associated with the Dynkin diagram $ G $, where every node is either a \emph{source} or a \emph{sink} as follows:
\begin{align}\label{eq: Dynkin}
    \begin{minipage}{0.46\linewidth}
        \begin{tabular}{cL{31ex}}
            $ A_{\mathrm{even}} $ & \includegraphics{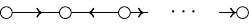} \\[1ex]
            $ A_{\mathrm{odd}} $ & \includegraphics{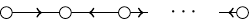} \\[1ex]
            $ D_{\mathrm{even}} $ & \includegraphics{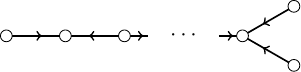} \\[1ex]
            $ D_{\mathrm{odd}} $ & \includegraphics{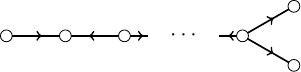}
        \end{tabular}
    \end{minipage}
    \hspace{1ex}
    \begin{minipage}{0.43\linewidth}
        \begin{tabular}{cL{31ex}}
            $ E_6 $ & \includegraphics{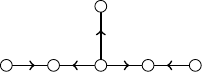} \\[1ex]
            $ E_7 $ & \includegraphics{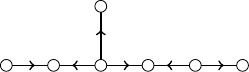} \\[1ex]
            $ E_8 $ & \includegraphics{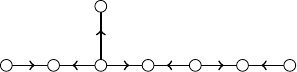}
        \end{tabular}
    \end{minipage}
\end{align}
The number of nodes in the BPS quiver $ G $ is equal to the rank of the Lie algebra $ \mathfrak{g} $.

\begin{figure}[tbp]
    \centering
    \begin{tabular}{C{25ex}cC{25ex}}
        \includegraphics{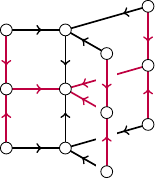} &
        $ \overset{\textcolor{purple}{\text{Flip}}}{\longrightarrow} $ &
        \includegraphics{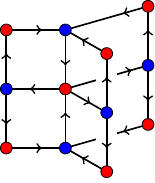} \\
        $ A_3 \otimes D_4 $ & & $ A_3 \sqprod D_4 $
    \end{tabular}
\caption{\label{fig: tensor} An example of the tensor product and the square product of the canonical quivers $A_3$ and $D_4$. The purple arrows in $A_3 \otimes D_4$ indicate the arrows belonging to the subquivers $\{I\}\otimes D_4$ and $A_3 \otimes \{J\}$, where $I$ and $J$ are sinks of $A_3$ and sources of $D_4$, respectively. By flipping these arrows, the square product quiver $A_3 \sqprod D_4$ is obtained whose nodes can be decomposed into two disjoint sets $\S_\pm$ as in \eqref{eq: disjoint set of square}. The nodes in $\S_+$ and $\S_-$ are colored red and blue, respectively.}
\end{figure}

To describe the BPS quivers of general $ (G, G') $ theories, we first introduce the tensor product of two Dynkin diagrams. If the BPS quivers $ G $ and $ G' $ have $ n $ and $ m $ nodes labeled by $ I $ and $ J $, respectively, then their tensor product quiver $ G \otimes G' $ has $ nm $ nodes labeled by pairs $ (I, J) $. The arrows of this quiver are determined by the Dirac pairing $ \langle \cdot, \cdot \rangle_{G\otimes G'} $ defined as
\begin{align}
    \langle \gamma_{(I_1,J_1)}, \gamma_{(I_2,J_2)} \rangle_{G \otimes G'} = \left\{ \begin{array}{ll}
            0 & \quad (I_1 \neq I_2 \text{ and } J_1 \neq J_2) \\
            \langle \gamma_{I_1}, \gamma_{I_2} \rangle_{G} & \quad (J_1=J_2) \\
            \langle \gamma_{J_1}, \gamma_{J_2} \rangle_{G'} & \quad (I_1=I_2) \, ,
    \end{array}\right.
\end{align}
where $ \gamma_{(I,J)} $ is the basis charge assigned to the $ (I, J) $-node of $ G\otimes G' $, and $ \langle \cdot, \cdot \rangle_{G} $ and $ \langle \cdot, \cdot \rangle_{G'} $ are the Dirac pairings associated with the BPS quivers $ G $ and $ G' $, respectively. Consequently, the subquivers $ G \otimes \{J\} $ for fixed $ J $ and $ \{I\} \otimes G' $ for fixed $ I $ inside $ G\otimes G' $ are isomorphic to $ G $ and $ G' $, respectively. An example of the tensor product quiver is illustrated on the left side of Figure~\ref{fig: tensor}.

\begin{figure}
    \centering
    \begin{subfigure}[b]{0.49\linewidth}
        \centering
        \includegraphics{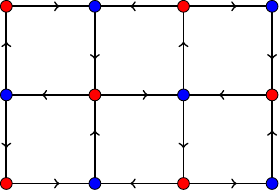}
        \caption{$ A_3 \sqprod A_4 $}
    \end{subfigure}
    \hfill
    \begin{subfigure}[b]{0.49\linewidth}
        \centering
        \includegraphics{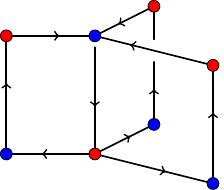}
        \caption{$ A_2 \sqprod D_4 $}
    \end{subfigure}
    \caption{Examples of the square product $G \sqprod G'$ of BPS quivers. The vertical and horizontal directions correspond to the Dynkin diagrams of $ G $ and $ G' $, respectively. Each of the four arrows surrounding a single plaquette circulates around it. In the depicted BPS quivers, the nodes $\sigma \in \Sigma_\pm$ are colored red and blue, respectively. Note that there are no arrows between any two nodes of the same color, which implies that the Dirac pairing between any two charges assigned to nodes of the same color is trivial.} \label{fig: BPS quiver of GG'} 
\end{figure}

The BPS quiver of the $ (G, G') $ Argyres-Douglas theory in the canonical chamber is given by the \emph{square product} of the two Dynkin diagrams $ G $ and $ G' $, denoted by
\begin{align}
    G \sqprod G' \, .
\end{align}
This quiver is obtained from the tensor product quiver $ G \otimes G' $ by inverting all arrows in the subquivers $\{I\}\otimes G'$ and $G\otimes\{J\}$ whenever the $I$-th node is a sink in $G$ and $J$-th node is a source in $G'$. Equivalently, $ G \sqprod G' $ is the quiver whose vertical and horizontal directions correspond to the Dynkin diagrams $ G $ and $ G' $, respectively, with arrows circulating around each square plaquette. An example comparing the tensor product quiver and the square product quiver is depicted in Figure~\ref{fig: tensor}, and additional examples of square product quivers are presented in Figure~\ref{fig: BPS quiver of GG'}.

Note that each node in $G \sqprod G'$ is either a sink of $G$ and a source of $G'$, or a source of $G$ and a sink of $G'$. Accordingly, the nodes can be decomposed into two disjoint sets as
\begin{align}
    \begin{aligned}
        \S+ &= \big\{ (I,J) \mid I \in \text{sink of } G , \ J \in \text{source of }G'  \big\} \, , \\
        \S_- &= \big\{ (I,J) \mid I \in \text{source of } G , \ J \in \text{sink of } G' \big\} \, .
    \end{aligned}
    \label{eq: disjoint set of square}
\end{align}
In the square product quivers given in Figure~\ref{fig: tensor} and Figure~\ref{fig: BPS quiver of GG'}, the nodes in $ \Sigma_+ $ and $ \Sigma_- $ are depicted in red and blue, respectively. Let us define two sequences of mutations as
\begin{align}
	\mathfrak{M}_+ \equiv \prod_{\s\in\S_+} \mathfrak{M}_\s \, , \qquad
	\mathfrak{M}_- \equiv \prod_{\s\in\S_-} \mathfrak{M}_\s \, .
\end{align}
Since there is no arrows between any two nodes within $ \Sigma_+ $ or within $ \Sigma_- $, the ordering of the product in each mutation sequence is irrelevant. By combining $ \mathfrak{M}_\pm $ in an alternating manner, we define the mutation $ \mathfrak{M}^{(G,G')} $ as
\begin{align} \label{eq: (G,G') mutation}
    \mathfrak{M}^{(G,G')}
    \equiv
    \underbrace{
        \cdots
        \mathfrak{M}_-
        \circ
        \mathfrak{M}_+
        \circ
        \mathfrak{M}_-
    }_{2|\Delta^+(\mathfrak{g})|\,/\,|\Pi(\mathfrak{g})|}  \, ,
\end{align}
where $|\D^+(\mathfrak{g})|$ and $|\Pi(\mathfrak{g})|$ are the numbers of positive roots and simple roots of the Lie algebra $\mathfrak{g}$ respectively, so that the total number of individual mutation steps is $|\D^+(\mathfrak{g})| \times |\P(\mathfrak{g}')|$. One can verify that this mutation $\mathfrak{M}^{(G,G')}$ corresponds to a $180^\circ$ rotation of the upper half $\CZ$-plane of the $(G,G')$ Argyres-Douglas theory in a chamber containing $|\D^+(\mathfrak{g})| \times |\P(\mathfrak{g}')|$ BPS particles.\footnote{There is another finite chamber containing $|\P(\mathfrak{g})| \times |\D^+(\mathfrak{g}')|$ BPS particles, which can be captured from a sequence of mutations constructed in a similar manner to that in \eqref{eq: (G,G') mutation}, but starting with $\mathfrak{M}_+$, and with the numbers of $\mathfrak{M}_\pm$ given by $2\frac{|\D^+(\mathfrak{g}')|}{|\P(\mathfrak{g}')|}$. This follows from the trivial isomorphism under the exchange $G \leftrightarrow G'$.} In particular, all the charge vectors $ \gamma_\sigma $ corresponds to the node of $G \sqprod G'$ flip their signs under the mutation,
\begin{align}
	\left\{
		\mathfrak{M}^{(G,G')} (\g_\s) \,\middle|\,
		\s \in \S_+ \cup  \S_- 
	\right\}
    =
    \left\{ -\g_\s \,\middle|\, \s \in \S_+\cup\S_- \right\} \, ,
\end{align}
and as illustrated in \eqref{eq: mutated charges}, the increasing order of the phases of the central charges can be determined by tracing the inverse sequence of individual mutations appearing in $\mathfrak{M}^{(G,G')}$.

\begin{figure}[tbp]
    \centering
    $ \displaystyle \vcenter{\hbox{\includegraphics{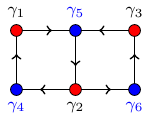}}}
    \overset{\mathfrak{M}_-}{\textcolor{blue}{\longrightarrow}} \!\!\!
    \vcenter{\hbox{\includegraphics{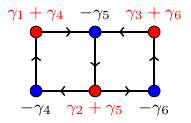}}} \!\!\!
    \overset{\mathfrak{M}_+}{\textcolor{red}{\longrightarrow}} \!\!\!
    \vcenter{\hbox{\includegraphics{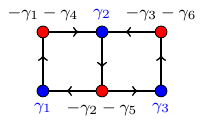}}} \!\!\!
    \overset{\mathfrak{M}_-}{\textcolor{blue}{\longrightarrow}}
    \vcenter{\hbox{\includegraphics{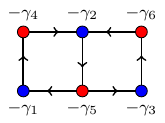}}} $ 
\caption{\label{fig: A2A3 mutation} An example for the mutation $\mathfrak{M}^{(A_2,A_3)}=\mathfrak{M}_-
    \circ
    \mathfrak{M}_+
    \circ
    \mathfrak{M}_-$ that rotates the upper-half $\CZ$-plane by $180^\circ$. We colored charges that are about to be mutated in blue for $\mathfrak{M}_-$ and red for $\mathfrak{M}_+$, respectively. As in the right-most quiver, all the signs of basis charges are flipped compared to the original basis charges in the left-most quiver. }
\end{figure}

As an example, consider the $(A_2,A_3)$ Argyres-Douglas theory whose BPS quiver is shown on the left-most side of Figure~\ref{fig: A2A3 mutation}. We assign the charges $\g_1,\g_2,\g_3$ and $\g_4,\g_5,\g_6$ to the nodes in the sets $\S_+$ and $\S_-$ respectively. We then define the mutations $ \mathfrak{M}_\pm $ as
\begin{align}
    \mathfrak{M}_+  = 
    \mathfrak{M}_1 \circ
    \mathfrak{M}_2 \circ
    \mathfrak{M}_3 \, , \quad
    \mathfrak{M}_-  = 
    \mathfrak{M}_4 \circ
    \mathfrak{M}_5 \circ
    \mathfrak{M}_6\,.
\end{align}
From \eqref{eq: (G,G') mutation}, the mutation defined as
\begin{align}
    \mathfrak{M}^{(A_2,A_3)}
    =
    \mathfrak{M}_-
    \circ
    \mathfrak{M}_+
    \circ
    \mathfrak{M}_- \, ,
\end{align}
flips all charges $ \gamma_i $ in the BPS quiver $ A_2 \sqprod A_3 $:
\begin{alignat}{3}
    \begin{aligned}
        &\mathfrak{M}^{(A_2,A_3)} (\g_1) = -\g_4 \, , \quad
        &&\mathfrak{M}^{(A_2,A_3)} (\g_2) = -\g_5 \, , \quad
        &&\mathfrak{M}^{(A_2,A_3)} (\g_3) = -\g_6 \, , \\
        &\mathfrak{M}^{(A_2,A_3)} (\g_4) = -\g_1 \, , \quad
        &&\mathfrak{M}^{(A_2,A_3)} (\g_5) = -\g_2 \, , \quad
        &&\mathfrak{M}^{(A_2,A_3)} (\g_6) = -\g_3 \, ,
    \end{aligned}
\end{alignat}
The details of the mutation procedure are illustrated in Figure~\ref{fig: A2A3 mutation}. This implies that the upper-half of the $\CZ$-plane is rotated by $180^\circ$. Thus, by reading the charges that are about to be mutated in the inverse order, which are shown in blue or red in the Figure~\ref{fig: A2A3 mutation}, we obtain the increasing order of the phases of the central charges as
\begin{gather}
    \begin{gathered}
        \arg(\CZ_{\g_1}) <
        \arg(\CZ_{\g_2}) <
        \arg(\CZ_{\g_3}) <
        \arg(\CZ_{\g_1+\g_4})
        <\arg(\CZ_{\g_2+\g_5}) \\
        <\arg(\CZ_{\g_3+\g_6}) <
        \arg(\CZ_{\g_4}) <
        \arg(\CZ_{\g_5}) <
        \arg(\CZ_{\g_6})\,.
    \end{gathered}
\end{gather}
Consequently, the trace formula for the $(A_2,A_3)$ Argyres-Douglas theory can be written as
\begin{align}
    \begin{aligned}
    Z_{S_b^3}^{(A_2,A_3)} = \Tr
    \Big(&
    \Phi_b(x_{\g_1})
    \Phi_b(x_{\g_2})
    \Phi_b(x_{\g_3})
    \Phi_b(x_{\g_1}+x_{\g_4})
    \Phi_b(x_{\g_2}+x_{\g_5}) \\
    &\times
    \Phi_b(x_{\g_3}+x_{\g_6})
    \Phi_b(x_{\g_4})
    \Phi_b(x_{\g_5})
    \Phi_b(x_{\g_6})
    \times (x_{\g_i} \to -x_{\g_i})
    \Big)\,.
    \end{aligned}
\end{align}
The detailed evaluation of the trace formula is presented in Appendix~\ref{app: embed}. For a general $(G,G')$ Argyres-Douglas theory with BPS quiver $G \sqprod G'$, the mutation $\mathfrak{M}^{(G,G')}$ provides the data required to write down the trace formula. We will utilize this mutation sequence in the next section.

\section{3D TFTs from 4D \texorpdfstring{$(G,G')$}{(G,G')} theories and their boundary VOAs} \label{sec: GG}

In this section, we discuss the 3d TFTs arising from the $ U(1)_r $ twisted compactification of the 4d $ (G, G') $ Argyres-Douglas theories, and their boundary VOAs. We begin by reviewing the 2d VOAs associated with the $ (G, G') $ Argyres-Douglas theories via the SCFT/VOA correspondence. We then present explicit examples of the 3d theories arising from these Argyres-Douglas theories using the trace formula. The ellipsoid partition function obtained from the trace formula provides a 3d $\CN=2$ Abelian Chern-Simons matter theory description, which in turn yields to a 3d TFT upon topological twisting. Finally, we extract the modular data of the boundary VOA using the 3d A-model techniques from the ellipsoid partition function.

\subsection{2d VOAs from \texorpdfstring{$ (G,G') $}{(G,G')} Argyres-Douglas theories}

There are many known results and conjectures concerning the VOAs corresponding to the Schur sectors of the Argyres-Douglas theories \cite{Xie:2016evu, Song:2017oew, Creutzig:2017qyf, Beem:2017ooy, Xie:2019yds, Xie:2019zlb, Xie:2019vzr}. In this section, we briefly summarize the VOAs associated with the $ (A_{k-1},G) $ Argyres-Douglas theories that are relevant in this paper. These theories belong to the family of Argyres-Douglas theories engineered by compactifying 6d $ \mathcal{N}=(2,0) $ $G$ theories on a Riemann sphere with an irregular singularity, and are also referred to as the $ G^{h^\vee}[k] $ theories \cite{Wang:2015mra}, where $ h^\vee $ is the dual Coxeter number of the Lie algebra $ \mathfrak{g} $ associated with the Dynkin diagram $ G $. In the case $ \gcd(k,h^\vee) = 1 $, the associated VOA is conjectured to be the W-algebra given by
\begin{align}
    W^{k'}(\mathfrak{g}) \, , \quad k' = -h^\vee + \frac{h^\vee}{h^\vee+k} \, .
\end{align}
This W-algebra is defined by the quantum Drinfeld-Sokolov reduction of the affine Kac-Moody algebra \cite{Drinfeld:1984qv, Feigin:1990pn, Kac:2003mjg}. A useful theorem \cite{Arakawa:2018iyk} for our purposes states that this W-algebra is a rational VOA and, moreover, admits a coset construction given by
\begin{align}\label{eq:Walg-coset}
    W^{k'}(\mathfrak{g}) \cong \frac{\hat{\mathfrak{g}}_l \oplus \hat{\mathfrak{g}}_1}{\hat{\mathfrak{g}}_{l+1}} \, , \quad
    l = -h^\vee + \frac{h^\vee}{k} \, ,
\end{align}
when $ \gcd(k,h^\vee)=1 $. Here, $ \hat{\mathfrak{g}}_l $ denotes the affine Kac-Moody algebra at level $ l $. When  $ l $ is a positive integer, $ \hat{\mathfrak{g}}_l $ corresponds to the vertex operator algebra of the WZW-model, which is unitary and rational. In contrast, a negative fractional level $ l $ of the form appearing in \eqref{eq:Walg-coset} is called an \emph{admissible level} and defines a non-rational, logarithmic VOA \cite{Kac:1988qc, Kac:1989, Gaberdiel:2001ny}. Nevertheless, the coset VOA in \eqref{eq:Walg-coset} is a (non-unitary) rational VOA with central charge given by
\begin{align}
    c_{2d} = \frac{l (l+2h^\vee+1) \dim \mathfrak{g}}{(l+h^\vee)(h^\vee+1)(l+h^\vee+1)} \, .
\end{align}
For example, for the $ (A_{k-1},A_1) $ theories, where $ \mathfrak{g}=\mathfrak{su}(2) $, the expression \eqref{eq:Walg-coset} realizes a coset representation of the Virasoro minimal models $ M(2,2+k) $. For the $ (A_{k-1}, A_{n-1}) $ theories, with $ \mathfrak{g}=\mathfrak{su}(n) $, it yields the coset representation of the W-algebra minimal model $ W_{n}(n,n+k) $. In case of the $ (A_{k-1},A_{n-1}) $ theories, as well as the $ (A_1, G) $ theories, this conjecture is supported by explicit computations of the Schur indices of the Argyres-Douglas theories and the vacuum characters of the corresponding W-algebras \cite{Song:2017oew, Xie:2019zlb}. A brief review of affine Kac–Moody algebras and the modular data of these coset models is provided in Appendix~\ref{app:voa}. The rationality of the associated VOAs is also expected from the SCFT/VOA correspondence, since the corresponding Argyres-Douglas theories have no flavor symmetry and possess a trivial Higgs branch when $ \gcd(k,h^\vee)=1 $.\footnote{More precisely, the absence of a Higgs branch indicates the associated variety of the corresponding VOA is trivial \cite{Beem:2017ooy, Song:2017oew}. Such a VOA is called \emph{$ C_2 $ cofinite}, or \emph{lisse} \cite{Zhu:1996gaq}.}

When $ \gcd(k, h^\vee) \neq 1 $, there are three possibilities. First, the theory may be related to another $ (A_{k'-1}, G') $ theory whose $ k' $ and the dual Coxeter number of $ G' $ are relatively prime, via a non-trivial 4d isomorphism. Such an isomorphism can be checked by comparing the defining hypersurface equations \eqref{eq: singularity} and their Milnor rings \eqref{eq: Milnor}. For instance, although the $ (A_1, E_6) $ theory does not satisfy the coprime condition $ \gcd(k,h^\vee) = \gcd(2,12) \neq 1 $, its defining hypersurface equation \eqref{eq: singularity} is identical to that of the $ (A_2, A_3) $ theory, which does satisfy $ \gcd(k,h^\vee)=1 $. In a later part of this section, we will further confirm that the trace formula and 3d TFT perspective are also consistent with such isomorphisms. In these cases, consequently, the conjecture \eqref{eq:Walg-coset} for the associated vertex operator algebra remains applicable up to the 4d isomorphism.

Second, the Argyres-Douglas theory may possess a non-trivial flavor symmetry. In fact, the $ (A_{k-1},G) $ theories have no flavor symmetry if
\begin{gather}
    \begin{gathered}\label{eq: noflavor}
        (A_{k-1},A_{n-1}) \ : \ \gcd(n, k) = 1 \, , \quad
        (A_{k-1},D_n) \ : \ k \notin 2\mathbb{Z} \, , \\
        (A_{k-1},E_6) \ : \ k \notin 3\mathbb{Z} \, , \quad
        (A_{k-1},E_7) \ : \ k \notin 2\mathbb{Z} \, , \quad
        (A_{k-1},E_8) \ : \ k \notin 30\mathbb{Z} \, .
    \end{gathered}
\end{gather}
When a non-trivial flavor symmetry is present, the associated VOA is no longer rational. For instance, the VOAs associated with $ (A_1, D_{2n+1}) $ theories are the affine Kac-Moody algebras $ \widehat{\mathfrak{su}}(2)_{-4n/(2n+1)} $, which are not rational \cite{Beem:2017ooy}. The non-rational logarithmic VOAs associated with $ (A_1, A_{2n+1}) $ and $ (A_1, D_{2n}) $ theories have also been identified \cite{Creutzig:2017qyf}, but the VOAs corresponding to general $ (A_{k-1},G) $ theories with a non-trivial flavor symmetry are still unknown. Although these VOAs are not rational, it is conjectured that the 2d VOAs associated with 4d $ \mathcal{N}=2 $ SCFTs are \emph{quasi-lisse} \cite{Beem:2017ooy}, meaning that they have finitely many characters of primary fields which transform covariantly under the $ \mathrm{SL}(2,\mathbb{Z}) $ transformation even if the VOA itself is non-rational. Affine Kac-Moody algebras at admissible levels are examples of the quasi-lisse VOAs. This implies that one can still define finite-dimensional modular $ S $- and $ T $-matrices for the VOAs associated with the Argyres-Douglas theories with a non-trivial Higgs branch. Hence, the trace formula and 3d TFT perspective are also applicable in this case.

Lastly, it is also possible that an Argyres-Douglas theory has a trivial Higgs branch but satisfies $ \gcd(k,h^\vee) \neq 1 $ and is not connected to any other $ (A_{k-1},G) $ type theories obeying the coprime condition through 4d isomorphisms. In this case, a naive application of conjecture \eqref{eq:Walg-coset} is unclear, since the level $ l $ of the affine Kac-Moody algebra appearing in the coset construction becomes non-admissible, and the representation theory of affine Kac-Moody algebras at non-admissible levels is not well understood. For certain special Argyres-Douglas theories, such as the $ (A_2, D_4) $ theory, the corresponding VOAs have been identified, but they turn out to be logarithmic rather than rational \cite{Buican:2016arp, Jiang:2024baj, Pan:2025gzh}. We observe that the $ (A_{k-1},G) $ Argyres-Douglas theories in this class contain a Coulomb branch operator with an integer conformal dimension, which complicates the application of the trace formula and the 3d TFT construction in our context. We will briefly comment on this point in the next subsection and will not discuss this case in detail.

\subsection{Examples} \label{sec: ex}

Now, we present examples of the 3d TFTs that interpolate between the 4d $\CN=2$ $(G,G')$ Argyres-Douglas theories and their 2d VOAs. Our machinery is the trace formula \eqref{eq: trace formula} that computes the 3d ellipsoid partition function of the 3d TFT, as discussed in the previous section. From the expression of the partition function, we read a 3d $\CN=2$ Abelian Chern-Simons matter (ACSM) theory description which gives rise to the 3d TFT upon monopole superpotential deformation and topological twisting. The general form of the partition function can be written as
\begin{align}
    Z_{S_b^3}^{(G,G')} \simeq
    \int d^l u\, e^{\pi i \, u^T L\,u } \prod_{I=1}^N \Phi_b\Big(Q_I^T\cdot u\Big ) \, ,
    \label{eq: general ellipsoid ptf}
\end{align}
where $u = (u_1 ,\cdots, u_l)^T$ is the 3d Coulomb branch parameter for the $U(1)^l$ gauge group, and $\simeq$ denotes equality up to a $b$-dependent overall phase that originates from the gravitational CS term and background CS level of the $U(1)_R$ symmetry, which we will ignore. One can also consider cases with background gauge fields for flavor symmetries by omitting the corresponding integrals.

For such a given integral expression \eqref{eq: general ellipsoid ptf}, the effective 3d $\CN=2$ description is characterized by a $l\times N$ charge matrix $Q$ for the $N$ chiral multiplets, where $Q_I \equiv (Q_{1I},\cdots,Q_{lI})^T$, and an effective CS level matrix,
\begin{align}
    K = L + QQ^T\,.
\end{align}
From these data , one can identify a finite set of half-BPS monopole operators that are turned on as superpotential deformations. After this deformation, the 3d $\CN=2$ ACSM theory is expected to flow either to a rank-0 SCFT or directly to a unitary 3d TFT. In the former case, the resulting SCFT has at least $\CN=4$ supersymmetry, which allow us to perform topological A/B-twists to obtain a pair of non-unitary 3d TFTs.

By employing the 3d A-model technique \cite{Closset:2018ghr}, we extract partial data of the modular $S$- and $T$-matrices of the resulting 3d TFTs,
\begin{align}
    \big\{
        |S_{0\a}|
    \big\}
    \, , \
    \big\{
        T_{\a\a}
    \big\} \, .
\end{align}
See Appendix~\ref{app: A-model} for a review of the detailed computation. For convenience, we introduce following notations:
\begin{align}\label{eq: sincos-def}
    \s_n^j \equiv \sin( \frac{\pi i }{n}j)\,,\quad
    \xi_n^j \equiv \cos( \frac{\pi i }{n}j)\,,\quad
    {\tt e}(s) \equiv \exp( 2\pi i s)\, , \quad
    \zeta_n \equiv \exp(\frac{2\pi i}{n}) \, .
\end{align}
Remarkably, we confirm that the extracted partial modular data are consistent with those of the expected VOAs associated with the $(G,G')$ Argyres-Douglas theories. In particular, the expected 2d VOA can be supported on the holomorphic boundary of the resulting 3d TFT. This provides highly non-trivial evidences for the existence of 3d TFTs that consistently interpolate between 4d SCFTs and 2d VOAs. Furthermore, the partial modular data for a previously unexplored case from the $(A_3,D_4)$ theory, which is isomorphic to the $(A_2,E_6)$ theory, can also be computed from the 3d TFT perspective, demonstrating the usefulness of our approach.
We summarize in Table~\ref{table: result} the Argyres-Douglas theories and their corresponding VOAs obtained from the topological A- and B-twists of the associated 3d $ \mathcal{N}=4 $ theories considered in this paper.
\begin{table}
    \centering
    \renewcommand{\arraystretch}{1.2}
    \begin{tabular}{c|C{28ex}|C{28ex}|c}
        $ (G, G') $ & VOA (A-twist) & VOA (B-twist) & Isomorphism \\ \hline
        $ (A_2, A_6) $ & $ W^{-\frac{27}{10}}(\mathfrak{su}(3)) = W_3(3,10) $ & $ \varphi_9 \in \Gal(\mathbb{Q}(\zeta_{10})/\mathbb{Q}) $ \\
        $ (A_2,A_7) $ & $ W^{-\frac{30}{11}}(\mathfrak{su}(3)) = W_3(3,11) $ & $ \varphi_5, \varphi_{27} \in \Gal(\mathbb{Q}(\zeta_{44})/\mathbb{Q}) $ \\
        $ (A_3,A_4) $ & $ W^{-\frac{32}{9}}(\mathfrak{su}(4)) = W_4(4,9) $ & $ \varphi_{13} \in \Gal(\mathbb{Q}(\zeta_{18})/\mathbb{Q}) $ & \\
        $ (A_3,A_6) $ & $ W^{-\frac{40}{11}}(\mathfrak{su}(4)) = W_4(4,11) $ & $ \varphi_9, \varphi_{31} \in \Gal(\mathbb{Q}(\zeta_{44})/\mathbb{Q}) $ \\
        $ (A_2,D_5) $ & $ W^{-\frac{80}{11}}(\mathfrak{so}(10)) $ & $ \varphi_5, \varphi_{27} \in \Gal(\mathbb{Q}(\zeta_{44})/\mathbb{Q}) $ \\
        $ (A_3,D_4) $ & \multicolumn{2}{c|}{(unknown, non-rational)} & $ (A_2,E_6) $ \\
        $ (A_2, E_8) $ & $ W^{-\frac{60}{11}}(\mathfrak{so}(8)) $ & $ \varphi_{15} \in \Gal(\mathbb{Q}(\zeta_{22})/\mathbb{Q}) $ & $ (A_4,D_4) $
    \end{tabular}
    \caption{Argyres-Douglas theories and their corresponding VOAs considered in this paper. The $ (A_3, D_4) $ theory has a non-trivial flavor symmetry, and its corresponding VOA is not yet known. In all other cases, the 4d theory has a trivial Higgs branch. The VOAs obtained from the A-twist correspond to \eqref{eq:Walg-coset}. The VOAs obtained from the B-twist are Galois conjugate of the VOAs appearing in the A-twist, and we present the Galois automorphism $ \varphi_n \in \Gal(\mathbb{Q}(\zeta_N)/\mathbb{Q}) $, $ \varphi_n(\zeta_N) = \zeta_N^n $ that relates the VOAs arising from A-twist and B-twist. Here, $ \Gal(\mathbb{Q}(\zeta_N)/\mathbb{Q}) $ is isomorphic to the multiplicative group $ \mathbb{Z}_N^\times=\{n \mid 0 \leq n < N, \, \gcd(n,N)=1\} $.}\label{table: result}
\end{table}

Let us make some comments. As noted in the previous section, it is technically crucial to sufficiently simplify the trace formula using the Faddeev's quantum dilogarithm identities. However, there is no known systematic or algorithmic procedure for doing this. Nonetheless, we have developed a powerful machinery, a {\tt Mathematica} code, that accomplishes this task and produces a drastically simplified integral expression for the partition function. We will utilize this machinery throughout the examples, and Appendix~\ref{app: code} provides detailed instructions for its use.

Secondly, from the construction of the $(G,G')$ theory \eqref{eq: singularity}, one can see an obvious isomorphism between a pair of Argyres-Douglas theories,
\begin{align}
    (G,G') \sim (G',G) \, ,
    \label{eq: trivial duality}
\end{align}
as well as more non-trivial isomorphisms such as \cite{Cecotti:2010fi,Xie:2019vzr}
\begin{gather}
    \begin{gathered}
        (A_2,A_2) \sim (A_1,D_4) \, , \quad
        (A_2,A_3) \sim (A_1,E_6) \, , \quad
        (A_2,A_4) \sim (A_1,E_8) \, ,\\
        (A_4,D_4) \sim (A_2,E_8) \, , \quad
        (A_3,E_8) \sim (A_4,E_6) \, ,
        \label{eq: AD isomorphism}
    \end{gathered}
\end{gather}
where $ \sim $ denotes an isomorphism. These can be verified from the defining hypersurface equation \eqref{eq: singularity} and corresponding Milnor rings \eqref{eq: Milnor}. For the rest of this section, we will not redundantly check both sides of the trivial isomorphism \eqref{eq: trivial duality}, but deal with one representative. We also omit the cases that are isomorphic to $(A_1,G)$ type theories, as these have already analyzed in \cite{Go:2025ixu}. Nevertheless, we provide explicit checks of the isomorphisms at the level of the ellipsoid partition function in Appendix~\ref{app: ex}. This is a non-trivial check because the trace formula expressions are quite different for the isomorphic pairs.

Lastly, we restrict out attention to cases in which all Coulomb branch operators have non-integer conformal dimensions. If there is a Coulomb branch operator with an integer conformal dimension, there will be a local operator in the resulting 3d TFT which survives under the $U(1)_r$ twisted reduction. The resulting 3d TFT is non-semisimple, and we currently lack the tools to handle it.

\subsubsection{\texorpdfstring{$(A_2,A_6)$}{(A2,A6)}}

Consider the 4d $\CN=2$ $(A_2,A_6)$ Argyres-Douglas theory whose central charge is $c_{4d} = \frac{31}{10}$. The BPS quiver is given as the square product of Dynkin diagrams $A_2 \sqprod A_6$ as discussed in the previous section:
\begin{align}
    \begin{aligned}
        \includegraphics{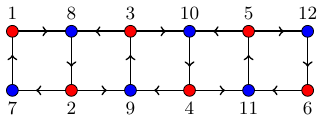}
    \end{aligned}
\end{align}
The sequence of mutations
\begin{align}
    \mathfrak{M}^{(A_2,A_6)} = 
    \mathfrak{M}_- \circ
    \mathfrak{M}_+ \circ
    \mathfrak{M}_-
\end{align}
captures a finite chamber containing 18 BPS particles together with their CPT conjugate anti-particles. The corresponding electromagnetic charges are enumerated in increasing order of their central charge phases $\arg(\CZ_\g)$ as
\begin{align}
    \begin{aligned}
        \G_{\text{BPS}} &= \big\{
            \g_1,\
            \g_2,\
            \g_3,\
            \g_4,\
            \g_5,\
            \g_6,\
            \g_1+\g_7,\
            \g_2+\g_8,\
            \g_3+\g_9,\
            \g_4+\g_{10}, \\
            &\qquad
            \g_5+\g_{11},\
            \g_6+\g_{12},\
            \g_7,\
            \g_8,\
            \g_9,\
            \g_{10},\
            \g_{11},\
            \g_{12},\
            (\g_i \to -\g_i)
        \big\} \, .
    \end{aligned}
\end{align}
Thus, the trace formula reads
\begin{align}
    \begin{aligned}
        Z_{S_b^3}^{(A_2,A_6)} &= \Tr \big(
            \Phi_b(x_1) \Phi_b(x_2) \cdots \Phi_b(-x_{11}) \Phi_b(-x_{12}) 
        \big) \\
        &= i^{1/2} e^{\frac{23 \pi i}{12}(b^2 + b^{-2}) } \int
        d^5 u \;
        e^{\pi i u^T L  u}
        \prod_{i=1}^5 \Phi_b(u_i) \, ,
        \label{eq: A2A6 int}
    \end{aligned}
\end{align}
where we use identities of the quantum dilogarithm and convert the trace into an integral expression by inserting the quantum mechanical completeness relation. This can be implemented systematically via our {\tt Mathematica} code \cite{code} whose instructions are provided in Appendix~\ref{app: code}. The CS term contributions in \eqref{eq: A2A6 int} can be organized as the symmetric matrix
\begin{align}
    L = \left(
    \begin{array}{ccccc}
        1 & 2 & -1 & 0 & -1 \\
        2 & 3 & -1 & 1 & -2 \\
        -1 & -1 & 1 & 0 & 2 \\
        0 & 1 & 0 & 0 & 1 \\
        -1 & -2 & 2 & 1 & 3
    \end{array}
    \right) \, ,
\end{align}
with five quantum dilogarithms. Hence, the resulting partition function $Z_{S_b^3}^{(A_2,A_6)}$ crresponds to a 3d $\CN=2$ gauge theory of $U(1)^5$ gauge group with 5 chiral multiplets, whose charge matrix $Q$ and CS level matrix $K$ are given by
\begin{align}
    Q={\bf 1}_{5\times 5} \, , \quad
	K= L + Q Q^T\,.
    \label{eq: A2A6 QK}
\end{align}
Since the 4d $(A_2,A_6)$ theory does not have non-trivial Higgs branch and all Coulomb branch operators have purely fractional conformal dimensions, the $U(1)_r$ twisted circle compactification lifts the Coulomb branch, resulting in a 3d rank-0 theory. Indeed, one can find four half-BPS monopole operators in the 3d ACSM description as
\begin{align}
	\phi_5^2\, V_{(-1,1,0,-1,0)} \, , \quad
    \phi_2^2\, V_{(0,0,-1,-1,1)} \, , \quad
    \phi_1 \phi_4\, V_{(0,0,2,0,-1)} \, , \quad
    \phi_3 \phi_4\, V_{(2,-1,0,0,0)} \, ,
    \label{eq: A2A6 monopole}
\end{align}
where $\phi_i$ is the scalar field in the $i$-th chiral multiplet charged under the $i$-th $U(1)$ gauge group, and $V_\mathfrak{m}$ is a bare-monopole operator with magnetic flux $\mathfrak{m}$. See Appendix~\ref{app: A-model} for the procedure to identify the half-BPS monopole operators.

Note that the monopole operators \eqref{eq: A2A6 monopole} are gauge invariant. Turning on a superpotential deformation with these operators breaks the $U(1)_{T_i}$ topological symmetries. As a result, only a single linear combination given by 
\begin{align}
	A = T_1 + 2 T_2 + T_3 + T_4 + 2 T_5 \,
\end{align}
unbroken in the infrared.\footnote{For the rank-0 theories from our construction, we cannot determine the overall sign of the $U(1)_A$ symmetry generator $A$. In this paper, we fix the sign in such a way that the choice of the mixing parameter $\n=-1$ of the R-symmetry $R+\n  A$ corresponds to the A-twist, i.e., the one that reproduces the modular data of the desired VOA.} This combination is identified with the $U(1)_A$ axial symmetry of the $\CN=4$ supersymmetry, defined as
\begin{align}
    A = J_3^C - J_3^H\,,
\end{align}
where $J_3^C$ and $J_3^H$ are the Cartan generators of the $\CN=4$ R-symmetry group $SO(4)_R = SU(2)_C \times SU(2)_H$ with normalization $J_3^{C},J_3^{H} \in \frac{1}{2}\mathbb{Z}$. As supporting evidence, we determine the conformal fixed point from the F-maximization computation, which fixes the mixing of the topological symmetries according the setup \eqref{eq: mu}, yielding $\m^* = (-1,-2,-1,-2,-2)$, and all the gauge invariant BPS states have integer quantized R-charges which is a strong signal of $\CN=4$ supersymmetry. 

The 3d superconformal index at this fixed point can be evaluated as
\begin{align}
    \begin{aligned}
        \CI_{S^2 \times S^1}^{(A_2,A_6)}
        (\eta,\n=0;q)
        &= 1 - q + \left( \eta^2 + \frac{1}{\eta^2}\right) q^2 + \left( \eta + \frac{1}{\eta} \right)q^{5/2} + q^3 -\left( \eta + \frac{1}{\eta} \right)q^{7/2} \\
        & +\left(\eta^4 - 4 \eta^2 - 6 - \frac{4}{\eta^2} + \frac{1}{\eta^4} \right) q^4 - \left( 13\eta + \frac{13}{\eta} \right) q^{9/2} + \mathcal{O}(q^5) \, ,
    \end{aligned}
\end{align}
where $\eta$ denotes the fugacity of the $U(1)_A$, and $\n$ is the parameter appearing the definition \eqref{eq: SCI def} as
\begin{align}
    R_\n \equiv R + \n A \,,
\end{align}
which controls the mixing between $U(1)_A$ and the $U(1)_R$ R-symmetry in the $\CN=2$ description with $R = J_3^C + J_3^H$. The value $\n=0$ corresponds to the conformal point. We observe that the extra supercurrent multiplet contribution $-(\eta + \eta^{-1})q^{3/2}$ required for the $\CN=4$ enhancement seems to be canceled by other contributions. Similar phenomena were previously observed for the $(A_1,E_6)$ and $(A_1,E_8)$ cases in \cite{Go:2025ixu}. By assuming $\CN=4$ enhancement, tuning $\n=\pm 1$ and $\eta = 1$ yields the Hilbert series that counts the Coulomb and Higgs branch operators respectively \cite{Razamat:2014pta}:
\begin{align}
    \CI_{S^2 \times S^1}^{(A_2,A_6)}
    (\eta = 1 , \n=\pm 1;q) = 1 \, ,
    \label{eq: A2A6 Hilbert}
\end{align}
where we have verified the equality up to $ q^5 $ order. This implies that both the Coulomb and Higgs branches of the resulting 3d theory are trivial, confirming that it is a rank-0 theory. Such theories have recently attracted attention, as they provide natural 3d bulk descriptions of non-unitary VOAs. See Appendix~\ref{app: rank-0} for a survey of the rank-0 theories.

The two values $\n=\pm 1$ coincide with the Cartan generators of the R-symmetry associated with the topological B- and A-twists, respectively. Thereby, the result \eqref{eq: A2A6 Hilbert} implies that there is {\it no} local operator in the topologically A/B-twisted sector, consequently, they become semisimple TFTs. The modular data in the topologically A-twisted sector can be extracted using the 3d A-model method reviewed in Appendix~\ref{app: A-model}, as
\begin{align}\label{eq: A2A6-Atwist}
    \{| S_{0\a} |\}&= \frac{1}{5}\{ \sigma_5^1 + 2 \sigma_5^2,
        \sigma_5^1,
        \sigma_5^1,
        2 \sigma_5^1 - \sigma_5^2,
        \sigma_5^2,
        \sigma_5^2,
        -\sigma_5^1 + 2 \sigma_5^2,
        2 \sigma_5^1,
        2 \sigma_5^1,
        2 \sigma_5^2,
        2 \sigma_5^2,
    2 \sigma_5^1 + \sigma_5^2 \} \, ,
    \nonumber\\
    \{T_{\a\a}\}&= \big\{
    {\tt e}(0),
    {\tt e}(0),
    {\tt e}(0),
    {\tt e}(\frac{2}{5}),
    {\tt e}(\frac{2}{5}),
    {\tt e}(\frac{2}{5}),
    {\tt e}(\frac{1}{2}),
    {\tt e}(\frac{3}{5}),
    {\tt e}(\frac{3}{5}),
    {\tt e}(\frac{4}{5}),
    {\tt e}(\frac{4}{5}),
    {\tt e}(\frac{9}{10})
    \big\} \, ,
\end{align}
which are ordered set. Thus $|S_{00}|=\frac{\sigma_5^1+2\sigma_5^2}{5}$ is the value of the vacuum module. These modular data are compatible with those of the $W_3(3,10)$ W-algebra minimal model which is the expected from the SCFT/VOA correspondence, satisfying the central charge relation $c_{2d} = -12 c_{4d} = -\frac{186}{5}$. On the other hand, we get another modular data at the topologically B-twisted sector as,
\begin{align}
    \{| S_{0\a} |\}&= \frac{1}{5}\{ \sigma_5^1 + 2 \sigma_5^2,
        \sigma_5^1,
        \sigma_5^1,
        2 \sigma_5^1 - \sigma_5^2,
        \sigma_5^2,
        \sigma_5^2,
        -\sigma_5^1 + 2 \sigma_5^2,
        2 \sigma_5^1,
        2 \sigma_5^1,
        2 \sigma_5^2,
        2 \sigma_5^2,
    2 \sigma_5^1 + \sigma_5^2 \} \, ,
    \nonumber\\
    \{T_{\a\a}\}&= \big\{
    {\tt e}(0),
    {\tt e}(0),
    {\tt e}(0),
    {\tt e}(\frac{3}{5}),
    {\tt e}(\frac{3}{5}),
    {\tt e}(\frac{3}{5}),
    {\tt e}(\frac{1}{2}),
    {\tt e}(\frac{2}{5}),
    {\tt e}(\frac{2}{5}),
    {\tt e}(\frac{1}{5}),
    {\tt e}(\frac{1}{5}),
    {\tt e}(\frac{1}{10})
    \big\} \, .
\end{align}
This modular data can be obtained by applying the Galois automorphism $ \zeta_{10} \mapsto \zeta_{10}^9 $ in the Galois group $ \operatorname{Gal}(\mathbb{Q}(\zeta_{10})/\mathbb{Q}) \cong \mathbb{Z}_{10}^\times $ to the modular data in \eqref{eq: A2A6-Atwist}.\footnote{The entries of the modular $ T $- and $ S $-matrices lie in $ \mathbb{Q}(\zeta_{10}) $, up to an overall central charge factor in the $ T $-matrix. Throughout this section, we ignore this overall central charge factor when discussing modular data and its Galois conjugates.} We therefore propose that the boundary VOA arising from the topological B-twist is a Galois conjugate of the $ W_3(3,10) $ W-algebra minimal model.

This example provides concrete evidence for the existence of an intermediate 3d TFT that bridges the two sides of the SCFT/VOA correspondence. It also confirms the trace formula, which serves as an efficient method for constructing the 3d TFT.

\subsubsection{\texorpdfstring{$(A_2,A_7)$}{(A2,A7)}}

Consider the $(A_2,A_7)$ theory whose 4d central charge is $c_{4d} = \frac{245}{66}$. The BPS quiver of the $(A_2,A_7)$ Argyres-Douglas theory is given by $A_2 \sqprod A_7$:
\begin{align}
    \begin{aligned}
        \includegraphics{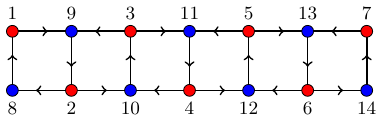}
    \end{aligned}
\end{align}
which admits a finite chamber with 21 BPS particles which can be captured by the sequence of mutations
\begin{align}
    \mathfrak{M}^{(A_2,A_7)}
    =
	\mathfrak{M}_-\circ
	\mathfrak{M}_+\circ
	\mathfrak{M}_-\,.
\end{align}
From this mutation, one can enumerate the electromagnetic charges of the BPS particles and their CPT conjugates in increasing order of their central charge phases as
\begin{align}
    \G_{\text{BPS}}
    &= \{ \g_1,\ \g_2,\ \g_3,\ \g_4,\ \g_5,\ \g_6,\ \g_7,\ \g_1+\g_8,\ \g_2+\g_9,\ \g_3+\g_{10},\ \g_4+\g_{11}, \\
    &\qquad \g_5+\g_{12},\ \g_6+\g_{13},\ \g_7+\g_{14},\ \g_{8},\ \g_{9},\ \g_{10},\ \g_{11},\ \g_{12},\ \g_{13},\ \g_{14},\ (\g_i \to -\g_i) \} \, . \nonumber
\end{align}
Consequently, the trace formula computes the ellipsoid partition function as
\begin{align}
	Z_{S_b^3}^{(A_2,A_7)}
	&= 
    i^{1/2} e^{\frac{7 \pi i}{3}(b^2 + b^{-2}) } \int
     d^7 u \;
    e^{\pi i u^T L u}
    \prod_{i=1}^7 \Phi_b(u_i) \, ,
\end{align}
where
\begin{align}
	L = 
	\left(
	\begin{array}{ccccccc}
        3 & 1 & -3 & 1 & -1 & 3 & 1 \\
        1 & 0 & 0 & 0 & 0 & 0 & 1 \\
        -3 & 0 & 2 & 0 & 2 & -3 & 0 \\
        1 & 0 & 0 & 0 & -1 & 1 & 1 \\
        -1 & 0 & 2 & -1 & 0 & -1 & -1 \\
        3 & 0 & -3 & 1 & -1 & 2 & 0 \\
        1 & 1 & 0 & 1 & -1 & 0 & 2
 	\end{array}
	\right)\,.
\end{align}
This result implies a 3d $\CN=2$ $U(1)^{7}$ gauge theory with 7 chiral multiplets, whose charge matrix $Q$ and CS level matrix $K$ are
\begin{align}
	Q = {\bf 1}_{7\times 7} \, , \quad
	K = L+QQ^T \, .
\end{align}
This theory admits six half-BPS monopole operators given by
\begin{gather}
    \begin{gathered}
        \phi_2\phi_5 V_{(-1,0,0,1,0,1,0)} \, , \quad
        \phi_4\phi_5 V_{(0,0,-1,0,0,-1,0)} \, , \quad
        \phi_3^2 V_{(0,0,0,-1,-1,0,0)} \, , \\
        \phi_1\phi_6 V_{(0,-1,0,-1,0,0,1)} \, , \quad
        \phi_1\phi_2 V_{(0,0,1,2,-1,0,-1)} \, , \quad
        \phi_7^2 V_{(1,-1,1,0,2,1,0)}\,.
    \end{gathered}
\end{gather}
Once these operators are turned on as superpotential terms, the theory flows to a fixed point with $\CN=4$ supersymmetry, at which a single surviving combination of the $U(1)_{T_i}$ topological symmetries given by
\begin{align}
	A = 5T_1 + T_2 - 3T_3 + 2T_4 - 2T_5 + 3T_6 + 3T_7 \, ,
\end{align}
becomes the $U(1)_A$ axial symmetry, and the mixing of the topological symmetries are fixed to be $\m^* = (-1,-2,-2,-1,-1,0,-2)$. The 3d superconformal index at this fixed point reads
\begin{align}
    \CI_{S^2\times S^1}^{(A_2,A_7)}(\eta,\n=0;q)
    &= 1 - q + \left( \eta^2 + \frac{1}{\eta^2} \right) q^2 + \left( \eta + \frac{1}{\eta} \right) q^{5/2} - \left( 3\eta + \frac{2}{\eta} \right) q^{7/2} \\
    &\quad + \left( \eta^4 - 5\eta^2 - 10 - \frac{3}{\eta^2} + \frac{1}{\eta^4} \right) q^4 - \left( 17\eta + \frac{14}{\eta} - \frac{1}{\eta^3} \right) q^{9/2} + \mathcal{O}(q^5) \, , \nonumber
\end{align}
and the Hilbert series of the Coulomb and Higgs branches become trivial:
\begin{align}
    \CI_{S^2\times S^1}^{(A_2,A_7)} (\eta = 1 , \n=\pm 1;q) = 1 \, ,
\end{align}
where we have checked the equality up to $ q^5 $ order. This indicates that the theory is a 3d rank-0 SCFT. This is precisely the expected result: the original 4d $\CN=2$ $(A_2,A_7)$ Argyres–Douglas theory has no Higgs branch, and all Coulomb branch operators carry purely fractional scaling dimensions. As a consequence, they are lifted under the $U(1)_r$ twisted circle compactification, leading to a 3d rank-0 SCFT.

We also extract the modular data of the topologically A-twist sector as
\begin{align}\label{eq: A2A7-Atwist}
\{|S_{0\a}|\} &= \frac{8}{11} \{
        \xi_{22}^1 (\xi_{22}^5)^2,
        (\sigma_{11}^1)^2 \sigma_{11}^2,
        (\sigma_{11}^2)^2 \xi_{22}^3,
        \sigma_{11}^1 \xi_{22}^1 \xi_{22}^3,
        \sigma_{11}^1 \xi_{22}^1 \xi_{22}^3,
        \sigma_{11}^2 \xi_{22}^1 \xi_{22}^5,
        \sigma_{11}^2 \xi_{22}^1 \xi_{22}^5,\nonumber \\
        &\ \sigma_{11}^2 \xi_{22}^1 \xi_{22}^3,
        \sigma_{11}^2 \xi_{22}^1 \xi_{22}^3,
        (\xi_{22}^3)^2 \xi_{22}^5,
        \sigma_{11}^1 \xi_{22}^3 \xi_{22}^5,
        \sigma_{11}^1 \xi_{22}^3 \xi_{22}^5,
        \sigma_{11}^1 (\xi_{22}^1)^2,
        \sigma_{11}^1 \sigma_{11}^2 \xi_{22}^5,
        \sigma_{11}^1 \sigma_{11}^2 \xi_{22}^5
    \},
    \nonumber\\
    \{T_{\a\a}\} &=
    \big\{
    {\tt e}(0),
    {\tt e}(\frac{1}{11}),
    {\tt e}(\frac{2}{11}),
    {\tt e}(\frac{3}{11}),
    {\tt e}(\frac{3}{11}),
    {\tt e}(\frac{4}{11}),
    {\tt e}(\frac{4}{11}),
    \nonumber\\
    &\qquad
    {\tt e}(\frac{5}{11}),
    {\tt e}(\frac{5}{11}),
    {\tt e}(\frac{6}{11}),
    {\tt e}(\frac{7}{11}),
    {\tt e}(\frac{7}{11}),
    {\tt e}(\frac{9}{11}),
    {\tt e}(\frac{10}{11}),
    {\tt e}(\frac{10}{11})
    \big\},
\end{align}
which are compatible with those of the desired $W_3(3,11)$ W-algebra minimal model, whose central charge is $c_{2d} = -12 c_{4d} = -\frac{490}{11}$. In contrast, the modular data obtained from the topologically B-twisted sector yields
\begin{align}
\{|S_{0\a}|\} &= \frac{8}{11}\{
        (\xi_{22}^3)^2 \xi_{22}^5,
        \xi_{22}^1 (\xi_{22}^5)^2,
        \sigma_{11}^2 \xi_{22}^1 \xi_{22}^3,
        \sigma_{11}^2 \xi_{22}^1 \xi_{22}^3,
        \sigma_{11}^1 \sigma_{11}^2 \xi_{22}^5,
        \sigma_{11}^1 \sigma_{11}^2 \xi_{22}^5,
        \sigma_{11}^2 \xi_{22}^1 \xi_{22}^5,\nonumber \\
        &\ \sigma_{11}^2 \xi_{22}^1 \xi_{22}^5,
        \sigma_{11}^1 (\xi_{22}^1)^2,
        \sigma_{11}^1 \xi_{22}^1 \xi_{22}^3,
        \sigma_{11}^1 \xi_{22}^1 \xi_{22}^3,
        (\sigma_{11}^2)^2 \xi_{22}^3,
        \sigma_{11}^1 \xi_{22}^3 \xi_{22}^5,
        \sigma_{11}^1 \xi_{22}^3 \xi_{22}^5,
        (\sigma_{11}^1)^2 \sigma_{11}^2
    \},
    \nonumber\\
    \{T_{\a\a}\} &=
    \big\{
    {\tt e}(0),
    {\tt e}(\frac{1}{11}),
    {\tt e}(\frac{2}{11}),
    {\tt e}(\frac{2}{11}),
    {\tt e}(\frac{3}{11}),
    {\tt e}(\frac{3}{11}),
    {\tt e}(\frac{4}{11}),
    \nonumber\\
    &\qquad
    {\tt e}(\frac{4}{11}),
    {\tt e}(\frac{5}{11}),
    {\tt e}(\frac{6}{11}),
    {\tt e}(\frac{6}{11}),
    {\tt e}(\frac{8}{11}),
    {\tt e}(\frac{9}{11}),
    {\tt e}(\frac{9}{11}),
    {\tt e}(\frac{10}{11})
    \big\} \, .
\end{align}
This modular data can be obtained by applying the Galois automorphisms $ \zeta_{44} \mapsto \zeta_{44}^5 $ or $ \zeta_{44} \mapsto \zeta_{44}^{27} $ in the Galois group $ \Gal(\mathbb{Q}(\zeta_{44})/\mathbb{Q}) \cong \mathbb{Z}_{44}^\times $ to the modular data \eqref{eq: A2A7-Atwist}, where the two conjugations differ only by an overall sign in the $ S $-matrix. Thus, we propose that the VOA arising from the B-twist is a Galois conjugate of the $ W_3(3,11) $ W-algebra minimal model.

\subsubsection{\texorpdfstring{$(A_3,A_4)$}{(A3,A4)}}

Consider the $(A_3,A_4)$ Argyres-Douglas theory whose central charge is $c_{4d} = \frac{29}{9}$ and the BPS quiver is given by $A_3 \sqprod A_4$:
\begin{align}
    \begin{aligned}
        \includegraphics{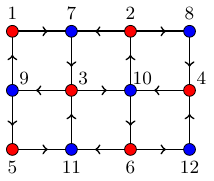}
    \end{aligned}
\end{align}
This quiver admits a finite chamber with 24 BPS particles and their CPT conjugates, captured from the sequence of mutations given by
\begin{align}
    \mathfrak{M}^{(A_3,A_4)}
    =
	\mathfrak{M}_+\circ
    \mathfrak{M}_-\circ
	\mathfrak{M}_+\circ
	\mathfrak{M}_-\,.
\end{align}
From this mutation, one can read off the electromagnetic charges of the BPS particles in increasing order of their central charge phases as
\begin{align}
    \begin{aligned}
        \G_{\text{BPS}}
        &= \{
            \g_1,\
            \g_2,\
            \g_3,\
            \g_4,\
            \g_5,\
            \g_6,\
            \g_3+\g_7,\
            \g_4+\g_8,\
            \g_1+\g_5+\g_9,\
            \g_2+\g_6+\g_{10}, \\
            &\qquad
            \g_3+\g_{11},\
            \g_4+\g_{12},\
            \g_5+\g_9,\
            \g_6+\g_{10},\
            \g_3+\g_7+\g_{11},\
            \g_4+\g_8+\g_{12}, \\
            &\qquad
            \g_1+\g_{9},\
            \g_2+\g_{10},\
            \g_7,\
            \g_8,\
            \g_9,\
            \g_{10},\
            \g_{11},\
            \g_{12},\
            (\g_i \to -\g_i)
        \} \, .
    \end{aligned}
\end{align}
Hence, the trace formula evaluates the ellipsoid partition function as
\begin{align}
	Z_{S_b^3}^{(A_3,A_4)}
	&= 
    i e^{2\pi i (b^2 + b^{-2}) } \int
     d^6 u \;
    e^{\pi i u^T L  u}
    \prod_{i=1}^6 \Phi_b(u_i) \, ,
    \label{eq: A3A4 int}
\end{align}
with
\begin{align}
    L = \left(
    \begin{array}{cccccc}
        1 & -1 & 0 & -1 & 1 & 1 \\
        -1 & 1 & 1 & 0 & 1 & -1 \\
        0 & 1 & 1 & 1 & 0 & 0 \\
        -1 & 0 & 1 & 1 & -1 & 1 \\
        1 & 1 & 0 & -1 & 0 & 1 \\
        1 & -1 & 0 & 1 & 1 & 0
    \end{array}
    \right) \, .
\end{align}
The resulting expression suggests a 3d $\CN=2$ gauge theory description of the $U(1)_r$ twisted compactification of the 4d $(A_3,A_4)$ Argyres-Douglas theory, which is the $U(1)^{6}$ gauge theory with 6 chiral multiplets, whose charge matrix $Q$ and CS level matrix $K$ are given by
\begin{align}
	Q = {\bf 1}_{6\times 6} \, , \quad
	K = L+QQ^T\,.
\end{align}
This gauge theory admits five half-BPS monopole operators given by
\begin{align}
	\phi_1^2 V_{(0,1,-1,1,0,0)},\
	\phi_2^2 V_{(1,0,0,0,-1,0)},\
	\phi_4^2 V_{(1,0,0,0,0,-1)},\
	\phi_5^2 V_{(0,-1,1,0,0,-1)},\
	\phi_6^2 V_{(0,0,1,-1,-1,0)}\, ,
    \nonumber\\
\end{align}
which, when turned on as superpotential deformations, the theory flows to a fixed point with $ \mathcal{N}=4 $ supersymmetry. The single surviving linear combination of the $U(1)_{T_i}$ topological symmetries upon the deformation is
\begin{align}
	A = T_1 + T_2 + 2 \, T_3 + T_4 + T_5 + T_6 \, ,
\end{align}
and it becomes the $U(1)_A$ symmetry of the $\CN=4$ supersymmetry. The mixing parameter of the topological symmetries, is fixed to be $\m^* = (-1,-1,-2,-1,-2,-2)$. The 3d superconformal index at the fixed point can be computed as
\begin{align}
	&\CI_{S^2\times S^1}^{(A_3,A_4)}(\eta, \nu=0; q)
    = 1 - q - \eta q^{3/2} - q^2 + \left( \eta - \frac{1}{\eta^3} \right) q^{5/2} + \left( 2\eta^2 + 1 - \frac{2}{\eta^2} \right) q^3 \\
    &\qquad + \left( 2\eta - \frac{4}{\eta} - \frac{1}{\eta^3} \right) q^{7/2} - \left( 5 + \frac{5}{\eta^2} - \frac{1}{\eta^4} \right) q^4 - \left( \eta^3 + 4\eta + \frac{13}{\eta} - \frac{2}{\eta^3} \right) q^{9/2} + \mathcal{O}(q^5) \, , \nonumber
\end{align}
from which one can extract the Hilbert series of the Coulomb and Higgs branches as
\begin{align}
    \CI_{S^2\times S^1}^{(A_3,A_4)} (\eta=1, \nu=\pm 1; q) = 1 \, ,
\end{align}
where we have checked the equality up to $ q^6 $ order. This indicates the resulting 3d theory is a rank-0 SCFT. Indeed, the original 4d $(A_3,A_4)$ Argyres-Douglas theory does not have non-trivial Higgs branch, and its Coulomb branch is lifted under the $U(1)_r$ twisted compactification, resulting in a 3d rank-0 SCFT.

The modular data at the topological A-twisted sector of the rank-0 theory can be extracted as,
\begin{align}\label{eq: A3A4-Atwist}
    &\{|S_{0\a}|\} = \frac{4}{9}\Big\{
        \frac{\sigma_{9}^1 (\xi_{18}^1)^2}{\sigma_{9}^2},
        \sigma_{9}^2 \xi_{18}^1,
        \sigma_{9}^2 \xi_{18}^1,
        \xi_{18}^1 \xi_{18}^3,
        \frac{(\sigma_{9}^1)^2 \sigma_{9}^2}{\xi_{18}^1},
        \sigma_{9}^1 \xi_{18}^1,
        \sigma_{9}^1 \xi_{18}^1,
        \sigma_{9}^2 \xi_{18}^3, \\
        &\qquad \qquad \qquad \frac{3}{4},
        \frac{3}{4},
        \sigma_{9}^1 \sigma_{9}^2,
        \sigma_{9}^1 \sigma_{9}^2,
        \frac{(\sigma_{9}^2)^2 \xi_{18}^1}{\sigma_{9}^1},
        \sigma_{9}^1 \xi_{18}^3
    \Big\} \, , \nonumber \\
    &\{T_{\a\a}\} =
    \big\{
    {\tt e}(0),
    {\tt e}(0),
    {\tt e}(0),
    {\tt e}(\frac{1}{9}),
    {\tt e}(\frac{1}{3}),
    {\tt e}(\frac{1}{3}),
    {\tt e}(\frac{1}{3}),
    {\tt e}(\frac{4}{9}),
    {\tt e}(\frac{5}{9}),
    {\tt e}(\frac{5}{9}),
    {\tt e}(\frac{2}{3}),
    {\tt e}(\frac{2}{3}),
    {\tt e}(\frac{2}{3}),
    {\tt e}(\frac{7}{9})
    \big\}.
     \nonumber
\end{align}
These are compatible with those of the $W_4(4,9)$ W-algebra whose central charge satisfies $c_{2d} = -12 c_{4d} = -\frac{116}{3}$. This VOA is precisely the one expected from the SCFT/VOA correspondence. On the other hand, the modular data of the topological B-twisted sector is
\begin{align}
    &\{|S_{0\a}|\}= \frac{4}{9}\Big\{
        \frac{(\sigma_{9}^2)^2 \xi_{18}^1}{\sigma_{9}^1},
        \sigma_{9}^1 \sigma_{9}^2,
        \sigma_{9}^1 \sigma_{9}^2,
        \xi_{18}^1 \xi_{18}^3,
        \frac{3}{4},
        \frac{3}{4},
        \frac{\sigma_{9}^1 (\xi_{18}^1)^2}{\sigma_{9}^2}, \\
        &\qquad \qquad \qquad \sigma_{9}^2 \xi_{18}^1,
        \sigma_{9}^2 \xi_{18}^1,
        \sigma_{9}^2 \xi_{18}^3,
        \frac{(\sigma_{9}^1)^2 \sigma_{9}^2}{\xi_{18}^1},
        \sigma_{9}^1 \xi_{18}^1,
        \sigma_{9}^1 \xi_{18}^1,
        \sigma_{9}^1 \xi_{18}^3
    \Big\} \, , \nonumber \\
    &\{T_{\a\a}\} =
    \big\{
    {\tt e}(0),
    {\tt e}(0),
    {\tt e}(0),
    {\tt e}(\frac{1}{9}),
    {\tt e}(\frac{2}{9}),
    {\tt e}(\frac{2}{9}),
    {\tt e}(\frac{1}{3}),
    {\tt e}(\frac{1}{3}),
    {\tt e}(\frac{1}{3}),
    {\tt e}(\frac{4}{9}),
    {\tt e}(\frac{2}{3}),
    {\tt e}(\frac{2}{3}),
    {\tt e}(\frac{2}{3}),
    {\tt e}(\frac{7}{9})
    \big\}. \nonumber
\end{align}
These modular data are compatible with the Galois automorphism $ \zeta_{18} \mapsto \zeta_{18}^{13} $ in the Galois group $ \Gal(\mathbb{Q}(\zeta_{18})/\mathbb{Q}) \cong \mathbb{Z}_{18}^\times $ of the modular data \eqref{eq: A3A4-Atwist}. We thus propose that this VOA is a Galois conjugate of the $ W_4(4,9) $ W-algebra minimal model.

\subsubsection{\texorpdfstring{$(A_3,A_6)$}{(A3,A6)}}

Consider the $(A_3,A_6)$ Argyres-Douglas theory whose central charge is $c_{4d} = \frac{117}{22}$ and the BPS quiver is given by $A_3 \sqprod A_6$:
\begin{align}
    \begin{aligned}
        \includegraphics{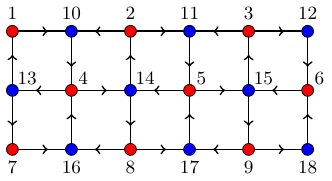}
    \end{aligned}
\end{align}
This quiver encodes a finite chamber with 36 BPS particles and their CPT conjugates, which can be captured by a sequence of mutations given by
\begin{align}
    \mathfrak{M}^{(A_3,A_6)}
    =
	\mathfrak{M}_+\circ
    \mathfrak{M}_-\circ
	\mathfrak{M}_+\circ
	\mathfrak{M}_- \, .
\end{align}
This sequence of mutations determines the electromagnetic charges of the BPS particles in increasing order of their central charge phases as
\begin{align}
    \begin{aligned}
        \G_{\text{BPS}}
        &= \{
            \g_1,\
            \g_2,\
            \cdots,\
            \g_8,\
            \g_9,\
            \g_{4}+\g_{10},\
            \g_{5}+\g_{11},\
            \g_{6}+\g_{12},\
            \g_{1}+\g_{7}+\g_{13},
            \\
            &\qquad
            \g_{2}+\g_{8}+\g_{14},\
            \g_{3}+\g_{9}+\g_{15},\
            \g_{4}+\g_{16},\
            \g_{5}+\g_{17},\
            \g_{6}+\g_{18},\
            \g_{7}+\g_{13},
            \\
            &\qquad
            \g_{8}+\g_{14},\
            \g_{9}+\g_{15},\
            \g_{4}+\g_{10}+\g_{16},\
            \g_{5}+\g_{11}+\g_{17},\
            \g_{6}+\g_{12}+\g_{18},
            \\
            &\qquad
            \g_{3}+\g_{15},\
            \g_{2}+\g_{14},\
            \g_{1}+\g_{13},\
            \g_{10},\
            \g_{11},\
            \cdots,\
            \g_{17},\
            \g_{18},\
            (\g_i \to -\g_i)
        \}\,.
    \end{aligned}
\end{align}
Therefore, the trace formula computes the ellipsoid partition function of the 3d theory arising from the $U(1)_r$ twisted compactification of the 4d $(A_3,A_6)$ theory as
\begin{align}
	Z_{S_b^3}^{(A_3,A_6)}
	&= i^{1/2} 
    e^{\frac{41\pi i}{12} (b^2 + b^{-2}) } \int
     d^9 u \;
    e^{\pi i u^T L u}
    \prod_{i=1}^9 \Phi_b(u_i) \, ,
    \label{eq: A3A6 int}
\end{align}
with
\begin{align}
    L = \left(
    \begin{array}{ccccccccc}
     0 & -1 & 0 & 0 & 1 & 0 & 1 & 1 & 0 \\
     -1 & 2 & -1 & -1 & -1 & 1 & -2 & 0 & 1 \\
     0 & -1 & 0 & -1 & -1 & -1 & 1 & -1 & 0 \\
     0 & -1 & -1 & -1 & 0 & -1 & -1 & 0 & 0 \\
     1 & -1 & -1 & 0 & 1 & -1 & -1 & 2 & 1 \\
     0 & 1 & -1 & -1 & -1 & 0 & -1 & -1 & 0 \\
     1 & -2 & 1 & -1 & -1 & -1 & 2 & 0 & -1 \\
     1 & 0 & -1 & 0 & 2 & -1 & 0 & 2 & 1 \\
     0 & 1 & 0 & 0 & 1 & 0 & -1 & 1 & 0 \\
    \end{array}
    \right)\,.
\end{align}
This implies a 3d $\CN=2$ $U(1)^{9}$ gauge theory with 9 chiral multiplets, whose charge matrix $Q$ and CS level matrix $K$ are given by
\begin{align}
	Q = {\bf 1}_{9\times 9} \, , \quad
	K = L+QQ^T \, .
\end{align}
This theory contains eight half-BPS monopole operators given by
\begin{gather}
    \begin{gathered}
        \phi_5 \phi_9 V_{(-1,0,-1,0,0,0,1,0,0)},\
        \phi_3 \phi_6 V_{(-1,0,0,0,0,0,0,1,-1)},\
        \phi_8^2 V_{(0,-1,1,-1,0,1,-1,0,0)}, \\
        \quad\phi_7^2 V_{(0,0,0,0,1,0,0,-1,1)},\
        \phi_4 \phi_6 V_{(0,0,1,0,0,0,0,0,1)},\
        \phi_1 \phi_5 V_{(0,1,0,0,0,-1,0,0,-1)}, \\
        \qquad\phi_3 \phi_4 V_{(1,0,0,0,0,1,0,0,0)},\
        \phi_2^2 V_{(1,0,0,0,1,0,0,-1,0)} .
    \end{gathered}
\end{gather}
Once these operators are turned on as superpotential terms, the theory flows to a fixed point with $\CN=4$ SUSY enhancement, leaving a single surviving combination of the $U(1)_{T_i}$ topological symmetries given by
\begin{align}
	A = T_1 - T_3 - 2 \, T_4 + T_5 - T_6 + T_8 + T_9\,,
\end{align}
which is identified to the $U(1)_A$ symmetry for the $\CN=4$ supersymmetry. The mixing parameter of the topological symmetries is fixed to be $\m^* = (-2,1,2,2,-1,2,1,-3,-2)$. The 3d superconformal index at this fixed point given by
\begin{align}
    \begin{aligned}
        \CI_{S^2\times S^1}^{(A_3,A_6)}(\eta, \nu=0; q)
        &= 1 - q + \left(\eta^2-1 \right) q^2 - \left( \frac{2}{\eta} + \frac{1}{\eta^3} \right)q^{5/2} - \left( 5 + \frac{2}{\eta^2} \right)q^3 \\
        &\quad - \left( 7\eta + \frac{8}{\eta} - \frac{1}{\eta^3} \right)q^{7/2} + \left( \eta^4 - 5\eta^2 - 17 + \frac{1}{\eta^4} \right)q^4 \\
        &\quad + \left( 2\eta^3 - 11\eta - \frac{1}{\eta} + \frac{6}{\eta^3} - \frac{1}{\eta^5} \right)q^{9/2} + \mathcal{O}(q^5) \, ,
    \end{aligned}
\end{align}
and the Hilbert series of the Coulomb and Higgs branches become trivial as
\begin{align}
    \CI_{S^2\times S^1}^{(A_3,A_6)}(\eta = 1 , \n=\pm 1;q) = 1 \, ,
\end{align}
indicating that the theory is a 3d rank-0 SCFT. We have checked this equality up to $ q^4 $ order. This result is expected, since the original 4d $(A_3,A_6)$ theory has trivial Higgs branch, while the Coulomb branch is lifted under the $U(1)_r$ twisted compactification.

The modular data at the topological A-twisted sector can be extracted as
\begin{align}\label{eq: A3A6-Atwist}
    \{|S_{0\a}|\} &= \frac{2}{11}\Big\{
        \frac{(\xi_{22}^3)^2 \xi_{22}^5}{\sigma_{11}^2 \xi_{22}^1},
        \sigma_{11}^1,
        \sigma_{11}^1,
        \frac{\xi_{22}^1 \xi_{22}^3}{\xi_{22}^5},
        \frac{\xi_{22}^3 \xi_{22}^5}{\sigma_{11}^2},
        \frac{\xi_{22}^3 \xi_{22}^5}{\sigma_{11}^2},
        \frac{(\sigma_{11}^1)^2 \sigma_{11}^2}{\xi_{22}^1 \xi_{22}^3},
        \xi_{22}^5,
        \xi_{22}^5,
        \frac{\sigma_{11}^2 \xi_{22}^5}{\xi_{22}^3},
        \frac{\sigma_{11}^2 \xi_{22}^3}{\sigma_{11}^1},\nonumber \\
        &\quad \frac{\sigma_{11}^2 \xi_{22}^3}{\sigma_{11}^1},
        \frac{\sigma_{11}^1 \xi_{22}^5}{\xi_{22}^1},
        \frac{\xi_{22}^1 \xi_{22}^5}{\xi_{22}^3},
        \frac{\xi_{22}^1 \xi_{22}^5}{\xi_{22}^3},
        \frac{\sigma_{11}^1 \xi_{22}^1}{\xi_{22}^5},
        \frac{\sigma_{11}^1 \xi_{22}^1}{\xi_{22}^5},
        \frac{\sigma_{11}^2 \xi_{22}^1}{\sigma_{11}^1},
        \frac{\sigma_{11}^1 (\xi_{22}^1)^2}{\sigma_{11}^2 \xi_{22}^5},
        \xi_{22}^3,
        \xi_{22}^3,
        \sigma_{11}^2,
        \sigma_{11}^2, \nonumber \\
        &\quad \frac{\xi_{22}^1 (\xi_{22}^5)^2}{\sigma_{11}^1 \xi_{22}^3},
        \xi_{22}^1,
        \xi_{22}^1,
        \frac{(\sigma_{11}^2)^2 \xi_{22}^3}{\sigma_{11}^1 \xi_{22}^5},
        \frac{\sigma_{11}^1 \xi_{22}^3}{\sigma_{11}^2},
        \frac{\sigma_{11}^1 \sigma_{11}^2}{\xi_{22}^1},
        \frac{\sigma_{11}^1 \sigma_{11}^2}{\xi_{22}^1}
    \Big\} \, ,
    \nonumber\\
    \{T_{\a\a}\} &=
    \big\{
    {\tt e}(0),
    {\tt e}(0),
    {\tt e}(0),
    {\tt e}(\frac{2}{11}),
    {\tt e}(\frac{2}{11}),
    {\tt e}(\frac{2}{11}),
    {\tt e}(\frac{3}{11}),
    {\tt e}(\frac{3}{11}),
    {\tt e}(\frac{3}{11}),
    {\tt e}(\frac{4}{11}),
    {\tt e}(\frac{4}{11}),
    {\tt e}(\frac{4}{11}),
    \nonumber\\
    & \quad
    {\tt e}(\frac{5}{11}),
    {\tt e}(\frac{5}{11}),
    {\tt e}(\frac{5}{11}),
    {\tt e}(\frac{6}{11}),
    {\tt e}(\frac{6}{11}),
    {\tt e}(\frac{6}{11}),
    {\tt e}(\frac{7}{11}),
    {\tt e}(\frac{7}{11}),
    {\tt e}(\frac{7}{11}),
    \nonumber\\
    & \quad
    {\tt e}(\frac{8}{11}),
    {\tt e}(\frac{8}{11}),
    {\tt e}(\frac{8}{11}),
    {\tt e}(\frac{9}{11}),
    {\tt e}(\frac{9}{11}),
    {\tt e}(\frac{9}{11}),
    {\tt e}(\frac{10}{11}),
    {\tt e}(\frac{10}{11}),
    {\tt e}(\frac{10}{11})
    \big\} \, ,
\end{align}
which are compatible with those of the $W_4(4,11)$ W-algebra minimal model whose central satisfies $c_{2d} = -12 c_{2d} = -\frac{702}{11}$. This VOA is the expected from the SCFT/VOA correspondence. The topologcal B-twisted sector, on the other hand, gives
\begin{align}
    \{|S_{0\a}|\} &= \frac{2}{11}\Big\{
        \frac{\xi_{22}^1 (\xi_{22}^5)^2}{\sigma_{11}^1 \xi_{22}^3},
        \sigma_{11}^2,
        \sigma_{11}^2,
        \frac{\sigma_{11}^2 \xi_{22}^1}{\sigma_{11}^1},
        \frac{\sigma_{11}^1 \xi_{22}^1}{\xi_{22}^5},
        \frac{\sigma_{11}^1 \xi_{22}^1}{\xi_{22}^5},
        \frac{\sigma_{11}^2 \xi_{22}^3}{\sigma_{11}^1},
        \frac{\sigma_{11}^2 \xi_{22}^3}{\sigma_{11}^1},
        \frac{\sigma_{11}^2 \xi_{22}^5}{\xi_{22}^3},
        \frac{\xi_{22}^3 \xi_{22}^5}{\sigma_{11}^2},\nonumber \\
        &\quad \frac{\xi_{22}^3 \xi_{22}^5}{\sigma_{11}^2},
        \frac{\xi_{22}^1 \xi_{22}^3}{\xi_{22}^5},
        \frac{(\xi_{22}^3)^2 \xi_{22}^5}{\sigma_{11}^2 \xi_{22}^1},
        \sigma_{11}^1,
        \sigma_{11}^1,
        \frac{(\sigma_{11}^2)^2 \xi_{22}^3}{\sigma_{11}^1 \xi_{22}^5},
        \xi_{22}^1,
        \xi_{22}^1,
        \xi_{22}^3,
        \xi_{22}^3,
        \frac{\sigma_{11}^1 (\xi_{22}^1)^2}{\sigma_{11}^2 \xi_{22}^5},
        \frac{\xi_{22}^1 \xi_{22}^5}{\xi_{22}^3},\nonumber \\
        &\quad \frac{\xi_{22}^1 \xi_{22}^5}{\xi_{22}^3},
        \frac{\sigma_{11}^1 \xi_{22}^5}{\xi_{22}^1},
        \xi_{22}^5,
        \xi_{22}^5,
        \frac{(\sigma_{11}^1)^2 \sigma_{11}^2}{\xi_{22}^1 \xi_{22}^3},
        \frac{\sigma_{11}^1 \xi_{22}^3}{\sigma_{11}^2},
        \frac{\sigma_{11}^1 \sigma_{11}^2}{\xi_{22}^1},
        \frac{\sigma_{11}^1 \sigma_{11}^2}{\xi_{22}^1}
    \Big\} \, ,
    \nonumber\\
    \{T_{\a\a}\} &=
    \big\{
    {\tt e}(0),
    {\tt e}(0),
    {\tt e}(0),
    {\tt e}(\frac{1}{11}),
    {\tt e}(\frac{1}{11}),
    {\tt e}(\frac{1}{11}),
    {\tt e}(\frac{2}{11}),
    {\tt e}(\frac{2}{11}),
    {\tt e}(\frac{2}{11}),
    {\tt e}(\frac{3}{11}),
    {\tt e}(\frac{3}{11}),
    {\tt e}(\frac{3}{11}),
    \nonumber\\
    &\quad
    {\tt e}(\frac{4}{11}),
    {\tt e}(\frac{4}{11}),
    {\tt e}(\frac{4}{11}),
    {\tt e}(\frac{5}{11}),
    {\tt e}(\frac{5}{11}),
    {\tt e}(\frac{5}{11}),
    {\tt e}(\frac{6}{11}),
    {\tt e}(\frac{6}{11}),
    {\tt e}(\frac{6}{11}),
    \nonumber\\
    &\quad
    {\tt e}(\frac{7}{11}),
    {\tt e}(\frac{7}{11}),
    {\tt e}(\frac{7}{11}),
    {\tt e}(\frac{8}{11}),
    {\tt e}(\frac{8}{11}),
    {\tt e}(\frac{8}{11}),
    {\tt e}(\frac{10}{11}),
    {\tt e}(\frac{10}{11}),
    {\tt e}(\frac{10}{11})
    \big\} \,.
\end{align}
This modular data are compatible with the Galois automorphisms $ \zeta_{44} \mapsto \zeta_{44}^9 $ or $ \zeta_{44} \mapsto \zeta_{44}^{31} $ in the Galois group $ \Gal(\mathbb{Q}(\zeta_{44})/\mathbb{Q}) \cong \mathbb{Z}_{44}^\times $ of the modular data \eqref{eq: A3A6-Atwist}. The actions of these two automorphisms differ only by an overall sign in the $ S $-matrix. Therefore, we propose that the VOA arising from the topological B-twisted sector is a Galois conjugate of the W-algebra minimal model $ W_4(4,11) $.

\subsubsection{\texorpdfstring{$(A_2,D_5)$}{(A2,D5)}}

Let us now consider the $(A_2,D_5)$ Argyres-Douglas theory whose central charge is $c_{4d} = \frac{175}{66}$, and the BPS quiver is given by $A_2 \sqprod D_5$:
\begin{align}
    \begin{aligned}
        \includegraphics{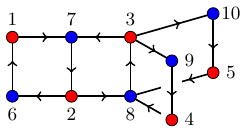}
    \end{aligned}
\end{align}
This quiver encodes a finite chamber with 15 BPS particles together with their CPT conjugate anti-particles, captured by the mutation sequence
\begin{align}
    \mathfrak{M}^{(A_2,D_5)}
    =
    \mathfrak{M}_-\circ
	\mathfrak{M}_+\circ
	\mathfrak{M}_-\,.
\end{align}
This sequence of mutations also determines the increasing ordering of the central charge phases. The electromagnetic charges of the particles arranged in this order are
\begin{align}
    \begin{aligned}
        \G_{\text{BPS}}
        &= \{
            \g_1,\
            \g_2,\
            \g_3,\
            \g_4,\
            \g_5,\
            \g_1+\g_6,\
            \g_2+\g_7,\
            \g_3+\g_8, \\
            &\qquad
            \g_4+\g_9,\
            \g_5+\g_{10},\
            \g_6,\
            \g_7,\
            \g_8,\
            \g_9,\
            \g_{10},\
            (\g_i\to -\g_i)
        \} \, ,
    \end{aligned}
\end{align}
from which the trace formula yields
\begin{align}
	Z_{S_b^3}^{(A_2,D_5)}
	&= i \,
    e^{\frac{5\pi i}{3} (b^2 + b^{-2}) } \int d^8 u \;
    e^{\pi i u^T L u}
    \prod_{i=1}^8 \Phi_b(u_i) \, ,
    \label{eq: A2D5 int}
\end{align}
with
\begin{align}
    L=
    \left(
    \begin{array}{cccccccc}
     0 & -1 & 0 & -1 & 0 & -1 & 0 & 1 \\
 -1 & -4 & 2 & -3 & -2 & 1 & 0 & 1 \\
 0 & 2 & -1 & 0 & 0 & 0 & 0 & 0 \\
 -1 & -3 & 0 & -4 & -2 & 1 & 2 & 1 \\
 0 & -2 & 0 & -2 & -1 & 1 & 0 & 1 \\
 -1 & 1 & 0 & 1 & 1 & 0 & 0 & 0 \\
 0 & 0 & 0 & 2 & 0 & 0 & -1 & 0 \\
 1 & 1 & 0 & 1 & 1 & 0 & 0 & -1 \\
    \end{array}
    \right)\,.
\end{align}
The result \eqref{eq: A2D5 int} indicates a 3d $\CN=2$ $U(1)^{8}$ gauge theory with 8 chiral multiplets, whose charge matrix $Q$ and CS level matrix $K$ are given by
\begin{align}
	Q = {\bf 1}_{8\times 8} \, , \quad
	K = L+QQ^T\,.
    \label{eq: QK A2D5}
\end{align}
There are seven half-BPS monopole operators given by
\begin{gather}
    \begin{gathered}
        \phi_1^2 V_{(0,0,0,0,0,1,0,-1)},\
        \phi_2^2 V_{(0,0,-1,0,0,0,0,0)},\
        \phi_3^2 V_{(0,-1,0,0,1,0,0,-1)},\
        \phi_4^2 V_{(0,0,0,0,0,0,-1,0)},\
        \\
        \qquad\phi_7^2 V_{(0,0,0,-1,1,0,0,-1)},\
        \phi_5 \phi_8 V_{(-1,0,0,0,0,-1,0,0)},\
        \phi_6 \phi_8 V_{(0,0,-1,0,-1,0,-1,0)}\,.
    \end{gathered}
\end{gather}
Once these operators are turned on as superpotential terms, the theory flows to a fixed point with enhanced $\CN=4$ supersymmetry, and a single unbroken combination of $U(1)_{T_i}$ topological symmetries given by
\begin{align}
	A = - T_1 - T_2 - T_4 + T_6 + T_8\,
\end{align}
becomes the $U(1)_A$ symmetry of the $\CN=4$ supersymmetry. The mixing of the topological symmetries is fixed as $\m^* = (0,4,-2,4,2,-2,-2,-3)$, and the superconformal index at this fixed point yields
\begin{align}
    \CI_{S^2\times S^1}^{(A_2,D_5)}(\eta, \nu=0; q)
    &= 1 - q - \left( \eta + \frac{1}{\eta} \right) q^{3/2} - 2q^2 + \left( 1 + \frac{1}{\eta^2}\right)q^3 - \left(\eta^3 - \eta - \frac{2}{\eta} \right)q^{7/2} \nonumber \\
    &\quad - \left( 2\eta^2 - 3 - \frac{1}{\eta^2}\right) q^4 - \left( 2\eta^3 - \eta - \frac{5}{\eta} \right)q^{9/2} + \mathcal{O}(q^5) \, .
\end{align}
The Hilbert series of the Coulomb and Higgs branches become trivial:
\begin{align}
    \CI_{S^2\times S^1}^{(A_2,D_5)}(\eta = 1 , \n=\pm 1;q)  = 1\,,
\end{align}
which implies that the resulting theory is a 3d rank-0 SCFT. We have checked the equality up to $ q^5 $ order. Consequently, one can extract the modular data of the topologically A-twisted sector of this theory as
\begin{align}\label{eq: A2D5-Atwist}
    \begin{aligned}
        \{|S_{0\a}|\} &= \frac{1}{\sqrt{11}} \{ 2\sigma_{22}^3, 1, 1, 2\xi_{11}^1, 2\sigma_{22}^1, 2\sigma_{22}^5, 2\xi_{11}^2  \} \, , \\
        \{T_{\a\a}\} &=
        \big\{
            {\tt e}(0),
            {\tt e}(\frac{1}{11}),
            {\tt e}(\frac{1}{11}),
            {\tt e}(\frac{3}{11}),
            {\tt e}(\frac{7}{11}),
            {\tt e}(\frac{8}{11}),
            {\tt e}(\frac{9}{11})
        \big\} \, ,
    \end{aligned}
\end{align}
which are consistent with those of the $ W^{-80/11}(\mathfrak{so}(10)) $ W-algebra as expected from the SCFT/VOA correspondence. The central charge of the VOA satisfies $c_{2d} = -12 c_{4d} = -\frac{350}{11}$. In contrast, the B-twisted sector produces,
\begin{align}
    \begin{aligned}
        \{|S_{0\a}|\} &= \frac{1}{\sqrt{11}} \{ 2\xi_{11}^2, 2\xi_{11}^1, 2\sigma_{22}^5, 2\sigma_{22}^1, 1, 1, 2\sigma_{22}^3 \} \, , \\
        \{T_{\a\a}\} &=
        \big\{
            {\tt e}(0),
            {\tt e}(\frac{1}{11}),
            {\tt e}(\frac{2}{11}),
            {\tt e}(\frac{4}{11}),
            {\tt e}(\frac{5}{11}),
            {\tt e}(\frac{5}{11}),
            {\tt e}(\frac{7}{11})
        \big\}\,.
    \end{aligned}
\end{align}
This modular data can be obtained from the Galois automorphisms $ \zeta_{44} \mapsto \zeta_{44}^5 $ or $ \zeta_{44} \mapsto \zeta_{44}^{27} $ in the Galois group $ \Gal(\mathbb{Q}(\zeta_{44})/\mathbb{Q}) $ of the modular data \eqref{eq: A2D5-Atwist}. These automorphisms act on $ \sqrt{11} $ as $ \sqrt{11}\mapsto \sqrt{11} $ and $ \sqrt{11} \mapsto -\sqrt{11} $, respectively. The resulting $ S $-matrix differ only by an overall sign. We thus propose that the VOA obtained from the B-twist is a Galois conjugation of the W-algebra $ W^{-80/11}(\mathfrak{so}(10)) $.

\subsubsection{\texorpdfstring{$(A_3,D_4)$}{(A3,D4)}}

Consider the $(A_3,D_4)$ Argyres-Douglas theory whose central charge is $c_{4d} = \frac{97}{30}$. This theory has a rank-2 flavor symmetry and admits a non-trivial Higgs branch. The BPS quiver is given by $A_3 \sqprod D_4$:
\begin{align}
    \begin{aligned}
        \includegraphics{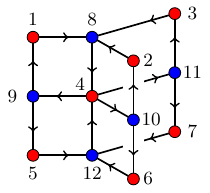}
    \end{aligned}
\end{align}
This quiver encodes a finite chamber with 24 BPS particles and their CPT conjugate anti-particles that can be captured by the mutation sequence
\begin{align}
    \mathfrak{M}^{(A_3,D_4)}
    =
    \mathfrak{M}_+\circ
    \mathfrak{M}_-\circ
	\mathfrak{M}_+\circ
	\mathfrak{M}_-\,.
\end{align}
This mutation sequence determines the increasing ordering of the central charge phases. The electromagnetic charges of the particles arranged in this order are
\begin{align}
    \begin{aligned}
	\G_{\text{BPS}}
	&= \{
    \g_1,\
    \g_2,\
    \g_3,\
    \g_4,\
    \g_5,\
    \g_6,\
    \g_7,\
    \g_1+\g_5+\g_9,\
    \g_2+\g_6+\g_{10},\
    \g_3+\g_7+\g_{11}, \\
    &\qquad
    \g_4+\g_8,\
    \g_4+\g_{12},\
    \g_4+\g_8+\g_{12},\
    \g_1+\g_9,\
    \g_2+\g_{10},\
    \g_3+\g_{11}, \\
    &\qquad
    \g_5+\g_9,\
    \g_6+\g_{10},\
    \g_7+\g_{11},\
    \g_{8},\
    \g_{9},\
    \g_{10},\
    \g_{11},\
    \g_{12},\
    (\g_i \to -\g_i)
	\} \, ,
    \end{aligned}
\end{align}
from which the trace formula yields
\begin{align} \label{eq: A3D4 int}
	Z_{S_b^3}^{(A_3,D_4)}
    &= -i^{3/2} e^{ \frac{7\pi i}{3}(b^2+b^{-2})}
    e^{\pi i ( m_1^2 + m_2^2 )}
    \int du_1 du_2\, e^{\pi i (-u_1^2 + 4 u_1 u_2 - u_2^2)}
    \Phi_b(u_1)\Phi_b(u_2) \\
    &\times
    \int du_3 du_4 du_5\,
    e^{\pi i(-3 u_3^2-8 u_3 u_4+4 u_3 u_5-7 u_4^2-3
   u_5^2+12 u_4 u_5 -4 m_2 u_3+2 m_1 u_4-4 m_2 u_4+6 m_2 u_5 )} , \nonumber
\end{align}
where $m_1$ and $m_2$ are real mass parameters arising from the flavor symmetry of the $ (A_3,D_4) $ theory. Note that the partition function splits into two parts. The first part in the first line, corresponds to a 3d $\CN=2$ $U(1)^2$ gauge theory with two chiral multiplets, whose charge and CS level matrices are
\begin{align}
    Q = {\bf 1}_{2\times 2} \, , \quad
    K =
    \left(
    \begin{array}{cc}
        0 & 2 \\
        2 & 0
    \end{array}
    \right)\,.
\end{align}
This theory has two half-BPS monopole operators given by
\begin{align}
    \phi_1^2 V_{(0,-1)} \, , \quad
    \phi_2^2 V_{(-1,0)} \, .
\end{align}
A superpotential deformation involving these operators yields a unitary 3d TFT whose modular data can be extracted as
\begin{align}
    \{|S_{0\a}|\} &= \Big\{
    \frac{2}{\sqrt{5}} \sin\big( \frac{\pi}{5} \big) ,
    \frac{2}{\sqrt{5}} \sin\big( \frac{2\pi}{5} \big)
    \Big\} \, , \quad
    \{T_{\a\a}\} = \big\{
    {\tt e}(0),
    {\tt e}(\frac{3}{5})
    \big\}\,.
\end{align}
These modular data are consistent with those of the conjugate Fibonacci modular tensor category. The same theory, together with its modular data also appears in the second power of the monodromy operator of the $(A_1,A_2)$ Argyres-Douglas theory \cite{Go:2025ixu}.

On the other hand, the second integral in \eqref{eq: A3D4 int} corresponds to a pure $U(1)^3$ CS theory coupled to two background gauge fields associated with the flavor symmetry. Surprisingly, this CS theory gives rise to a partial modular data compatible with those of the affine Kac-Moody algebra $\widehat{\mathfrak{su}}(3)_{-\frac{12}{5}}$ at admissible level. Consequently, the partial modular data of the 3d TFT arising from the $U(1)_r$ twisted compactification of the $(A_3,D_4)$ theory are given by the tensor product:
\begin{align}
    \begin{aligned}
        \{|S_{0\a}|\} &= \Big\{
            \frac{2}{\sqrt{5}} \sin\big( \frac{\pi}{5} \big) ,
            \frac{2}{\sqrt{5}} \sin\big( \frac{2\pi}{5} \big)
        \Big\}
        \otimes
        \Big\{ \overbrace{\frac{1}{5},\frac{1}{5},\cdots,\frac{1}{5}}^{25} \Big \} \\
        \{T_{\a\a}\} &= \big\{
            {\tt e}(0),
            {\tt e}(\frac{3}{5})
        \big\}
        \otimes
        \big\{
            {\tt e}(0),
            {\tt e}(\frac{1}{5})^{\otimes 6},
            {\tt e}(\frac{2}{5})^{\otimes 6},
            {\tt e}(\frac{3}{5})^{\otimes 6},
            {\tt e}(\frac{4}{5})^{\otimes 6}
        \big\}\,.
        \label{eq: A3D4 MTC}
    \end{aligned}
\end{align}
Moreover, the superconformal index turns out to be trivial:
\begin{align}
    \CI_{S^2\times S^1}^{(A_3,D_4)} = 1 \, ,
\end{align}
indicating that the resulting 3d TFT is unitary. However, the central charge of the tensor product VOA built from the conjugate Fibonacci MTC and $ \widehat{\mathfrak{su}}(3)_{-\frac{12}{5}} $ does not satisfy the relation $ c_{2d}=-12c_{4d} $. As far as we are aware, the associated VOA of the $(A_3,D_4)$ theory is not known. Although we could not match the central charge relation since the modular data currently available do not fully characterize the VOA, we expect that the associated VOA should have modular data compatible with \eqref{eq: A3D4 MTC}.

\subsubsection{\texorpdfstring{$(A_2,E_6)$}{(A2,E6)}}

Let us consider the $(A_2,E_6)$ Argyres-Douglas theory whose central charge is $c_{4d} =\frac{97}{30}$, and the BPS quiver is given by $A_2 \sqprod E_6$:
\begin{align}
    \begin{aligned}
        \includegraphics{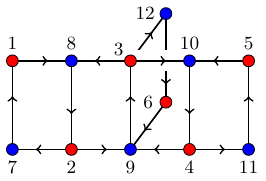}
    \end{aligned}
\end{align}
From the sequence of mutations
\begin{align}
    \mathfrak{M}^{(A_2,E_6)}
    =
    \mathfrak{M}_-\circ
	\mathfrak{M}_+\circ
	\mathfrak{M}_-\,,
\end{align}
we obtain the spectrum of a finite chamber containing 18 BPS particles whose electromagnetic charges can be arranged as,
\begin{align}
    \begin{aligned}
        \G_{\text{BPS}}
        &= \{
            \g_1,\
            \g_2,\
            \g_3,\
            \g_4,\
            \g_5,\
            \g_6,\
            \g_1+\g_{7},\
            \g_2+\g_{8},\
            \g_3+\g_{9},\
            \g_4+\g_{10}, \\
            &\qquad 
            \g_5+\g_{11},\
            \g_6+\g_{12},\
            \g_{7},\
            \g_{8},\
            \g_{9},\
            \g_{10},\
            \g_{11},\
            \g_{12},\
            (\g_i \to -\g_i)
        \} \, ,
    \end{aligned}
\end{align}
ordered by increasing central charge phases. Then, the trace formula yields
\begin{align}
	Z_{S_b^3}^{(A_2,E_6)}
	&= i^{1/2}
    e^{\frac{7\pi i}{3} (b^2 + b^{-2}) }
    e^{\pi i ( 4 m_1^2 - 4 m_1 m_2 - 2 m_2^2)}
    \int du_1 du_2 \,
    e^{\pi i(-u_1^2 + 4 u_1 u_2 - u_2^2)}\Phi_b(u_1)\Phi_b(u_2)
    \nonumber\\
    &\times\int du_3 du_4 du_5\,
    e^{\pi i(
    -2 u_3^2 - 5 u_4^2 + 2 u_3 u_5 - 3 u_5^2 + 4 u_3 m_1 - 8 u_4 m_1 + 4 u_5 m_1  + 4 u_3 m_2 - 16 u_4 m_2 -2 u_5 m_2 
    )} .
    \label{eq: A2E6 int}
\end{align}
Note that, as in the previous example of the $(A_3,D_4)$ theory, the partition function splits into two integrals where the first one gives the modular data of the conjugate Fibonacci MTC. In fact, the $(A_2,E_6)$ theory is dual to the $(A_3,D_4)$ theory:
\begin{align}
    (A_2,E_6) \sim (A_3,D_4) \, ,
    \label{eq: A2E6 dual A3D4}
\end{align}
and we have checked that the integral in the second line of \eqref{eq: A2E6 int} corresponding to a $U(1)^3$ CS theory, yields modular data compatible with that of the affine Kac-Moody algebra $\widehat{\mathfrak{su}}(3)_{-\frac{12}{5}}$ at admissible level. Thus, the partial modular data are coincide with those in \eqref{eq: A3D4 MTC}. Although the details of the pure CS theories inferred from the partition functions $Z_{S_b^3}^{(A_3,D_4)}$ and $Z_{S_b^3}^{(A_2,E_6)}$ are different, they encode the same 3d TFT as expected from the dual relation \eqref{eq: A2E6 dual A3D4}. This example illustrates the consistency of our 3d TFT construction.

\subsubsection{\texorpdfstring{$(A_2,E_8)$}{(A2,E8)}}

Lastly, let us consider the $(A_2,E_8)$ Argyres-Douglas theory whose central charge is $c_{4d}=\frac{164}{33}$. The BPS quiver is given by $A_2 \sqprod E_8$:
\begin{align}
    \begin{aligned}
        \includegraphics{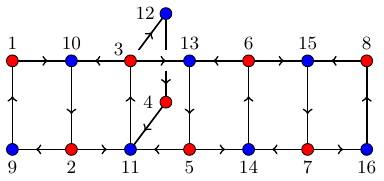}
    \end{aligned}
\end{align}
This quiver admits a finite chamber with 24 BPS particles and their CPT conjugate anti-particles, which are captured by a sequence of mutations
\begin{align}
    \mathfrak{M}^{(A_2,E_8)}
    =
    \mathfrak{M}_-\circ
	\mathfrak{M}_+\circ
	\mathfrak{M}_-\,.
\end{align}
This mutation determines the increasing ordering of the central charge phases. The electromagnetic charges of the particles arranged in this order are
\begin{align}
    \begin{aligned}
        \G_{\text{BPS}}
        &= \{
            \g_1,\
            \g_2,\
            \cdots,\
            \g_7,\
            \g_8,\
            \g_1+\g_9,\
            \g_2+\g_{10},\
            \g_3+\g_{11},\
            \g_4+\g_{12},\
            \g_5+\g_{13},
            \nonumber\\
            &\qquad
            \g_6+\g_{14},\
            \g_7+\g_{15},\
            \g_8+\g_{16},\
            \g_{9},\
            \g_{10},\
            \cdots,\
            \g_{15},\
            \g_{16},
            (\g_i \to -\g_i)
        \}\,.
    \end{aligned}
\end{align}
Consequently, the trace formula yields
\begin{align}
	Z_{S_b^3}^{(A_2,E_8)}
    &= e^{\frac{10 \pi i}{3} (b^2 + b^{-2}) } \int
     d^8 u \;
    e^{\pi i u^T L u}
    \prod_{i=1}^8 \Phi_b(u_i) \, ,
    \label{eq: A2E8 int}
\end{align}
with
\begin{align}
    L=
    \left(
        \begin{array}{cccccccc}
            0 & -1 & 0 & 1 & 1 & 1 & 0 & 0 \\
            -1 & 1 & 1 & -1 & -1 & 0 & 1 & 1 \\
            0 & 1 & 0 & -1 & 1 & 1 & 0 & 0 \\
            1 & -1 & -1 & 1 & -1 & 0 & 1 & 1 \\
            1 & -1 & 1 & -1 & 1 & 0 & -1 & 1 \\
            1 & 0 & 1 & 0 & 0 & 0 & 1 & 1 \\
            0 & 1 & 0 & 1 & -1 & 1 & 0 & 0 \\
            0 & 1 & 0 & 1 & 1 & 1 & 0 & -1 \\
        \end{array}
    \right) \, .
\end{align}
This indicates a 3d $\CN=2$ $U(1)^{8}$ gauge theory with 8 chiral multiplets, whose charge matrix $Q$ and CS level matrix $K$ are given by
\begin{align}
	Q = {\bf 1}_{8\times 8} \, , \quad
	K = L+QQ^T\,.
    \label{eq: QK A2E8}
\end{align}
This 3d theory has seven half-BPS monopole operators given by
\begin{gather}
    \begin{gathered}
        \phi_2^2 V_{(0,0,0,1,0,0,-1,0)},\
        \phi_6^2 V_{(0,0,0,0,1,0,-1,0)},\
        \phi_7^2 V_{(0,-1,1,0,0,-1,0,-1)},\
        \phi_8^2 V_{(1,0,0,0,0,0,-1,0)}, \\
        \phi_1 \phi_3 V_{(0,0,0,-1,-1,0,1,1)},\
        \phi_3 \phi_4 V_{(-1,1,0,0,-1,0,1,0)},\
        \phi_3 \phi_5 V_{(-1,0,0,-1,0,1,1,0)} \, .
    \end{gathered}
\end{gather}
Once these operators are turned on as superpotential terms, the theory flows to a fixed point with $\CN=4$ supersymmetry. A surviving combination of the $U(1)_{T_i}$ topological symmetries given by
\begin{align}
	A = T_1 + T_2 + 3 T_3 + T_4 + T_5 + T_6 + T_7 + T_8 \, ,
\end{align}
becomes the $U(1)_A$ symmetry of the $\CN=4$ algebra, and the mixing of the topological symmetries is fixed to be $\m^* = (-2,-1,-2,-1,-1,-2,-2,-3)$. The 3d superconformal index at this fixed point yields
\begin{align}
    \CI_{S^2\times S^1}^{(A_2,E_8)}(\eta, \nu=0; q)
    &= 1 - q - \left(\eta+\frac{1}{\eta}\right) q^{3/2} - \left(2 + \frac{2}{\eta^2}\right) q^2 + \left( 2\eta - \frac{3}{\eta} - \frac{2}{\eta^3}\right) q^{5/2} \nonumber \\
    &\quad + \left( 3\eta^2 - 1 - \frac{4}{\eta^2} \right)q^3 - \left( 2\eta + \frac{10}{\eta}\right)q^{7/2} - \left( 6\eta^2 + 20 + \frac{6}{\eta^2} - \frac{2}{\eta^4}\right) q^4 \nonumber \\
    &\quad - \left( 4\eta^3 + 18\eta + \frac{18}{\eta} - \frac{5}{\eta^3} \right) q^{9/2} + \mathcal{O}(q^5) \, ,
\end{align}
and the Hilbert series of the Coulomb and Higgs branches become trivial as
\begin{align}
    \CI_{S^2\times S^1}^{(A_2,E_8)}(\eta=1, \nu=\pm 1; q) = 1 \, ,
\end{align}
where we have checked the equality up to $ q^5 $ order. This indicates the resulting 3d theory is a 3d rank-0 SCFT. Hence, we can perform the topological A-twist of this rank-0 theory to extract the modular data as
\begin{align}\label{eq: A2E8-Atwist}
    \{|S_{0\a}|\} &= \frac{2}{11} \big\{
        1 + \sigma_{22}^3 + \sigma_{22}^5 - \xi_{11}^2, 
        2 \sigma_{11}^1 \sigma_{11}^2, 
        2 \sigma_{11}^1 \sigma_{11}^2, 
        2 \sigma_{11}^1 \sigma_{11}^2, 
        1 - \sigma_{22}^1 + \sigma_{22}^5 + \xi_{11}^1, 
        2 \xi_{22}^1 \xi_{22}^5, \nonumber \\
        &\quad 2 \xi_{22}^1 \xi_{22}^5, 
        2 \xi_{22}^1 \xi_{22}^5, 
        1 + \sigma_{22}^1 - \sigma_{22}^3 - \sigma_{22}^5, 
        -1 - \sigma_{22}^1 + \xi_{11}^1 + \xi_{11}^2,
        1 - \sigma_{22}^3 + \xi_{11}^1 + \xi_{11}^2, \nonumber \\
        &\quad 2 \xi_{22}^3 \xi_{22}^5, 
        2 \xi_{22}^3 \xi_{22}^5, 
        2 \xi_{22}^3 \xi_{22}^5, 
        2 \sigma_{11}^2 \xi_{22}^3, 
        2 \sigma_{11}^2 \xi_{22}^3, 
        2 \sigma_{11}^2 \xi_{22}^3, 
        2 \sigma_{11}^1 \xi_{22}^1, 
        2 \sigma_{11}^1 \xi_{22}^1, 
        2 \sigma_{11}^1 \xi_{22}^1
    \}
    \nonumber\\
    \{T_{\a\a}\} &=
    \big\{
    {\tt e}(0),
    {\tt e}(\frac{1}{11}),
    {\tt e}(\frac{1}{11}),
    {\tt e}(\frac{1}{11}),
    {\tt e}(\frac{3}{11}),
    {\tt e}(\frac{4}{11}),
    {\tt e}(\frac{4}{11}),
    {\tt e}(\frac{4}{11}),
    {\tt e}(\frac{5}{11}),
    {\tt e}(\frac{6}{11}),
    \nonumber\\
    &
    {\tt e}(\frac{7}{11}),
    {\tt e}(\frac{8}{11}),
    {\tt e}(\frac{8}{11}),
    {\tt e}(\frac{8}{11}),
    {\tt e}(\frac{9}{11}),
    {\tt e}(\frac{9}{11}),
    {\tt e}(\frac{9}{11}),
    {\tt e}(\frac{10}{11}),
    {\tt e}(\frac{10}{11}),
    {\tt e}(\frac{10}{11})
    \big\}
\end{align}
which are compatible with those of the $ W^{-60/11}(\mathfrak{so}(8)) $ W-algebra as expected from the SCFT/VOA correspondence, whose central charge also satisfies the relation $c_{2d} = -12 c_{2d} = -\frac{656}{11}$. Similarly, performing the topological B-twist yields the modular data given by
\begin{align}
    \{|S_{0\a}|\} &= \frac{2}{11}\{
        1 - \sigma_{22}^1 + \sigma_{22}^5 + \xi_{11}^1, 
        1 - \sigma_{22}^3 + \xi_{11}^1 + \xi_{11}^2, 
        1 + \sigma_{22}^3 + \sigma_{22}^5 - \xi_{11}^2, 
        2 \xi_{22}^1 \xi_{22}^5, 
        2 \xi_{22}^1 \xi_{22}^5, \nonumber \\
        &\quad 2 \xi_{22}^1 \xi_{22}^5, 
        2 \xi_{22}^3 \xi_{22}^5, 
        2 \xi_{22}^3 \xi_{22}^5, 
        2 \xi_{22}^3 \xi_{22}^5, 
        2 \sigma_{11}^1 \sigma_{11}^2, 
        2 \sigma_{11}^1 \sigma_{11}^2, 
        2 \sigma_{11}^1 \sigma_{11}^2, 
        1 + \sigma_{22}^1 - \sigma_{22}^3 - \sigma_{22}^5, \nonumber \\
        &\quad 2 \sigma_{11}^2 \xi_{22}^3, 
        2 \sigma_{11}^2 \xi_{22}^3, 
        2 \sigma_{11}^2 \xi_{22}^3, 
        -1 - \sigma_{22}^1 + \xi_{11}^1 + \xi_{11}^2, 
        2 \sigma_{11}^1 \xi_{22}^1, 
        2 \sigma_{11}^1 \xi_{22}^1, 
        2 \sigma_{11}^1 \xi_{22}^1
    \}
    \nonumber\\
    \{T_{\a\a}\} &=
    \big\{
    {\tt e}(0),
    {\tt e}(\frac{1}{11}),
    {\tt e}(\frac{2}{11}),
    {\tt e}(\frac{3}{11}),
    {\tt e}(\frac{3}{11}),
    {\tt e}(\frac{3}{11}),
    {\tt e}(\frac{4}{11}),
    {\tt e}(\frac{4}{11}),
    {\tt e}(\frac{4}{11}),
    {\tt e}(\frac{5}{11}),
    {\tt e}(\frac{5}{11}),
    \nonumber\\
    &
    {\tt e}(\frac{5}{11}),
    {\tt e}(\frac{6}{11}),
    {\tt e}(\frac{7}{11}),
    {\tt e}(\frac{7}{11}),
    {\tt e}(\frac{7}{11}),
    {\tt e}(\frac{9}{11}),
    {\tt e}(\frac{10}{11}),
    {\tt e}(\frac{10}{11}),
    {\tt e}(\frac{10}{11})
    \big\}\,.
\end{align}
These are compatible with the Galois automorphism $ \zeta_{22} \mapsto \zeta_{22}^{15} $ in the Galois group $ \Gal(\mathbb{Q}(\zeta_{22})/\mathbb{Q}) \cong \mathbb{Z}_{22}^\times $ of the modular data \eqref{eq: A2E8-Atwist}. Therefore, we propose that the VOA arising from the topological B-twist is a Galois conjugate of the W-algebra $ W^{-60/11}(\mathfrak{so}(8)) $. It is known that the $(A_4,D_4)$ theory is dual to the $(A_2,E_8)$ theory and we have verified that the trace formula indeed produces the same result, up to a permutation of the integral variables. See Appendix~\ref{app: ex} for the explicit check.

\section{Discussion} \label{sec: discussion}

In this work, we construct topologically twisted 3d $ \mathcal{N}=4 $ SCFTs arising from the $U(1)_r$ twisted circle compactification of the 4d $\CN=2$ SCFTs, in particular for the $(G,G')$ Argyres-Douglas theories, by employing a wall-crossing invariant that computes the ellipsoid partition function of the resulting 3d theories. We extract partial modular data of the 3d TFTs using A-model techniques and confirm that they are compatible with the modular data of the expected VOAs suggested by the 4d SCFT/2d VOA correspondence. We emphasize that our procedure uses 3d TFTs to study VOAs, thereby allowing us to employ TFT techniques in the analysis. With this advantage, we were able to compute, for the first time, the partial modular data of the previously unexplored VOA associated with the $(A_3,D_4)\sim (A_2,E_6)$ Argyres-Douglas theories. We anticipate that our construction will provide new avenues for investigating the 4d SCFT/2d VOA correspondence. We conclude by suggesting several future directions.

\paragraph{Higher power of monodromy} As discussed in \cite{Kim:2024dxu,Go:2025ixu}, one can consider higher powers of the monodromy operator in the trace formula. They correspond to multi-wrappings of the Janus-like loop and give rise to a family of TFTs related by Galois conjugations for a given 4d Argyres-Douglas theory. However, a practical difficulty arises when we increase the powers, due to the increasing number of quantum dilogarithms to be simplified. This difficulty can be efficiently resolved using our {\tt Mathematica} code. Indeed, we examined several $(G,G')$ theories at higher powers, explicitly verifying the {\it periodicity}, whose power of monodromy operator becomes the identity. This opens a promising direction to explore the full families of TFTs arising from $(G,G')$ Argyres-Douglas theories and it is interesting to ask how many semisimple TFTs can be realized in this manner.

\paragraph{Schur index computation} Since we have explicitly determined the BPS spectra of the general $(G,G')$ Argyres-Douglas theories, it is straightforward to apply them to compute the Schur index using the IR formula \cite{Cordova:2015nma}, which requires exactly the same input. Moreover, the quantum dilogarithm identities implementing the wall-crossing phenomena are universal, allowing our {\tt Mathematica} code to simplify the expressions. In this setting, the vacuum character of the corresponding VOA, rather than the modular $S$- and $T$-matrices, is characterized, which enables a more precise determination of the VOA.

\paragraph{Insertion of line defects} As discussed in \cite{Cordova:2016uwk,Cordova:2017ohl,Cordova:2017mhb,Cirafici:2017iju}, one can also incorporate insertion of defects in the wall-crossing invariant formula. Especially, the supersymmetric line defects in 4d $\CN=2$ theories can be systematically constructed from the framed BPS quiver, and the Schur index formula with such insertions yields linear combinations of the non-vacuum characters of the associated VOA. We expect these line defects in the 4d theory are mapped to {\it simple lines} \cite{Gang:2024loa} in the 3d theory obtained from our trace formula. Thus, by characterizing them, one can compute the full modular $S$-matrix following the map introduced in \cite{Cho:2020ljj}. Furthermore, one can also read the Verlinde algebra from the fusion of the line defects \cite{Neitzke:2017cxz}.

\paragraph{Extension to non-semisimple TFT} Among the $(G,G')$ Argyres-Douglas theories, our analysis does not apply to those with Coulomb branch operators of integer conformal dimensions. These operators survive under the $U(1)_r$ twisted circle compactification, so the resulting 3d TFT contains them, becoming a non-semisimple TFT. At present, we lack the tools to handle this situation. One peculiar observation is that our ellipsoid partition function typically involves free integrals, making the partition function diverges. Since these free integrals solely contribute to the divergence, it is natural to ask whether the remaining convergent part of the partition function might still encode some semisimple TFT data, thereby capturing part of the full theory. We wonder that our trace formula may offer a way to investigate non-semisimple TFTs.

\section*{Acknowledgements}

We would like to thank Cyril Closset, Dongmin Gang, Saebyeok Jeong, Hee-Cheol Kim, Heeyeon Kim, Sungjay Lee, and Jaewon Song for valuable discussions and comments. The research of MK is supported by a KIAS Individual Grant QP097501. The research of SK is supported by a KIAS Individual Grant PG09102 at Korea Institute for Advanced Study.

\appendix

\section{Kac-Moody algebras and their coset models} \label{app:voa}

In this appendix, we review basics of affine Kac-Moody algebras and their modular data at admissible levels. We also review coset models of affine Kac-Moody algebras, which are related to the W-algebra \eqref{eq:Walg-coset} appearing in the VOAs associated with the $ (G, G') $ Argyres-Douglas theories. Throughout this appendix, we denote $ \mathbb{Z} $, $ \mathbb{Z}_+ $ and $ \mathbb{N} $ as the set of integers, positive integers and non-negative integers, respectively, and $ \zeta_n = e^{2\pi i/n} $.

\subsection{Affine Kac-Moody algebras}

We begin by fixing the notation for Lie algebras. Let $ \mathfrak{g} $ be a finite-dimensional semisimple Lie algebra with generators $ J^a $ satisfying the commutation relations $ [J^a, J^b] = if^{ab}_{\;\;\;c}J^c $, where $ f^{ab}_{\;\;\;c} $ are the structure constants. Let $ \mathfrak{h} $ be the Cartan subalgebra of $ \mathfrak{g} $ and $ \mathfrak{h}^* $ the dual space of $ \mathfrak{h} $. We denote by $ \Pi = \{ \alpha_1, \cdots, \alpha_\ell \} \subset \mathfrak{h}^* $ the set of simple roots and by $ \Pi^\vee = \{ \alpha_1^\vee, \cdots, \alpha_\ell^\vee \} \subset \mathfrak{h} $ the set of simple coroots of $ \mathfrak{g} $, where $ \ell=\dim\mathfrak{h} $ is the rank of $ \mathfrak{g} $. We also denote by $ \Delta $ and $ \Delta_+ $ the sets of all roots and positive roots, respectively. The pairing $ \mathfrak{h}^* \times \mathfrak{h} \to \mathbb{C} $ induces a non-degenerate symmetric bilinear form $ ( \cdot, \cdot ) $ on $ \mathfrak{h} $, which provides an isomorphism between $ \mathfrak{h} $ and $ \mathfrak{h}^* $. We shall identify $ \mathfrak{h} $ and $ \mathfrak{h}^* $ via this non-degenerate bilinear form. More generally, there exists a non-degenerate symmetric bilinear form on a semisimple Lie algebra $ \mathfrak{g} $ that extends the bilinear form $ ( \cdot, \cdot ) $ on $ \mathfrak{h} $, called the \emph{Killing form}. The roots and coroots are related by 
\begin{align}\label{eq:coroot-root}
    \alpha^\vee = \frac{2 \alpha}{( \alpha, \alpha )} \, ,
\end{align}
where we normalize $ ( \alpha, \alpha ) = 2 $ for a long root $ \alpha \in \Delta $. The Cartan matrix $ C $ of $ \mathfrak{g} $ is given by $ C_{ij} = ( \alpha_i^\vee, \alpha_j ) $. A convenient basis $ \{ \omega_1, \cdots, \omega_\ell\} $ of $ \mathfrak{h}^* $ is called the \emph{fundamental weights}, defined as the dual of the simple coroots:
\begin{align}
    ( \omega_i, \alpha_j^\vee ) = \delta_{ij} \, .
\end{align}
The simple roots can be expressed as $ \alpha_i = \sum_j C_{ij} \omega_j $. We define two distinguished elements in $ \mathfrak{h}^* $: the \emph{highest root} $ \theta $ and the \emph{Weyl vector} $ \rho $. The highest root is the unique root $ \theta \in \Delta_+ $ such that $ \theta + \alpha_i \notin \Delta $ for all $ \alpha_i \in \Pi $, while the Weyl vector is the weight vector $ \rho \in \mathfrak{h}^* $ satisfying $ ( \rho, \alpha_i^\vee ) = 1 $ for all $ \alpha_i \in \Pi $. They can be represented as
\begin{align}
    \theta = \sum_{i=1}^\ell a_i \alpha_i = \sum_{i=1}^r a_i^\vee \alpha_i^\vee \, , \quad
    \rho = \sum_{i=1}^\ell \omega_i = \frac{1}{2}\sum_{\alpha\in \Delta_+} \alpha \, ,
\end{align}
where the coefficients $ a_i = 2a_i^\vee / ( \alpha_i, \alpha_i ) $ and $ a_i^\vee $ are called the marks and comarks, respectively. The \emph{dual Coxeter number} is defined by
\begin{align}
    h^\vee = 1+\sum_{i=1}^\ell a_i^\vee \, ,
\end{align}
and its value for simply-laced Lie algebras is listed in Table~\ref{table:group-const}. We also define the \emph{weight lattice} $ P $, the \emph{root lattice} $ Q $ and the \emph{coroot lattice} $ Q^\vee $ as
\begin{align}
    P = \bigoplus_{i=1}^\ell \mathbb{Z} \omega_i \, , \quad
    Q = \bigoplus_{i=1}^\ell \mathbb{Z} \alpha_i \, , \quad
    Q^\vee = \bigoplus_{i=1}^\ell \mathbb{Z} \alpha_i^\vee \, .
\end{align}
The root lattice is a sublattice of the weight lattice, and $ |P/Q| = \det C $. For a simply-laced Lie algebra $ \mathfrak{g} $, we have $ Q=Q^\vee $.

For a root $ \alpha \in \Delta $, the Weyl reflection $ r_\alpha $ acting on a weight $ \lambda \in \mathfrak{h}^* $ is defined by
\begin{align}
    r_\alpha(\lambda) = \lambda - ( \alpha^\vee, \lambda ) \alpha \, ,
\end{align}
which is a reflection of $ \lambda $ with respect to the hyperplane orthogonal to $ \alpha $. The \emph{Weyl group} $ W $ of $ \mathfrak{g} $ is the group generated by all Weyl reflections $ r_{i} \equiv r_{\alpha_i} $ associated with the simple roots $ \alpha_i \in \Pi $. Every element $ w \in W $ can be expressed as a product of the generators $ r_{i} $; we denote by $ \ell(w) $ the minimum number of generators in such a decomposition. The \emph{signature} of $ w $ is defined as $ \epsilon(w) = (-1)^{\ell(w)} $. For $ w \in W $, the \emph{shifted Weyl reflection} is defined by
\begin{align}
    w \cdot \lambda = w(\lambda+\rho) - \rho \, ,
\end{align}
where $ \rho $ is the Weyl vector.

\begin{table}
    \centering
    \begin{tabular}{c|ccccc}
        & $ A_n $ & $ D_n $ & $ E_6 $ & $ E_7 $ & $ E_8 $ \\ \hline
        $ \dim \mathfrak{g} $ & $ (n+1)^2-1 $ & $ n(2n-1) $ & $ 78 $ & $ 133 $ & $ 248 $ \\
        $ h^\vee $ & $ n+1 $ & $ 2n-2 $ & 12 & 18 & 30 \\
        $ |W| $ & $ (n+1)! $ & $ 2^{n-1}n! $ & $ 51840 $ & $ 2903040 $ & $ 696729600 $ \\
        $ |P/Q| $ & $ n+1 $ & $ 4 $ & $ 3 $ & $ 2 $ & $ 1 $
    \end{tabular}
    \caption{Group theoretic constants for the simply-laced Lie algebras.} \label{table:group-const}
\end{table}

\begin{figure}[t]
	\centering
	\begin{subfigure}[b]{0.2\linewidth}
		\centering
        \includegraphics{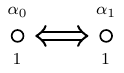}
        \caption{$ \widehat{\mathfrak{su}}(2) $}
		\vspace{3ex}
	\end{subfigure}
	\begin{subfigure}[b]{0.35\linewidth}
		\centering
        \includegraphics{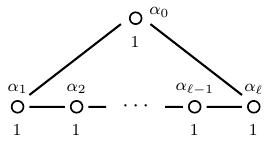}
        \caption{$ \widehat{\mathfrak{su}}(\ell+1) \ (\ell \geq 2) $}
		\vspace{3ex}
	\end{subfigure}
	\begin{subfigure}[b]{0.4\linewidth}
		\centering
        \includegraphics{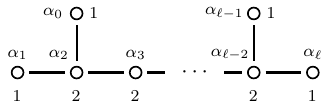}
        \caption{$ \widehat{\mathfrak{so}}(2\ell) \ (\ell \geq 4) $}
		\vspace{3ex}
	\end{subfigure}
    \hfill	
    \begin{subfigure}[b]{0.4\textwidth}
		\centering
        \includegraphics{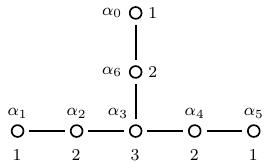}
        \caption{$ \hat{\mathfrak{e}}_{6} $}
		\vspace{3ex}
	\end{subfigure}
	\begin{subfigure}[b]{0.5\textwidth}
		\centering
        \includegraphics{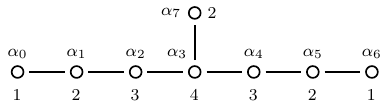}
        \caption{$ \hat{\mathfrak{e}}_{7} $}
		\vspace{3ex}
	\end{subfigure}
	\hfill
	\begin{subfigure}[b]{0.9\textwidth}
		\centering
        \includegraphics{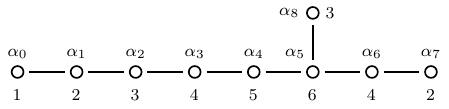}
        \caption{$ \hat{\mathfrak{e}}_{8} $}
		\vspace{3ex}
	\end{subfigure}
    \caption{Dynkin diagrams of (untwisted) affine Lie algebras $ \hat{\mathfrak{g}} $ associated to the simply-laced Lie algebras $ \mathfrak{g} $. The number for each node is the comarks $ a_i^\vee $, which is same with the marks $ a_i $ in the case of the simply-laced Lie algebras. The Dynkin diagrams without the affine node $ \alpha_0 $ reduce to the Dynkin diagrams of $ \mathfrak{g} $.} 
	\label{fig:dynkin}
\end{figure}

We now review the affine extension of the simple Lie algebra. Mathematical details can be found in \cite{Kac:1990gs}. The (untwisted) \emph{affine Lie algebra} (or the \emph{affine Kac-Moody algebra}) associated with $ \mathfrak{g} $ is defined as the central extension of the loop algebra $ \mathfrak{g}[t,t^{-1}] = \mathfrak{g} \otimes \mathbb{C}[t,t^{-1}] $ given by
\begin{align}
    \hat{\mathfrak{g}} = \mathfrak{g}[t,t^{-1}] \oplus \mathbb{C}K \oplus \mathbb{C}d \, .
\end{align}
Here, $ K $ is a central element, $ d = t \frac{d}{dt} $ and $ \mathfrak{g}[t,t^{-1}] $ is an infinite-dimensional Lie algebra whose generators are given by $ J^a_n = J^a \otimes t^n $, where $ J^a $ is a generator of $ \mathfrak{g} $ and $ n \in \mathbb{Z} $. If the generators $ J^a $ are chosen to be orthonormal with respect to the Killing form of $ \mathfrak{g} $, then the generators of $ \hat{\mathfrak{g}} $ satisfy the following commutation relations:
\begin{align}
    [J^a_n, J^b_m] &= if^{ab}_{\;\;\;c} J^c_{n+m} + K n \delta_{ab} \delta_{n+m,0} \, , \quad
    [d, J^a_n] = n J^a_{n} \, .
\end{align}
The Cartan subalgebra of $ \hat{\mathfrak{g}} $ is $ \hat{\mathfrak{h}} = \mathfrak{h} \oplus \mathbb{C}K \oplus \mathbb{C}d $. The non-degenerate symmetric bilinear form on $ \mathfrak{h} $ is extended to $ \hat{\mathfrak{h}} $ by
\begin{align}
    ( \mathfrak{h}, \mathbb{C}K + \mathbb{C}L_0 ) = ( K,K ) = ( d,d )= 0  \, , \quad
    ( K, d ) = 1 \, .
\end{align}
This bilinear form identifies the dual space $ \hat{\mathfrak{h}}^* $ and $ \hat{\mathfrak{h}} $. The sets of simple roots and coroots of $ \hat{\mathfrak{g}} $ are $ \widehat{\Pi} = \{ \alpha_0, \alpha_1, \cdots, \alpha_\ell \} \subset \hat{\mathfrak{h}}^* $ and $ \widehat{\Pi}^\vee = \{\alpha_0^\vee, \alpha_1^\vee, \cdots, \alpha_\ell^\vee\} \subset \hat{\mathfrak{h}} $, respectively, and roots and coroots are related by \eqref{eq:coroot-root}. The Dynkin diagram of $ \hat{\mathfrak{g}} $ associated with the simply-laced algebra $ \mathfrak{g} $ is shown in Figure~\ref{fig:dynkin}, and the Cartan matrix $ C_{ij} = ( \alpha_i^\vee, \alpha_j ) $ of $ \hat{\mathfrak{g}} $ can be read off from the diagrams. In the root system of $ \hat{\mathfrak{g}} $, there is a special root called the \emph{imaginary root} given by
\begin{align}
    \delta = \sum_{i=0}^\ell a_i \alpha_i \, ,
\end{align}
where $ a_0 = 2a_0^\vee / ( \alpha_0, \alpha_0 ) $ for the 0th comark $ a_0^\vee=1 $, and satisfies $ ( \delta, \alpha_i^\vee ) = ( \delta, \delta ) = 0 $. Under the isomorphism between $ \hat{\mathfrak{h}} $ and $ \hat{\mathfrak{h}}^* $, the central element $ K $ is identified with the imaginary root $ \delta $. The 0th simple root $ \alpha_0 $ can be expressed as $ \alpha_0 = -\theta+\delta $, where $ \theta $ is the highest root of $ \mathfrak{g} $. The full root system of $ \hat{\mathfrak{g}} $ is given by
\begin{align}
    \hat{\Delta} = \hat{\Delta}^{\mathrm{re}} \cup \hat{\Delta}^{\mathrm{im}} \, , \quad
    \hat{\Delta}^{\mathrm{re}} = \{ \alpha + n\delta \mid \alpha \in \Delta, n \in \mathbb{Z} \} \, , \quad
    \hat{\Delta}^{\mathrm{im}} = \{ n\delta \mid n \in \mathbb{Z}^\times \} \, ,
\end{align}
where the elements in $ \hat{\Delta}^{\mathrm{re}} $ and $ \hat{\Delta}^{\mathrm{im}} $ are referred to as the real roots and imaginary roots, respectively, and $ \mathbb{Z}^\times = \mathbb{Z} \setminus \{0\} $. We also denote $ \hat{\Delta}_+ = \Delta_+ \cup \{ \alpha+n \delta \mid \alpha \in \Delta, n\in \mathbb{Z}_+ \} $ as the set of positive roots.

Analogously to the finite-dimensional Lie algebra $ \mathfrak{g} $, we introduce the fundamental weights $ \{\widehat{\omega}_0, \widehat{\omega}_1, \cdots, \widehat{\omega}_\ell \} $ as the dual basis of simple coroots: $ ( \widehat{\omega}_i, \alpha_j^\vee ) = \delta_{ij} $. The affine fundamental weights $ \widehat{\omega}_i $ and the fundamental weights $ \omega_i $ of $ \mathfrak{h} $ are related by
\begin{align}
    \widehat{\omega_i} = a_i^\vee \widehat{\omega}_0 + \omega_i \, .
\end{align}
Under the isomorphism between $ \hat{\mathfrak{h}} $ and $ \hat{\mathfrak{h}}^* $, the derivation $ d $ is identified with $ a_0 \omega_0 $. An affine weight $ \hat{\lambda} \in \hat{\mathfrak{h}}^* $ can be expanded as
\begin{align}
    \hat{\lambda} = \sum_{i=0}^\ell \lambda_i \widehat{\omega}_i + l \delta \, ,
\end{align}
where $ l \in \mathbb{R} $ and the coefficients $ \lambda_i $ are called Dynkin labels. For simplicity, we will also denote $ \hat{\lambda} $ as $ \hat{\lambda}=[\lambda_0,\lambda_1,\cdots,\lambda_\ell] $ using the Dynkin labels when the $ \delta $-component is not relevant. The Dynkin labels of the simple roots $ \alpha_i $ are the entries of the Cartan matrix $ C_{ij} $. The \emph{affine Weyl vector} is defined as
\begin{align}
    \hat{\rho} = \sum_{i=0}^\ell \widehat{\omega}_i \, .
\end{align}
If all the Dynkin labels are non-negative integer, then $ \hat{\lambda} $ is called \emph{dominant}. The \emph{level} of $ \hat{\lambda} $ is defined by
\begin{align}
    k = ( K, \hat{\lambda} ) = \sum_{i=0}^\ell a_i^\vee \lambda_i \, .
\end{align}
We denote $ P_+ $ and $ P_+^k $ as the sets of dominant weights and dominant weights of level $ k $, respectively:
\begin{align}
    P_+ = \{ \hat{\lambda} \in \hat{\mathfrak{h}}^* \mid ( \hat{\lambda}, \alpha_i^\vee ) \geq 0 ,\  \forall \alpha_i \in \widehat{\Pi}^\vee \} \, , \quad
    P_+^k = \{ \hat{\lambda} \in P_+ \mid ( \lambda, K ) = k \} \, .
\end{align}
For a fixed level $ k \in \mathbb{Z}_+ $, there are finitely many \emph{dominant highest weight representations}, namely, the highest weight representations whose highest weights are dominant weights. We will denote the affine Kac-Moody algebra at level $ k $ by $ \hat{\mathfrak{g}}_k $.

The affine Kac-Moody algebra can be realized in physics through the Wess-Zumino-Witten (WZW) model. The fundamental aspects of WZW models and their relation to affine Lie algebras can be found in \cite{DiFrancesco:1997nk}. A WZW model whose target space is the Lie group associated with the Lie algebra $ \mathfrak{g} $ at level $ k\in \mathbb{Z}_+ $ has a conserved current whose component $ j^a $ satisfies the \emph{current algebra}
\begin{align}
    j^a(z) j^b(w) + \frac{k \delta^{ab}}{(z-w)^2} + \frac{if^{ab}_{\;\;\;c} j^c(w)}{z-w} + \mathcal{O}\left((z-w)^0\right) \, .
\end{align}
The energy-momentum tensor can be expressed in terms of the currents $ j^a $ via the Sugawara construction. The OPE of the energy-momentum tensor determines the central charge of the WZW model as
\begin{align}\label{eq:aff-c}
    c = \frac{k \dim \mathfrak{g}}{k + h^\vee} \, .
\end{align}
The modes $ J^a_n $ of the conserved currents $ j^a(z) = \sum J^a_n z^{-n-1} $ and the Virasoro generators $ L_n $ satisfy the following commutation relations:
\begin{align}
    \begin{aligned}
        \relax [L_n, L_m] &= (n-m)L_{n+m} + \frac{c}{12}(n^3-n) \delta_{n+m,0} \, , \\
        \relax [L_n, J^a_m] &= -m J^a_{n+m} \, , \quad
        \relax [J^a_n, J^b_m] = if^{ab}_{\;\;\;c} J^c_{n+m} + K n \delta_{ab} \delta_{n+m,0} \, .
    \end{aligned}
\end{align}
This structure is the Virasoro algebra and the affine Lie algebra with the identification $ d=-L_0 $. The primary fields of the WZW model correspond to dominant highest weight states, whose conformal weights are given by
\begin{align}
    h_{\hat{\lambda}} = \frac{( \lambda, \lambda+2\rho )}{2(k+h^\vee)} \, ,
\end{align}
where $ \lambda = \sum_{i=1}^r \lambda_i \omega_i $ is a weight in $ \mathfrak{h}^* $. The characters $ \chi_{\hat{\lambda}}(\tau) = \Tr e^{2\pi i \tau (L_0-c/24)} $ transform under the modular transformations as
\begin{align}
    \chi_{\hat{\lambda}}(\tau+1) = \sum_{\hat{\mu}} T_{\hat{\lambda}\hat{\mu}} \chi_{\hat{\mu}} \, , \quad
    \chi_{\hat{\lambda}}(-1/\tau) = \sum_{\hat{\mu}} S_{\hat{\lambda}\hat{\mu}} \chi_{\hat{\mu}} \, ,
\end{align}
where the $ T $- and $ S $-matrices are given by
\begin{align}
    T_{\hat{\lambda}\hat{\mu}}
    &= \delta_{\hat{\lambda}\hat{\mu}} e^{2\pi i (h_{\hat{\lambda}}-c/24)}
    = \delta_{\hat{\lambda}\hat{\mu}} \exp\left[2\pi i \left( \frac{( \lambda+\rho, \lambda+\rho )}{2(k+h^\vee)} - \frac{( \rho, \rho )}{2h^\vee} \right) \right] \, , \label{eq:affine-T} \\
    S_{\hat{\mu}\hat{\nu}}
    &= i^{|\Delta_+|} |P/Q^\vee|^{-1/2} (k+h^\vee)^{-\ell/2} \sum_{w\in W} \epsilon(w) \exp\left(- \frac{2\pi i}{k+h^\vee} ( w(\mu+\rho), \nu+\rho ) \right) \, .
\end{align}
These matrices are unitary, $ TT^\dagger = SS^\dagger = 1 $.

\paragraph{Example 1}\label{eg:su2-1}
As an example, let us consider the modular matrices of $ \hat{\mathfrak{g}}_k = \widehat{\mathfrak{su}}(2)_1 $, whose central charge \eqref{eq:aff-c} is $ c=1 $. The simple root of the associated finite-dimensional algebra $ \mathfrak{g}=\mathfrak{su}(2) $ is $ \alpha_1=2\omega_1 $, and the bilinear form is given by $ ( \omega_1, \omega_1 ) = 1/2 $. The Weyl group of $ \mathfrak{g} $ is $ W = \{1, r_1\} $, where $ r_1(\lambda) = -\lambda $ for $ \lambda \in \mathfrak{h}^* $. From the Dynkin diagram shown in Figure~\ref{fig:dynkin}(a), the set of simple roots of $ \hat{\mathfrak{g}} $ is $ \widehat{\Pi} = \{ \alpha_0 = 2\widehat{\omega}_0-2\widehat{\omega}_1 + \delta, \alpha_1 = -2\widehat{\omega}_0+2\widehat{\omega}_1 \} $. There are two dominant highest weights at level $ 1 $, given by $ \hat{\lambda} = \hat{\omega}_0 $ and $ \hat{\omega}_1 $. The modular matrices are given by
\begin{align}
    T = e^{-\pi i/12} \begin{pmatrix} 1 & 0 \\ 0 & i \end{pmatrix} \, , \quad
    S = \frac{1}{\sqrt{2}} \begin{pmatrix} 1 & 1 \\ 1 & -1 \end{pmatrix} \, .
\end{align}

\paragraph{Example 2}
We next consider the $ \hat{\mathfrak{g}}_k = \widehat{\mathfrak{su}}(3)_1 $ as another example, whose central charge is $ c=2 $. The simple roots of the associated finite-dimensional Lie algebra $ \mathfrak{g}=\mathfrak{su}(3) $ are $ \Pi=\{\alpha_1=2\omega_1-\omega_2, \alpha_2=-\omega_1+2\omega_2\} $, while the set of positive roots is $ \Delta_+=\{\alpha_1,\alpha_2,\alpha_1+\alpha_2\} $. The bilinear form is given by $ ( \omega_i, \omega_i ) = \frac{2}{3} $ and $ ( \omega_1, \omega_2 ) = \frac{1}{3} $. The Weyl group of $ \mathfrak{g} $ is $ W = \{1,s_1,s_2, s_1s_2, s_2s_1, s_1s_2s_1\} $, where $ s_1 $ and $ s_2 $ are the Weyl reflections with respect to the simple roots $ \alpha_1 $ and $ \alpha_2 $, respectively. The dominant highest weights at level $ 1 $ are $ \hat{\lambda} = \widehat{\omega}_0, \widehat{\omega}_1, \widehat{\omega}_2 $. The modular matrices of $ \widehat{\mathfrak{su}}(3)_1 $ are
\begin{align}
    T = e^{-\pi i/6} \begin{pmatrix} 1 & 0 & 0 \\ 0 & \zeta_3 & 0 \\ 0 & 0 &\zeta_3 \end{pmatrix} \, , \quad
    S = \frac{1}{\sqrt{3}} \begin{pmatrix} 1 & 1 & 1 \\ 1 & \zeta_3 & \zeta_3^2 \\ 1 & \zeta_3^2 & \zeta_3 \end{pmatrix} \, .
\end{align}

\subsection{Admissible representations} \label{sec:admissible}

In this section, we consider affine Kac-Moody algebras at fractional levels. Although the WZW action is not well-defined for non-integer levels, the corresponding vertex operator algebra can be still defined through the Sugawara construction. Moreover, for a certain special fractional levels $ k \in \mathbb{Q} $, known as \emph{admissible levels}, there exists a finite number of primary fields called \emph{admissible representations}, whose characters transform covariantly under the modular $ \mathrm{SL}(2,\mathbb{Z}) $ transformations \cite{Kac:1988qc, Kac:1989}. The affine Kac-Moody algebra at an admissible level is an example of non-rational but quasi-lisse and logarithmic VOA \cite{Feigin:1995xp, Arakawa:2016hkg, Gaberdiel:2001ny}.
 
For a given weight $ \hat{\lambda} \in \hat{\mathfrak{h}}^* $, let $ \hat{\Delta}^{\hat{\lambda}} = \{ \alpha \in \hat{\Delta}^{\mathrm{re}} \mid ( \hat{\lambda}, \alpha^\vee ) \in \mathbb{Z} \} $. The weight $ \hat{\lambda} $ is called an admissible weight if
\begin{align}
    ( \hat{\lambda} + \hat{\rho}, \alpha^\vee ) \notin \mathbb{Z}_{\leq 0} \ \text{for all } \alpha \in \hat{\Delta}_+ \, , \quad
    \mathbb{Q} \hat{\Delta}^{\hat{\lambda}} = \mathbb{Q} \hat{\Delta} \, .
\end{align}
The level $ k=( K, \hat{\lambda} ) $ of an admissible weight is a rational number with denominator $ u \in \mathbb{Z}_+ $ satisfying
\begin{align}
    k + h^\vee \geq \frac{h^\vee}{u} \, , \quad
    \gcd(u,h^\vee) = \gcd(u,r^\vee) = 1 \, ,
\end{align}
where $ r^\vee=1 $ for $ \mathfrak{g} $ of type $ A,D,E $; $ r^\vee=2 $ for types $ B,C,F $; and $ r^\vee=3 $ for $ G_2 $. It is possible to decompose an admissible highest weight $ \hat{\lambda} $ into two integral weights $ \hat{\lambda}^I $ and $ \hat{\lambda}^{F,y} $ as \cite{Mathieu:1990dy}
\begin{align}
    \hat{\lambda} = y \cdot \left( \hat{\lambda}^I - (k+h^\vee) \hat{\lambda}^{F,y} \right) \, ,
\end{align}
where $ y \in W $. Here, the levels of $ \hat{\lambda}^I $ and $ \hat{\lambda}^{F,y} $ are given by
\begin{align}\label{eq:admissible-levels}
    k^I = u(k+h^\vee) - h^\vee \, , \quad
    k^F = u - 1 \, ,
\end{align}
and they are non-negative integers for an admissible weight $ \hat{\lambda} $. In addition, $ \hat{\lambda}^{F,y} $ satisfies
\begin{align}\label{eq:frac-weight-cond}
    \lambda_j^{F,y} \in \frac{a_j}{a_j^\vee} \mathbb{Z} \, , \quad
    \lambda_j^{F,y} \sum_{i=0}^\ell a_i^\vee \alpha_i^\vee + y(\alpha_j^\vee) \in \widehat{Q}_+^\vee \, ,
\end{align}
where $ \widehat{Q}_+^\vee = \bigoplus_{i=0}^\ell \mathbb{N} \alpha_i^\vee \setminus \{0\} $. We note that $ a_j/a_j^\vee $ is always an integer, and equals one when $ \mathfrak{g} $ is a simply-laced Lie algebra. We denote the set of admissible highest weights $ \hat{\lambda} $ at a level $ k $ for a fixed $ y $ as $ P_y^k $.

Not all elements $ y\in W $ yield independent admissible highest weights $ \hat{\lambda} $. It turns out that considering $ y \in W/W' $ is sufficient to construct all admissible highest weights, where $ W' $ is a subgroup of the Weyl group isomorphic to the outer automorphism group $ \operatorname{Out}(\hat{\mathfrak{g}}) $ of $ \hat{\mathfrak{g}} $. The subgroup $ W' $ can be identified as follows. Let $ A \in \operatorname{Out}(\hat{\mathfrak{g}}) $ whose action on the generators for a weight of $ \hat{\mathfrak{g}} $ is given in Table~\ref{table:outauto}. For each element $ A $, we associate an element $ w_A \in W $ satisfying
\begin{align}
    A \hat{\lambda} = k(A-1) \widehat{\omega}_0 + w_A \hat{\lambda} \, .
\end{align}
Such an element $ w_A $ is given by $ w_A = w_i w_0 $, where $ w_0 $ is the longest element of $ W $, and $ w_i $ is the longest element of the subgroup of $ W $ generated by all Weyl reflections $ r_{j\neq i} $ for which $ A\widehat{\omega}_0 = \widehat{\omega}_i $. Finally, $ W' $ is defined as the subgroup generated by all $ w_A $ for $ A \in \operatorname{Out}(\hat{\mathfrak{g}}) $.

\begin{table}
    \centering
    \begin{tabular}{ccl}
        $ \mathfrak{g} $ & $ \operatorname{Out}(\hat{\mathfrak{g}}) $ & Action of generators on $ \hat{\lambda} \in \hat{\mathfrak{h}}^* $ \\ \hline
        $ A_n $ & $ \mathbb{Z}_{n+1} $ & $ [\lambda_0, \lambda_1, \cdots, \lambda_n] \mapsto [\lambda_n, \lambda_0, \lambda_1, \cdots, \lambda_{n-1}] $ \\
        $ D_{n=\mathrm{even}} $ & $ \mathbb{Z}_2 \times \mathbb{Z}_2 $ & $ [\lambda_0, \lambda_1, \cdots, \lambda_n] \mapsto [\lambda_1, \lambda_0, \lambda_2, \cdots, \lambda_{n-2}, \lambda_n, \lambda_{n-1}] $ \\
        & & $ [\lambda_0, \lambda_1, \cdots, \lambda_n] \mapsto [\lambda_n, \lambda_{n-1}, \lambda_{n-2}, \cdots, \lambda_0] $ \\
        $ D_{n=\mathrm{odd}} $ & $ \mathbb{Z}_4 $ & $ [\lambda_0, \lambda_1, \cdots, \lambda_n] \mapsto [\lambda_{n-1}, \lambda_{n}, \lambda_{n-2}, \cdots, \lambda_2, \lambda_1, \lambda_0] $ \\
        $ E_6 $ & $ \mathbb{Z}_3 $ & $ [\lambda_0, \lambda_1, \cdots, \lambda_6] \mapsto [\lambda_1, \lambda_5, \lambda_4, \lambda_3, \lambda_6, \lambda_0, \lambda_2] $ \\
        $ E_7 $ & $ \mathbb{Z}_2 $ & $ [\lambda_0, \lambda_1, \cdots, \lambda_7] \mapsto [\lambda_6, \lambda_5, \lambda_4, \lambda_3, \lambda_2, \lambda_1, \lambda_0, \lambda_7] $ \\
        $ E_8 $ & $ \{1\} $ & trivial
    \end{tabular}
    \caption{Outer automorphisms of $ \hat{\mathfrak{g}} $ associated with a simply-laced Lie algebra $ \mathfrak{g} $}\label{table:outauto}
\end{table}

The characters of the admissible representations form a finite-dimensional representation of $ \mathrm{SL}(2,\mathbb{Z}) $. The modular $ T $-matrix takes the same form as in \eqref{eq:affine-T}, while the $ S $-matrix is given by
\begin{align}
    \begin{aligned}
        S_{\hat{\lambda}\hat{\mu}}
        &= i^{|\Delta_+|} |P/Q^\vee|^{-1/2} (u^{2}(k+h^\vee))^{-\ell/2} \epsilon(yy') \\
        &\quad \cdot \exp\left(2\pi i \left( ( \lambda^I+\rho, \mu^F ) + ( \lambda^F, \mu^I + \rho ) - (k+h^\vee) ( \lambda^F, \mu^F ) \right) \right) \\
        &\quad \cdot \sum_{w\in W} \epsilon(w) \exp\left(- \frac{2\pi i}{k+h^\vee} ( w(\lambda^I+\rho), \mu^I+\rho ) \right) \, ,
    \end{aligned}
\end{align}
where $ \lambda^I $ and $ \lambda^F $ are the finite parts of the affine weights $ \hat{\lambda}^I $ and $ \hat{\lambda}^F = y(\hat{\lambda}^{F,y}) $, respectively.

\paragraph{Example 1}
We first consider the case of $ \hat{\mathfrak{g}} = \widehat{\mathfrak{su}}(2) $. Since $ W=W' $ for the $ \mathfrak{su}(2) $ algebra, the only possible $ y $ is the identity element, and the conditions \eqref{eq:frac-weight-cond} yield $ \lambda_j^{F,1} \in \mathbb{Z} $ and $ \lambda_j^{F,1} \geq 0 $. Now, let us consider the admissible level $ k = -\frac{4}{3} $ which has central charge $ c=-6 $. From \eqref{eq:admissible-levels}, we have $ (k^I, k^F)=(0,2) $, and the possible choices of $ \hat{\lambda}^I $ and $ \hat{\lambda}^{F,1} $ are
\begin{align}
    \hat{\lambda}^I = [0,0] , \quad
    \hat{\lambda}^{F,1} = [2,0] , \, [1,1] , \, [0,2] \, .
\end{align}
Thus, there are three admissible highest weights given by
\begin{align}\label{eq:su2-4/3}
    \hat{\lambda} = -\frac{4}{3}[1,0] , \, -\frac{2}{3} [1,1], \, -\frac{4}{3}[0,1] \, ,
\end{align}
and their modular matrices are
\begin{align}
    T = i \begin{pmatrix} 1 & 0 & 0 \\ 0 & \zeta_3^2 & 0 \\ 0 & 0 & \zeta_3^2 \end{pmatrix} \, , \quad
    S = -\frac{1}{\sqrt{3}} \begin{pmatrix} 1 & -1 & 1 \\ -1 & \zeta_3^2 & -\zeta_3 \\ 1 & -\zeta_3 & \zeta_3^2 \end{pmatrix} \, .
\end{align}

\paragraph{Example 2}
We next consider $ \hat{\mathfrak{g}} = \widehat{\mathfrak{su}}(3) $. The outer automorphism group of $ \hat{\mathfrak{g}} $ is $ W'=\{1,r_1r_2,r_2r_1\} \cong \mathbb{Z}_3 $, and consequently, $ W/W' = \{1,r_1\} $. The conditions \eqref{eq:frac-weight-cond} for $ \lambda_j^{F,y} $ are $ \lambda_j^{F,y} \in \mathbb{Z} $ and
\begin{align}
    \lambda_j^{F,1} \geq 0 \, , \quad
    \lambda_{j \neq 1}^{F,r_1} \geq 0 \, , \quad
    \lambda_1^{F,r_1} \geq 1 \, .
\end{align}
Now, let us consider the admissible level $ k = -\frac{3}{2} $, which has central charge $ c=-8 $. From \eqref{eq:admissible-levels}, we find $ (k^I,k^F)=(0,1) $. Hence, $ \hat{\lambda}^I=[0,0] $, and there are four possible choices of $ \hat{\lambda}^{F,y} $ given by
\begin{align}
    \hat{\lambda}^{F,1} = [1,0,0] , \, [0,1,0] , \, [0,0,1] , \quad
    \hat{\lambda}^{F,s_1} = [0,1,0] \, .
\end{align}
For each case, the admissible highest weight is
\begin{align}
    \hat{\lambda} = -\frac{3}{2}[1,0,0] , \, 
    -\frac{3}{2}[0,1,0] , \,
    -\frac{3}{2}[0,0,1] , \,
    -\frac{1}{2}[1,1,1] \, .
\end{align}
The corresponding modular matrices are
\begin{align}
    T = e^{2\pi i/3} \begin{pmatrix} 1 & 0 & 0 & 0 \\ 0 & -1 & 0 & 0 \\ 0 & 0 & -1 & 0 \\ 0 & 0 & 0 & -1 \end{pmatrix} \, , \quad
    S = -\frac{1}{2} \begin{pmatrix} 1 & 1 & 1 & -1 \\ 1 & 1 & -1 & 1 \\ 1 & -1 & 1 & 1 \\ -1 & 1 & 1 & 1 \end{pmatrix} \, .
\end{align}

\paragraph{Example 3}
We next consider a more non-trivial example, $ \hat{\mathfrak{g}} = \widehat{\mathfrak{so}}(8) $. We first identify the subgroup $ W' $ of the Weyl group $ W $. The longest element of $ W $ is
\begin{align}
    w_0 = (423124123121) \equiv r_4r_2r_3 \cdots r_2r_1 \, .
\end{align}
Two generators $ A_1 $ and $ A_2 $ of $ \operatorname{Out}(\hat{\mathfrak{g}}) $ act as $ A_1 \widehat{\omega}_0 = \widehat{\omega}_1 $ and $ A_2 \widehat{\omega}_0 = \widehat{\omega}_4 $. The longest elements of the corresponding subgroups are $ w_1=(324232) $ and $ w_4 = (123121) $. Therefore, the subgroup $ W' $ is given by
\begin{align}
    W' = \{ 1, w_{A_1}, w_{A_2}, w_{A_1} w_{A_2} \}
    = \{ 1, (124321), (423124), (324123) \} \cong \mathbb{Z}_2 \times \mathbb{Z}_2 \, .
\end{align}
Consequently, the quotient group $ W/W' $ has 48 elements given by
\begin{align}
    W/W' &= \{ 1, (1), (2), (3), (4), (21), (31), (41), (12), (32), (42), (23), (43), (2 4), (1 2 1), (3 2 1), \nonumber \\
        & \quad (2 3 1), (4 3 1), (2 4 1), (3 1 2), (4 1 2), (2 3 2), (4 3 2), (2 4 2), (1 2 3), (2 4 3), (1 2 4), (3 1 2 1), \nonumber \\
        &\quad (2 3 2 1), (1 2 3 1), (1 2 4 3), (2 4 3 1), (1 2 4 1), (2 3 1 2), (4 3 1 2), (2 4 1 2), (1 2 3 2), (2 4 3 2),  \nonumber \\
        &\quad (1 2 4 2), (2 3 1 2 1), (1 2 3 2 1), (1 2 4 3 1), (1 2 3 1 2), (1 2 4 3 2), (2 4 3 1 2), (1 2 4 1 2), \nonumber \\
    &\quad (1 2 3 1 2 1), (1 2 4 3 1 2) \} .
\end{align}
Using this data, one can find constraints on $ \hat{\lambda}^{F,y} $ from \eqref{eq:frac-weight-cond}.

\subsection{Coset models}

We now consider coset models of affine Kac-Moody algebras of the form
\begin{align}
    \frac{\hat{\mathfrak{g}}_{k_1} \oplus \hat{\mathfrak{g}}_{k_2}}{\hat{\mathfrak{g}}_{k_1+k_2}} \, ,
\end{align}
where $ k_1 $ is an admissible level and $ k_2 \in \mathbb{N} $. The central charge of this coset model is
\begin{align}
    c = c(\hat{\mathfrak{g}}_{k_1}) + c(\hat{\mathfrak{g}}_{k_2}) - c(\hat{\mathfrak{g}}_{k_1+k_2}) = \frac{k_1k_2 (k_1+k_2+2h^\vee) \dim \mathfrak{g}}{(k_1+h^\vee)(k_2+h^\vee)(k_1+k_2+h^\vee)} \, ,
\end{align}
where the central charge of the affine Kac-Moody algebra is given in \eqref{eq:aff-c}. This coset defines a non-unitary rational VOA. The primary fields of the coset model are labelled by $ \Lambda = \{\hat{\lambda}, \hat{\mu}, \hat{\nu}\} $, where $ \hat{\lambda} \in P_y^{k_1} $, $ \hat{\mu} \in P_+^{k_2} $ and $ \hat{\nu} \in P_{y'}^{k_1+k_2} $. The characters $ \chi_{\Lambda} $ of the coset primary fields satisfy the decomposition
\begin{align}
    \chi_{\hat{\lambda}}^{(k_1)} \chi_{\hat{\mu}}^{(k_2)} = \sum_{\hat{\nu}} \chi_\Lambda \chi_{\hat{\nu}}^{(k_1+k_2)} \, .
\end{align}
The coset primary characters can be non-vanishing if
\begin{align}\label{eq:coset-selection}
    \lambda + \mu - \nu \in Q \, , \quad
    y = y' \, , \quad
    \lambda^{F,y} = \nu^{F,y'} \, ,
\end{align}
where $ Q $ is the root lattice of $ \mathfrak{g} $, and $ \lambda $, $ \lambda^I $ and $ \lambda^{F,y} $ represent the finite parts of the affine weight $ \hat{\lambda} $, $ \hat{\lambda}^I $ and $ \hat{\lambda}^{F,y} $, respectively. Moreover, not all coset primary fields $ \Lambda $ are independent: many coset fields share the same characters and are therefore indistinguishable. We identify two coset primary fields $ \{\hat{\lambda}_1, \hat{\mu}_1, \hat{\nu}_1 \} $ and $ \{\hat{\lambda}_2, \hat{\mu}_2, \hat{\nu}_2 \} $ in the following cases \cite{Mathieu:1991fz}. First, two primaries are identified if they are related by an outer automorphism as
\begin{align}\label{eq:coset-identify-1}
    \{\hat{\lambda}_2, \hat{\mu}_2, \hat{\nu}_2\} = \{ A \hat{\lambda}_1, A \hat{\mu}_1, A \hat{\nu}_1 \} \, , \quad (A \in \operatorname{Out}(\hat{\mathfrak{g}}) ) \, .
\end{align}
Second, two primaries are identified if they satisfy
\begin{align}\label{eq:coset-identify-2}
    \lambda_1^I = \lambda_2^I \, , \quad
    \mu_1 = \mu_2 \, , \quad
    \nu_1^I = \nu_2^I \, , \quad
    \lambda_1^{F,y} = \lambda_2^{F,y} \bmod Q^\vee \, ,
\end{align}
where $ Q^\vee $ is the coroot lattice of $ \mathfrak{g} $. Third, two primaries are identified if they are related by the Weyl group as
\begin{align}\label{eq:coset-identify-3}
    \{\hat{\lambda}_2, \hat{\mu}_2, \hat{\nu}_2\} = \{ w \cdot \hat{\lambda}_1, \hat{\mu}_1, w \cdot \hat{\nu}_1 \} \, ,
\end{align}
where $ w \in W/W^\lambda $ for the \emph{associated Weyl group} $ W^{\lambda} = \{ w \in W \mid ( w, \alpha^\vee ) \in \mathbb{Z}, \, \forall \alpha \in \Delta_+ \} $. The modular matrices of the coset model are given by
\begin{align}\label{eq:coset-TS}
    T_{\Lambda_1 \Lambda_2} = T_{\hat{\lambda}_1 \hat{\lambda}_2}^{(k_1)} T_{\hat{\mu}_1 \hat{\mu}_2}^{(k_2)} \left( T_{\hat{\nu}_1\hat{\nu}_2}^{(k_1+k_2)}\right)^* \, , \quad
    S_{\Lambda_1 \Lambda_2} = N S_{\hat{\lambda}_1 \hat{\lambda}_2}^{(k_1)} S_{\hat{\mu}_1 \hat{\mu}_2}^{(k_2)} \left( S_{\hat{\nu}_1\hat{\nu}_2}^{(k_1+k_2)}\right)^* \, ,
\end{align}
where $ T^{k_i} $ and $ S^{k_i} $ are $ T $- and $ S $-matrices of the affine Kac-Moody algebra at level $ k_i $, and $ N $ is the number of coset primaries identified by the conditions \eqref{eq:coset-identify-1}-\eqref{eq:coset-identify-3}.

\paragraph{Example}
One simple example is the coset realization of the Lee-Yang VOA, which is a non-unitary Virasoro minimal model:
\begin{align}
    \text{Lee-Yang} = \frac{\widehat{\mathfrak{su}}(2)_{-\frac{4}{3}} \oplus \widehat{\mathfrak{su}}(2)_1}{\widehat{\mathfrak{su}}(2)_{-\frac{1}{3}}} \, .
\end{align}
This is the VOA associated with the $ (A_1, A_2) $ AD theory. There are two primary fields in $ \widehat{\mathfrak{su}}(2)_1 $, while $ \widehat{\mathfrak{su}}(2) $ has three admissible highest weights \eqref{eq:su2-4/3}. Applying the method in section~\ref{sec:admissible}, onecan find 12 admissible highest weights in $ \widehat{\mathfrak{su}}(2)_{-\frac{1}{3}} $ from two integer levels $ (k^I, k^F) = (3,2) $. The central charge of the coset model is
\begin{align}
    c = -6 + 1 - \left(- \frac{3}{5} \right) = -\frac{22}{5} \, ,
\end{align}
which is the central charge of the Lee-Yang CFT. Among $ 3\times 2 \times 12 $ coset primary fields $ \{\hat{\lambda}, \hat{\mu}, \hat{\nu}\} $, only 12 primaries survive under the condition \eqref{eq:coset-selection}. Moreover, by applying the field identification conditions \eqref{eq:coset-identify-1}-\eqref{eq:coset-identify-3}, only two primary fields
\begin{align}
    \Lambda_1 = \{ [-\tfrac{4}{3},0], [1,0], [-\tfrac{1}{3},0] \} \, , \quad
    \Lambda_2 = \{ [-\tfrac{4}{3},0], [1,0], [-\tfrac{7}{3},2] \}
\end{align}
are independent. Using these primary fields, one can find the modular matrices as
\begin{align}
    T = e^{-\frac{11\pi i}{30}} \begin{pmatrix} 1 & 0 \\ 0 & e^{-\frac{2\pi i}{5}} \end{pmatrix} \, , \quad
    S = \sqrt{\frac{4}{5}} \begin{pmatrix} -\sin(\frac{2\pi}{5}) & \sin(\frac{4\pi}{5}) \\ \sin(\frac{4\pi}{5}) & \sin(\frac{\pi}{5}) \end{pmatrix} \, ,
\end{align}
where we use $ N=6 $ in \eqref{eq:coset-TS}.

\subsection{Modular data of coset models}

We now list the $ S $-matrices of the coset VOAs of the form
\begin{align}
    \frac{\hat{\mathfrak{g}}_k \oplus \hat{\mathfrak{g}}_1}{\hat{\mathfrak{g}}_{k+1}} \, ,
\end{align}
which are the VOAs associated with the family of Argyres-Douglas theories studied in this paper. We note that when $ \mathfrak{g}=\mathfrak{su}(n) $, this coset VOA is the W-algebra minimal model whose modular matrices can be also be computed using method in \cite{Beltaos:2010ka}.

\afterpage{
\begin{landscape}
        \centering
        \renewcommand{\arraystretch}{1.2}
\begin{longtable}{C{9ex}|C{7ex}|C{6ex}|L{103ex}}
            \caption{$ S $-matrices of the VOAs appearing in section~\ref{sec: ex}. The notation follows that of \eqref{eq: sincos-def}; in addition, we define $ \tau=i\frac{\sqrt{11}}{4} $ for the $ (A_2, A_7) $ theory. In case of the $ (A_3,A_6) $ theory, the exact expression is very complicated, so we only present the numerical values. The exact expressions for the first row are given in \eqref{eq: A3A6-Atwist}.} \label{table:data}\\
            \endfirsthead
            \caption*{\textbf{Table \thetable\ (Continued)}}\\
            \endhead
            \endfoot
            \endlastfoot
            $ (G,G') $ & $ \mathfrak{g} $ & $ k $ & \multicolumn{1}{c}{$ S $} \\
\hline
        \end{longtable}
\end{landscape}
}

\newpage
\section{Review of the 3d rank-0 SCFT and the 3d A-model method} \label{app: 3d review}

In this appendix, we survey recent developments in 3d rank-0 SCFTs. We also review the 3d A-model method that computes 3d $\CN=2$ supersymmetric partition functions on various Seifert manifolds, which we employ in the present work to extract the modular data of 3d TFTs.

\subsection{Survey on the 3d rank-0 SCFT \label{app: rank-0}}

Starting with the pioneering discovery by Gang and Yamazaki's minimal theory\footnote{The name {\it minimal theory} indicates that the three sphere free energy $F = - \log(|Z_{S^3}|)$ of the theory, i.e., a measure of degrees of freedom, is the minimum among the 3d $\CN=4$ SCFTs, which is smaller than that of a free hyper multiplet.} \cite{Gang:2018huc}, the 3d rank-0 SCFTs had been formulated in \cite{Gang:2021hrd} with many examples. The definition of the 3d rank-0 theory is that, the 3d superconformal field theories {\it without} Higgs and Coulomb branch. The theories look trivial, however, are still strongly interacting SCFTs. It is worth noting that while 3d rank-0 SCFTs typically enjoy $\CN=4$ supersymmetry, they do not admit $\CN=4$ manifest Lagrangian descriptions. Instead, they usually have $\CN=2$ UV descriptions and are believed to exhibit $\CN= 4$ or $5$ enhancement in the IR. Such enhancements have been extensively examined case by case for numerous examples. A definite proof of the $\CN=4$ enhancement for the rank-0 SCFTs is still an intriguing open problem. 

One reason this seemingly simple class of 3d SCFTs has attracted attention is precisely its apparent simplicity. Under the assumption that rank-0 SCFTs possess $\CN=4$ supersymmetry, one can consider the topological A- or B-twist. With generic 3d $\CN=4$ SCFTs, the A/B-twist produces 3d cohomological non-unitary TFTs which contain local operators from the Coulomb/Higgs branch, thus, they are not semisimple TFTs. However, for the rank-0 SCFTs, the resulting TFTs do not possess local operators, since there is no Coulomb and Higgs branch to begin with. Hence, they give rise to semisimple 3d non-unitary TFTs. 

A recent work \cite{Ferrari:2023fez} has established a connection between 2d VOAs and these 3d TFTs from the rank-0 theories by showing that the former arise as boundary theories of the latter under holomorphic boundary conditions.\footnote{See \cite{Costello:2018fnz} for original discussion on VOAs from topological twist of 3d $\CN=4$ theories. Also, see \cite{Nishinaka:2025nbe} for discussions on VOA arising from topological twist of 3d $\CN=4$ obtained from circle reduction of 4d $\CN=2$ Argyres-Douglas theories.} Meanwhile, the work \cite{Dedushenko:2023cvd} clarified a picture for understanding the 4d $\CN=2$ SCFT/2d VOA correspondence via intermediate 3d theory obtained by $U(1)_r$ twisted cigar circle compactification of the holomorphic-topological twist with omega deformation. If we focus on the Argyres-Douglas theories with Coulomb branch operators having purely fractional conformal dimensions and empty Higgs branch, the resulting 3d theory becomes the rank-0 theory. Upon the topological A-twist, the rank-0 theory supports a 2d VOA on its holomorphic boundary which is the desired VOA of the SCFT/VOA correspondence \cite{Gaiotto:2024ioj,ArabiArdehali:2024ysy,ArabiArdehali:2024vli,Kim:2024dxu,Go:2025ixu}.

Besides this, there are several additional ways to construct the rank-0 theories in other contexts, which we summarize below:
\begin{itemize}
    \item {\bf Gluing S-duality wall theories}: The S-duality wall theory denoted as $T[G]$ is the 3d $\CN=4$ SCFT with flavor symmetry group $G\times G_L$ that leaves on the S-duality wall of the 4d $\CN=4$ SYM theory with gauge group $G$ \cite{Gaiotto:2008ak}, where $G_L$ is the Langlands dual of $G$. By gluing a multiple of $T[G]$ theories via $\CN=3$ Chern-Simons gauge multiplets in such a way that there is no remaining flavor symmetry, it is expected that a 3d $\CN=4$ rank-0 SCFT is obtained \cite{Gang:2018wek,Assel:2022row,Garozzo:2019ejm,Jeong:2025xid}. While the explicit examples are primarily focused on the $T[SU(2)]$ cases for technical reason, they nonetheless yield a variety of significant insights. In \cite{Gang:2024tlp}, a 3d bulk description for the non-unitary $M(p,q)$ Virasoro minimal models was first constructed with full generality by gluing the $T[SU(2)]$'s. In the series of works \cite{Gang:2021hrd,Gang:2022kpe,Gang:2023ggt}, it has been checked that by diagonal gauging the two $SU(2)$'s of a single $T[SU(2)]$ theory with Chern-Simons level $|k|\geq 3$, an exotic non-unitary TFT arises upon topological twists whose modular data cover that of the Haagerup-Izumi RCFTs \cite{haagerup1994principal,Asaeda_1999,Izumi:2000qa}. It would be interesting to further work out the rank-0 theories from gluing $T[G]$'s for general $G$.
    
    \item {\bf 3d/3d correspondence}: The 3d/3d correspondence states that 3d $\CN=2$ SCFT $\CT[M]$ can be labeled by 3-manifold $M$ \cite{Terashima:2011qi,Dimofte:2011ju}, thus, one may wonder which 3-manifold $M$ would present the 3d rank-0 SCFT with $\CN=2$ description  $\CT[M]$ that has $\CN=4$ enhancement in the IR. Some clues were observed in \cite{Gang:2018gyt,Cho:2020ljj} and then systematically investigated in \cite{Choi:2022dju} with numerous examples. The claim is that, a 3d rank-0 SCFT arises from a closed 3-manifold $M$ with (1) the 3D index is a formal Laurant series, (2) all irreducible flat $SL(2,\mathbb{C})$ connections are real, and (3) the corresponding $\CT[M]$ has a subsector of non-unitary TFT. Note that the first two conditions filter the $M$ to be non-hyperbolic. This motivated the construction of the 3d bulk description for the Virasoro minimal model $M(p,q)$ from the Seifert fibered space \cite{Gang:2024tlp}. Similar 3d bulk constructions for 2d $\CN=1$ supersymmetric Virasoro minimal model $SM(p,q)$ as well as for the $W_N$-algebra minimal model $W_N(p,q)$ are also recently proposed \cite{Baek:2025uev}.
    
    \item {\bf HT-twist of 4d $\CN=2$ SCFTs}: As previously mentioned, for the 4d $\CN=2$ Argyres-Douglas theories with all the Coulomb branch operators having purely fractional conformal dimension and empty Higgs branch, the $U(1)_r$ twisted circle compactification lifts all the Coulomb branch and produces 3d rank-0 SCFTs \cite{Dedushenko:2023cvd}. This construction has recently been checked \cite{Gaiotto:2024ioj,ArabiArdehali:2024ysy,ArabiArdehali:2024vli,Kim:2024dxu,Go:2025ixu} and provided various interesting rank-0 theories. The $(A_1,A_{2n})$ Argyres-Douglas theory is one example whose corresponding rank-0 theory is given in \cite{Gang:2023rei}. See also \cite{Ferrari:2023fez} for its 3d $\CN=4$ mirror symmetry counterpart, and a recent discussion on the connection to the level/rank duality \cite{Creutzig:2024ljv}. We also observed several 3d theories arising from a more geometric engineering based 3d–4d system constructed from the Argyres–Douglas theories \cite{Kucharski:2025lcr} seem to give rise to the rank-0 theories.
    
    \item {\bf Abelian CS matter theories}: Many of the currently known examples of the rank-0 SCFTs arise in the form of 3d $\CN=2$ Abelian Chern-Simons matter theories(ACSM) \cite{Gang:2023rei,Creutzig:2024ljv,Baek:2024tuo,Go:2025ixu}. Motivated by the Nahm sum expressions of VOA characters, the work \cite{Gang:2024loa} initiated an investigation for the rank-0 theories of ACSM type. The ACSM description is rather simple, still, captures various non-trivial VOAs. Furthermore, the 3d theories appearing in the present paper are all of this type by construction from the Coulomb branch data. Note that in the 3d/3d correspondence, the triangulation of a 3-manifold $M$ with ideal tetrahedra also yields ACSM description for the corresponding 3d $\CN=2$ SCFT $\CT[M]$. It is a highly compelling question that how many rank-0 theories can be characterized in terms of the ACSM description.
\end{itemize}
To summarize, the 3d rank-0 SCFTs provide a primary 3d bulk description of non-unitary semisimple VOAs, thus allowing us to investigate the relationships between seemingly disparate subjects, including the SCFT/VOA correspondence, the connection between 3-manifolds and VOAs, and the interplay between mirror symmetry and level/rank duality.

\subsection{The 3d A-model method for partition function computation \label{app: A-model}}

Here, we briefly summarize the {\it 3d A-model} method, which computes the half-BPS partition functions of 3d $\CN=2$ gauge theories on any compact Seifert 3-manofold \cite{Closset:2018ghr} which makes use of the topologically A-twisted 2d $\CN=(2,2)$ theory on its base together with the Bethe/gauge correspondence \cite{Nekrasov:2009uh,Nekrasov:2009ui}. See \cite{Closset:2019hyt} for a review with examples, and \cite{Closset:2023vos,Closset:2023jiq,Closset:2023bdr,Closset:2023izb,Closset:2024sle,Closset:2025lqt} for recent applications. Let us first schematically explain the formula of the partition functions and then explicitly apply it to the ACSM theory.

The twisted partition function $Z_{\CM_{g,p}}$ on a degree $p$ bundle over a genus $g$ Riemann surface, 
\begin{align}
    S^1 \;\;\overset{p}{\to}\;\; \CM_{g,p} \;\;\to\;\; \S_g\;,
\end{align}
can be evaluated by a formula,
\begin{align}
    Z_{\CM_{g,p}} = \sum_{u^{(\a)} \in \CS} (\CH_\a)^{g-1} (\CF_\a)^p
    \label{eq: ptf Mgp}
\end{align}
where $\CH$ and $\CF$ are called the {\it handle gluing} and {\it fibering} operators respectively that can be computed from the 3d $\CN=2$ gauge theory description.\footnote{Note that $Z_{\CM_{0,1}} = Z_{S^3}$ computes the $S^3$ partition function and $Z_{\CM_{g,0}} = \CI_{\S_g}$ computes the twisted index on $\S_g \times S^1$ \cite{Gukov:2015sna,Benini:2015noa,Benini:2016hjo,Closset:2016arn}.} This formula is remarkable in that the path integral computation for the half-BPS partition function on arbitrary $\CM_{g,p}$ is reduced to a finite sum over the {\it Bethe vacua}, $\CS$. Consequently, what we need to apply the formula \eqref{eq: ptf Mgp} are simply $\CH$, $\CF$, and $\CS$. Let us explain how to calculate them.

For a given 3d $\CN=2$ gauge theory with gauge group $G$ and matter content, one can write down the 3d twisted superpotential $\CW$ and the effective dilaton $\Omega$ by summing over all the massive fluctuations and Kaluza-Klein modes along the $S^1$ as,
\begin{align}
    \CW(u,m) \, , \quad
    \Omega(u,m)
\end{align}
which characterizes the Coulomb branch low-energy dynamics with the gauge and flavor symmetry parameters,
\begin{align}
    u = (u_1,\cdots,u_{\rank(G)}) \, , \quad
    m = (m_1,\cdots,m_{\rank(F)})\,,
\end{align}
where $F$ it the flavor symmetry group. Next, we define gauge and flavor symmetry flux operators as,
\begin{align}
    \Pi_j(u,m) \equiv
    \exp(
    2\pi i \frac{\partial \CW}{\partial u_j} ) \, , \quad
    \Pi_a^f(u,m) \equiv
    \exp(
    2\pi i \frac{\partial \CW}{\partial m_a}
    )\,.
\end{align}
The Bethe vacua, then, are the solutions of the set of equations called {\it Bethe equations} divided by the Weyl group of $G$,
\begin{align}
    \CS = \left\{ u^{(\a)} \,\middle| \,
    \Pi_j(u,m) = 1 \, , \ 
    j=1,\cdots,\rank(G) \, , \
    w(u) \neq u \, , \
    \forall w\in W_G
    \middle\} \right/ W_G \, .
\end{align}
Note that the equations for the Bethe vacua essentially reduce to polynomial equations when the gauge and flavor symmetry parameters are rewritten in terms of fugacities,
\begin{align}
    z_i \equiv e^{2\pi i u_i} \, , \quad
    t_a \equiv e^{2\pi i m_a}\, ,
\end{align}
so that the number of Bethe vacua $|\CS|$ is finite. Observe that this number is the same as the Witten index by considering the formula \eqref{eq: ptf Mgp} with base $\CM_{g=1,p=0}$. On the other hand, the handle gluing and fibering operators can be evaluated from $\CW$ and $\Omega$ as,
\begin{align}
    \begin{aligned}
        \CH(u,m) &= e^{2\pi i \Omega} 
        \det(
        \frac{\partial^2 \CW}{\partial u_i \partial u_j}
        ) \, , \\
        \CF(u,m) &= 
        \exp(
        2\pi i \left(
            \CW 
            - u \cdot \frac{\partial \CW}{\partial u}
            - m \cdot \frac{\partial \CW}{\partial m}
        \right)
        )\,.
    \end{aligned}
    \label{eq: H and F}
\end{align}
Therefore, what we mean by $\CH_\a$ and $\CF_\a$ in \eqref{eq: ptf Mgp} are the values of $\CH$ and $\CF$ evaluated at the Bethe vacua,
\begin{align}
    \CH_\a \equiv \CH(u^{(\a)},m) \, , \quad
    \CF_\a \equiv \CF(u^{(\a)},m)\,.
\end{align}

\paragraph{Rank-0 from ACSM}

Now, as a particular model, let us focus on a 3d $\CN=2$ ACSM theory with the gauge group $U(1)^r$ and $N$ chiral multiplets. This can be characterized by $r\times r$ CS level matrix $K$ and $r\times N$ charge matrix $Q$,
\begin{align}
    K = \left(
    \begin{array}{ccc}
         K_{11}  & \cdots & K_{1r} \\
         \vdots & \ddots & \vdots  \\
         K_{1r} & \cdots  & K_{rr}
    \end{array}
    \right) \, , \quad
    Q = \left(
    \begin{array}{cccc}
         Q_{11} &  \cdots &  & Q_{1N} \\
         \vdots & \ddots & & \vdots \\
         Q_{r1} & \cdots &  & Q_{rN}
    \end{array} \, ,
    \right)
\end{align}
where $K$ is symmetric and $Q_{iJ}$ is the electric charge of the $J$-th chiral multiplet under the $i$-th $U(1)$ gauge group. The theory, in general, has $U(1)^N$ flavor symmetry and if we assume $r>N$ this is realized as combinations of the $U(1)_{T_i}$ topological symmetries where $U(1)_{T_i}$ comes from the $i$-th $U(1)$ gauge group. For a convenient discussion, let us treat all the $r$ of $U(1)_{T_i}$ as if independent for a moment.\footnote{We will break most of the $U(1)_{T_i}$'s by turning on a certain monopole superpotential, leaving only a single combination that becomes $U(1)_A$ axial symmetry for $\CN=4$ algebra of the rank-0 theory.} Then, the twisted superpotential and the effective dilaton for this model read,
\begin{align}
    \CW &=
    \frac{1}{2}\sum_{i,j=1}^r K_{ij} u_i u_j
    +\frac{1}{2} \sum_{i=1}^r \big((1+2\n_R)K_{ii} + 2m_i\big) u_i + \frac{1}{(2\pi i)^2}\sum_{I=1}^N \Li_2 \big( e^{2\pi i u\cdot Q_I} \big)
    \nonumber\\
    \Omega &= \frac{1}{2\pi i} \sum_{I=1}^N \log \big(
    1-e^{2\pi i u\cdot Q_I}
    \big) + \frac{1}{2}k_{RR}\,,
\end{align}
where $\n_R \in \frac{(p\,\text{mod}\,2) + 1}{2}\mathbb{Z}$ and $k_{RR}$ are the $U(1)_R$ fugacity and CS level respectively, and we set all the R-charges of the chiral multiplets to be zero. Hence, from \eqref{eq: H and F}, the handle gluing and fibering operators can be computed as,
\begin{align}
    \CH &=(-1)^{k_{RR}} \prod_{I=1}^N \Big( 1- z^{Q_I} \Big)
    \det_{i,j}
    \Bigg(
    K_{ij} + \sum_{I=1}^N Q_{iI} Q_{jI} \frac{z^{Q_I}}{1-z^{Q_I}}
    \Bigg) \, ,
    \nonumber\\
    \CF &= 
    \prod_{I=1}^N
    \big(1-z^{Q_I}\big)^{u\cdot Q_I}
    \exp(
    \frac{1}{2\pi i}
    \sum_{I=1}^N
     \Li_2\big( z^{Q_I} \big)
     -\pi i\sum_{i,j=1}^r K_{ij}u_i u_j - 
     2\pi i \sum_{i=1}^r m_i u_i
    ) \, ,
    \label{eq: ACSM HF}
\end{align}
with a short-hand notation,
\begin{align}
    z^{Q_I} \equiv \prod_{i=1}^r z_i^{Q_{iI}}
    =
    e^{2\pi i \sum_{i=1}^r u_i Q_{iI} }
    \,,
\end{align}
and the Bethe equations are given by,
\begin{align}
    (-1)^{(1+2\n_R)K_{ii}} t_i
    \prod_{j=1}^r z_j^{K_{ij}} 
    =\prod_{I=1}^N \big(1-z^{Q_I}\big)^{Q_{iI}} \quad \text{for} \quad i=1,\cdots,r
    \label{eq: ACSM Bethe}
\end{align}
which are polynomial equations in $z$. Thus, by solving \eqref{eq: ACSM Bethe} for $z$ and plugging the solutions into $\CH$ and $\CF$ in \eqref{eq: ACSM HF}, one can compute the partition function $Z_{\CM_{g,p}}$ from the formula \eqref{eq: ptf Mgp}.

As mentioned in the previous section, many of the currently known rank-0 theories are realized as ACSM descriptions. More precisely, certain $\CN=2$ ACSM theories admit a superpotential deformation by half-BPS monopole operators, under which they flow to an $\mathcal{N}=4$ rank-0 SCFTs in the infrared. Suppose we start with such an ACSM theory. Then the half-BPS monopole operators can be written in the form,
\begin{align}
    V_{(d,\mathfrak{m})}
    \equiv
    \Big(\prod_{I=1}^{N} \phi_I^{d_I} \Big)
    V_{\mathfrak{m}}
    \label{eq: half BPS monopole}
\end{align}
where $\phi_I$ is a scalar field in the $I$-th chiral multiplet with gauge charge $Q_{jI}$ under the $j$-th $U(1)$ gauge group, $d_I \in \mathbb{Z}_{\geq 0}$ is a dressing number chosen such that all electric charges of the bare-monopole operator $V_{\mathfrak{m}}$ of magnetic flux $\mathfrak{m} = (\mathfrak{m}_1,\cdots,\mathfrak{m}_r)$ are canceled to become trivial. Namely, the electric charge of $V_\mathfrak{m}$ under the $i$-th $U(1)$ gauge group is given by,
\begin{align}
    e_i\big(V_{\mathfrak{m}}\big) = \sum_{j=1}^r K_{ij} \mathfrak{m}_j
    -\!\!\!
    \sum_{I \,|\, \mathfrak{m}\cdot Q_I > 0}
    (\mathfrak{m}\cdot Q_I) Q_{i I}
\end{align}
where $Q_{I} = (Q_{1I},\cdots,Q_{rI})$ is the charge of the $I$-th chiral multiplet under $U(1)^r$, therefore, the dressing numbers $d_I$ should be chosen such that the electric charge of $V_{(d,\mathfrak{m})}$ vanishes,
\begin{align}
    e_i\big(
    V_{(d,\mathfrak{m})}
    \big)
    =
    \sum_{I\,|\, \mathfrak{m}\cdot Q_I = 0} Q_{iI} d_I + e_i\big( V_\mathfrak{m} \big)
    =0 \, ,
\end{align}
where the condition in the sum $\mathfrak{m} \cdot Q_I = 0$ is due to the half-BPS condition.

Now, let us consider turning on the monopole superpotential. If we define the mixing of $U(1)_{T_i}$ symmetry with the R-symmetry as,
\begin{align}
    R_\m = R + \sum_{i=1}^r \m_i T_i
\end{align}
with a reference R-charge $R$ and mixing parameters $\m_i$, the half-BPS monopole operators have R-charge as,
\begin{align}
    R_\m \big(
    V_{(d,\mathfrak{m})}
    \big)
    =
    \sum_{I\,|\, \mathfrak{m}\cdot Q_I} \mathfrak{m}\cdot Q_I
    +
    \m\cdot \mathfrak{m}
\end{align}
where the magnetic flux $\mathfrak{m}_i$ is the conserved charge of the $U(1)_{T_i}$ topological symmetry. Suppose there are $r-1$ half-BPS monopole operators of magnetic fluxes $\mathfrak{m}^{(l)}$, $l = 1,\cdots, r-1$ whose superpotential deformation triggers an RG flow to a rank-0 theory in the infrared. Once these operators are turned on as superpotential terms, their R-charges are fixed to be 2,
\begin{align}
    \sum_{I\,|\, \mathfrak{m}^{(l)}\cdot\, Q_I} \mathfrak{m}^{(l)}\cdot Q_I
    +
    \m\cdot \mathfrak{m}^{(l)}
    =
    2
\end{align}
for $l=1,\cdots,r-1$ which solve the mixing parameter $\m$ as,
\begin{align}
    \m_i = \m_i^* + \n a_i\,.
    \label{eq: mu}
\end{align}
Here $a_i$'s parametrize the unbroken combination of $U(1)_{T_i}$ symmetries that becomes the $U(1)_A$ axial symmetry responsible for the $\CN=4$ enhancement, which is defined by
\begin{align}
    A = J_3^C - J_3^H\,,
\end{align}
where $J_3^C$ and $J_3^H$ are the Cartans of $SO(4)_R = SU(2)_C\times SU(2)_H$ respectively. The parameter $\n$ controls a mixing,
\begin{align}
    R_\n \equiv R + \n  A
\end{align}
between $A$ and $R = J_3^C + J_3^H$ where $R$ is the generator of $U(1)_R$ symmetry in $\CN=2$ description. The constant vector $\m_i^*$ is fixed such that $\n=0$ corresponds to the conformal R-symmetry, therefore, $\n=-1$ and $\n = 1$ correspond to the topological A- and B-twist respectively. This mixing shifts the flavor fugacities and fluxes, yielding a partition function given by,
\begin{align}
    Z_{\CM_{g,p}} = \sum_{u^{(\a)}\in\CS}
    (\CH_\a)^{g-1} (\CF_\a)^p z^{(g-1+\n_Rp)\m}
\end{align}
where the additional factor $z^{(g-1+\n_R p)\m}$ is due to the shift. By comparing it with the partition function of a semisimple 3d TFT,
\begin{align}
    Z_{\CM_{g,p}} = \sum_{\a} (S_{0\a})^{2-2g} (T_{\a\a})^{-p} \, ,
\end{align}
we find a relation between the modular $S$- and $T$-matrices and handle, fibering operators,
\begin{align}
    (S_{0\a})^{-2} = \CH(u^{(\a)},m) z^\m \, , \quad
    (T_{\a\a})^{-1} = \CF(u^{(\a)},m) z^{m}
\end{align}
with $m = \n_R \m$. Consequently, one can compute the modular data $\{ (S_{0\a})^2 ,\; T_{\a\a} \}$ of the semisimple TFT obtained from the A/B-twist of the rank-0 theory by employing the 3d A-model method.

\subsection{Superconformal index}

Another supersymmetric observable that we consider is the superconformal index which counts gauge invariant BPS operators. For 3d $\CN=4$ theories, it can be defined following the convention in \cite{Gang:2021hrd} as follows,
\begin{align}
    \CI_{S^2\times S^1} = \Tr_{\CH_{S^2}}
    (-1)^{R_\n} q^{\frac{R_\n}{2} + j_3} \eta^A
    \label{eq: SCI def}
\end{align}
where the trace is over the radially quantized Hilbert space on $S^2$, $j_3 \in \frac{\mathbb{Z}}{2}$ is the Lorentz spin of $SO(3)$ isometry on $S^2$. Note that $\n=0$ corresponds to conformal point, while $\n=-1$ and $\n=+1$ correspond to topological A- and B-twists respectively. If we focus on a rank-0 theory having ACSM description with CS level matrix $K$ and charge matrix $Q$, the index can be calculated as,
\begin{align}
    \CI_{S^2\times S^1}(\eta,\n;q) &= 
    \sum_{\mathfrak{m} \in \mathbb{Z}^r}
    \oint
    \prod_{i=1}^r \frac{dz_i}{2\pi i\,z_i}
    \prod_{i,j=1}^r z_i^{K_{ij} \mathfrak{m}_j}
    \prod_{i=1}^r \big(
    \eta^{a_i} (-q^{\frac{1}{2}})^{\m_i}
    \big)^{\mathfrak{m}_i}
    \prod_{I=1}^N
    \CI_{\D}(\mathfrak{m}\cdot Q_I,z^{Q_I};q)\,
    \nonumber\\
    \CI_\Delta(f,z;q) &= 
    \prod_{n=0}^\infty
    \frac{1-z^{-1} q^{n+1+\frac{f}{2}}}{1- z q^{n+\frac{f}{2}} }
\end{align}
where $\m_i$ is the mixing parameters given in \eqref{eq: mu} and $\eta$ is the fugacity for the $U(1)_A$ axial symmetry. The Hilbert series, which counts Higgs/Coulomb branch operators, can be obtained by tuning $\eta = 1, \n = \pm 1$ \cite{Razamat:2014pta}. Since rank-0 theories have trivial Higgs and Coulomb branches, the result should be,
\begin{align}
    \CI_{S^2 \times S^1} (\eta = 1,\n=\pm 1;q) = 1
\end{align}
which is a necessary condition for being 3d rank-0 SCFTs.

\section{Toolkit for the trace formula computation} \label{app: embed}

In this appendix, we present the computational details of the trace formula and provide several introductory examples illustrating its computation.

\subsection{Faddeev's quantum dilogarithm \label{app: QDL}}

The Faddeev's quantum dilogarithm is defined as \cite{Faddeev:1993rs}
\begin{align}
    \Phi_b(z) = \exp\left( \frac{1}{4} \int_{\mathbb{R}+i 0^+} \frac{dt}{t}\frac{e^{-2izt}}{\sinh(bt)\sinh(b^{-1}t)} \right)\, .
    \label{eq: QDL def}
\end{align}
Its infinite product representation is given by
\begin{align}
    \Phi_b(z) = \prod_{l=0}^{\infty}
    \frac{1+ e^{2\pi i b^2 (l+1/2)} e^{2\pi b z} }{1+ e^{-2\pi i b^{-2} (l+1/2)  }e^{2\pi b^{-1} z}} \, .
\end{align}
The function $ \Phi_b(z) $ possesses the symmetry properties
\begin{align}
    \Phi_b(z) = \Phi_{b^{-1}}(z) \, , \quad
    \Phi_b(z) = \Phi_{-b}(z) \, ,
\end{align}
and is a meromorphic function of $ z $ with
\begin{align}
    \text{poles: } \frac{i}{2}(b+b^{-1}) + i\mathbb{N}b + i\mathbb{N}b^{-1} \, , \quad
    \text{zeros: } -\frac{i}{2}(b+b^{-1}) - i\mathbb{N}b - i\mathbb{N}b^{-1} \, .
\end{align}
The quasi-periodicity of this function is
\begin{align}
    \Phi_b\Big(z-\frac{ib^{\pm 1}}{2}\Big) = \big( 1 + e^{2\pi b^{\pm 1} z} \big)
    \Phi_b\Big(z+\frac{ib^{\pm 1}}{2}\Big)\,.
\end{align}
Two important properties required for the computation of the trace formula \eqref{eq: trace formula} are
\begin{align} \label{eq: Phi identity}
    \boxed{\quad
    \begin{array}{cl}
    \text{Fusion :} & \displaystyle \Phi_b(x)\Phi_b(-x) = \Phi_b(0)^2 e^{\pi i x^2} \\
    \text{Pentagon relation :} &
    \displaystyle \Phi_b(\hat{p})\Phi_b(\hat{q}) = \Phi_b(\hat{q}) \Phi_b(\hat{q} + \hat{p}) \Phi_b(\hat{p}) \quad \text{if } [\hat{p},\hat{q}] = \frac{1}{2\pi i}
    \end{array}\quad
    }
\end{align}
where $\Phi_b(0) = e^{\frac{\pi i}{24}(b^2 + b^{-2})}$ is a constant. We also emphasize a quantum mechanical property of the Gaussian operator that is crucial in the trace formula computation:
\begin{equation} \label{eq: gaussian}
    \boxed{\quad
    \text{Shift :} \quad
    e^{\pi i \hat{x}^2} f(\hat y) =
    f(\hat y + c\hat x)\,
    e^{\pi i \hat{x}^2}  \quad \text{if }
    [\hat x , \hat y] = \frac{c}{2\pi i}
    \quad
    }
\end{equation}
for some function $f(\hat y)$. Namely, when $f(\hat y)$ passes through the Gaussian operator $e^{\pi i \hat x^2}$ from the right, the argument of $f$ is shifted by $\hat x$ proportional to the commutator with the operator. The boxed identities above are all that is required to simplify the trace formula, which will be useful for later technical computations.

\subsection{From BPS quivers to trace formula}

We now discuss computational scheme of the trace formula \eqref{eq: trace formula} from the BPS quivers of 4d $ \mathcal{N}=2 $ SCFTs. Let $\{  \g_i \}$ be a basis of the charge lattice $\G$ with $\rank\Gamma = N$; each node of the BPS quiver is labeled by one of these basis charges. Define an integer-valued $ N\times N $ anti-symmetric matrix $ C_{ij} $ from the Dirac pairing of the charges as,
\begin{align}
    C_{ij} = \langle \gamma_i, \gamma_j \rangle \, .
\end{align}
We denote the rank of this matrix as $ 2r=\rank(C_{ij}) $, since the rank of an anti-symmetric matrix is always even. The rank of the sublattice $ \Gamma_f $ associated with the flavor symmetry is given by the dimension of the kernel of $ C_{ij} $ as $ \rank \Gamma_f = N-2r $. We then introduce the Weyl algebra generated by $ N $ varaibles $ x_{\gamma_i} $ satisfying the commutation relations
\begin{align}
    [x_{\gamma_i}, x_{\gamma_j}] = \frac{1}{2\pi i} C_{ij} \, .
\end{align}
To simplify the trace formula, it is useful to introduce a new basis $ \{p_\alpha, q_\alpha, m_i \} $ satisfying the canonical commutation relations
\begin{align}\label{eq:canonical-comm}
    [p_\alpha, q_\beta] = \frac{1}{2\pi i} \delta_{\alpha\beta} \, ,
\end{align}
where $ \alpha, \beta = 1, \cdots, r $ and $ m_i $ are $ c $-numbers that parametrize the sublattice $ \Gamma_f $. After changing the basis from $ \{x_{\gamma_i}\} $ to $ \{p_\alpha, q_\alpha, m_i\} $, one can make use of the inner product and completeness of the eigenstates
\begin{align}\label{eq:completeness}
    \braket{q_\alpha}{p_\alpha} = e^{2\pi i q_\alpha p_\alpha} \, , \quad
    1 = \int_{-\infty}^{\infty} dp_\alpha \dyad{p_\alpha} = \int_{-\infty}^{\infty} dq_\alpha \dyad{q_\alpha} \, ,
\end{align}
as in ordinary quantum mechanics.

We now focus on the $ (G,G') $ Argyres-Douglas theories whose BPS quiver is given by the square product of two Dynkin diagrams, $ G \sqprod G' $. If the ranks of $ G $ and $ G' $ are $ n $ and $ n' $, respectively, then the charge lattice $ \Gamma $ has dimension $ N=nn' $ and is parametrized by the $ N $ charges $ \gamma_\sigma $. As in \eqref{eq: disjoint set of square}, the nodes of the BPS quiver $ G \sqprod G' $ are decomposed into two disjoint subsets $ \Sigma_\pm $, and we denote $ \Gamma_\pm $ as the sublattices generated by the charges $ \gamma_\sigma $ for $ \sigma \in \Sigma_\pm $. Their ranks $ d_\pm = \rank(\Gamma_\pm) $  satisfy $ d_+ + d_- = N $. Since the Dirac pairings between any two charges in $ \Gamma_+ $ (and likewise in $ \Gamma_- $) are trivial, the anti-symmetric matrix $ C_{ij} $ can be expressed as
\begin{align}\label{eq:Dirac-c}
    C = \left(
        \begin{array}{c|c}
            \mathbf{0} & c \\ \hline
            -c^T & \mathbf{0}
        \end{array}
    \right) \, ,
\end{align}
where $ \mathbf{0} $ is the zero matrix and $ c=(c_{i\bar{j}}) $ is a $ d_+ \times d_- $ matrix defined by
\begin{align}
    c_{i\bar{j}} = \langle \gamma_i, \gamma_{\bar{j}} \rangle \quad \text{for} \quad
    \gamma_i \in \Gamma_+ \, , \ 
    \gamma_{\bar{j}} \in \Gamma_- \, .
\end{align}
If $ \rank(C_{ij}) = 2r $, then $ \rank(c_{i\bar{j}}) = r $. Let us choose $ r $ basis vectors $ \{b_{\bar{j}\alpha}\}_{\alpha=1}^r $ of the row space of $ c_{i\bar{j}} $ and write the matrix $ c_{i\bar{j}} $ as
\begin{align}\label{eq: rank decomp}
    c_{i\bar{j}} = \sum_{\alpha=1}^r a_{i\alpha}\, b_{\bar{j}\alpha} \, ,
\end{align}
for some coefficients $ a_{i\alpha} $. Note that $ a_{i\alpha} $ and $ b_{\bar{j}\alpha} $ form $ d_+\times r $ and $ d_- \times r $ matrices, respectively. Then the variables of the Weyl algebra can be decomposed as
\begin{align}\label{eq:xTocanonical}
    x_{\gamma_i} = \sum_{\alpha=1}^r a_{i\alpha}\, p_\alpha + m_i \, , \quad
    x_{\gamma_{\bar{j}}} = \sum_{\alpha=1}^r b_{\bar{j}\alpha}\, q_\alpha + m_{\bar{j}} \, ,
\end{align}
where $ p_\alpha $ and $ q_\alpha $ are canonical variables satisfying \eqref{eq:canonical-comm}. Here, $ m_i $ and $ m_{\bar{j}} $ are $ c $-numbers satisfying
\begin{align} \label{eq: m kernel}
    \sum_{i=1}^{d_+} m_i a_{i\alpha} = 0 \, , \quad
    \sum_{\bar{j}=1}^{d_-} m_{\bar{j}} b_{\bar{j}\alpha} = 0 \, .
\end{align}
In total, there are $ \dim \Gamma_f = d_++d_--2r $ independent variables $ m_i $ and $ m_{\bar{j}} $, which serve as parameters for the flavor symmetry.

\subsection{Examples \label{app: ex}}

\begin{figure}
    \centering
    \begin{tabular}{cC{25ex}cC{33ex}c}
        $ A_2 \sqprod A_2 $ &
        \includegraphics{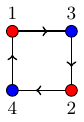} &
        $ \overset{\text{mutation}}{\longleftrightarrow} $ &
        \includegraphics{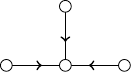} &
        $ D_4 $
        \\
        $ A_2 \sqprod A_3 $ & 
        \includegraphics{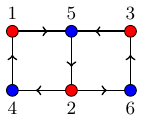} &
        $ \longleftrightarrow $ &
        \includegraphics{fig/A1E6.pdf} &
        $ E_6 $
        \\
        $ A_2 \sqprod A_4 $ & 
        \includegraphics{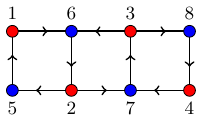} &
        $ \longleftrightarrow $ &
        \includegraphics{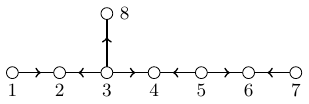} &
        $ E_8 $
    \end{tabular}
    \caption{BPS quivers of $(A_2,A_2)$, $(A_2,A_3)$, and $(A_2,A_4)$ Argyres-Douglas theories with respective isomorphisms to $(A_1,D_4)$, $(A_1,E_6)$, and $(A_1,E_8)$. These isomorphisms can be checked via quiver mutations.} \label{fig: iso for quiver}
\end{figure}

In this subsection, we present explicit examples of the trace formula computations. We consider four Argyres-Douglas theories, $(A_2,A_2)$, $(A_2,A_3)$, $(A_2,A_4)$, and $(A_4,D_4)$, which are isomorphic to the $(A_1,D_4)$, $(A_1,E_6)$, $(A_1,E_8)$, and $(A_2,E_8)$ theories respectively. These isomorphisms can be verified from the BPS quivers by performing a certain sequence of mutations, as illustrated in Figure~\ref{fig: iso for quiver}. We confirm that the resulting 3d theories of each isomorphism pair are identical by computing their trace formulae.

\paragraph{Example 1}
As a first example, let us consider the $(A_2,A_2)$ Argyres-Douglas theory. From the BPS quiver shown in the first line of Figure~\ref{fig: iso for quiver}, we can separate the charges into two sets as
\begin{align}
	\G_+ = \{ \g_1 , \g_2 \} \, , \quad
	\G_- = \{ \g_3,\g_4 \} \, .
\end{align}
The Weyl algebra is organized in terms of the Dirac pairing matrix $ C_{ij} $ and its submatrix $ c_{i\bar{j}} $ defined in \eqref{eq:Dirac-c} as
\begin{align}
    C = \left(
        \begin{array}{cc}
            \mathbf{0} & c \\
            -c^T & \mathbf{0}
        \end{array}
    \right) , \quad
    c = \left(
        \begin{array}{cc}
            1 & -1 \\
            -1 & 1
        \end{array}
    \right) .
\end{align}
The rank of the Dirac pairing matrix is $ \rank(C)=2 $; consequently, the rank of the flavor symmetry of the 4d theory is $ \rank \Gamma_f = 2 $. The submatrix $ c $ can be decomposed as $ c=ab^T $, as in \eqref{eq: rank decomp}, where $ a=b=(1,-1)^T $. From \eqref{eq:xTocanonical}, we parametrize the variables $ x_i \equiv x_{\gamma_i} $ as
\begin{align}
    x_1 = p_1 + m_1 \, , \quad
    x_2 = -p_1 + m_2 \, , \quad
    x_3 = q_1 + m_3 \ , \quad
    x_4 = -q_1 + m_4 \, ,
\end{align}
where $ [p_1, q_1] = (2\pi i)^{-1} $, and the parameters $ m_i $ satisfy
\begin{align}
    m_1-m_2 = 0 \, , \quad
    m_3-m_4 = 0 \, .
\end{align}
Following the arguments in section \ref{sec: mutation}, one can find a finite chamber with 6 BPS particles from the BPS quiver $A_2 \sqprod A_2$. Thus, the trace formula reads
\begin{align}\label{eq: A2A2 integral}
    Z_{S_b^3}^{(A_2,A_2)} &= \Tr \left( \Phi_b(x_1) \Phi_b(x_2) \Phi_b(x_2+x_3) \Phi_b(x_1+x_4) \Phi_b(x_3) \Phi_b(x_4) \times (x_i \to -x_i) \right) \, .
\end{align}
By using \eqref{eq: Phi identity} and \eqref{eq: gaussian}, one can reduce the number of quantum dilogarithms as
\begin{equation}\label{eq:(A2A2)-trace-2}
    Z_{S_b^3}^{(A_2,A_2)} = \Phi_b(0)^{12} \Tr\Big( e^{2\pi i ( q_1^2 + m_3^2 )} e^{2\pi i ( p_1^2 +m_1^2 )}  e^{\pi i ( p_1 - q_1 - m_1 - m_3 )^2}   e^{\pi i ( p_1 - q_1 +m_1+m_3 )^2} \Big) .
\end{equation}
This computation can be carried out using our \texttt{Mathematica} code \cite{code}, which is described in Appendix~\ref{app: code}. Finally, by using the inner product and completeness of the eigenstates of $ p_1 $ and $ q_1 $ given in \eqref{eq:completeness}, we obtain
\begin{align}\label{eq: (A2,A2) gaussian}
    Z_{S_b^3}^{(A_2,A_2)}
	&= -i \Phi_b(0)^{12} \int du_1 du_2\, e^{\pi i(-4u_1 u_2 + 4m_1^2 + 4m_3^2 + 4m_1m_3 )} \, .
\end{align}
Surprisingly, all quantum dilogarithms disappear in this case and the resulting ellipsoid partition function is given by a simple Gaussian integral. The result \eqref{eq: A2A2 integral} coincides with that of the $(A_1,D_4)$ theory computed in \cite{Go:2025ixu}, up to a rescaling of the flavor symmetry parameters. This confirms the isomorphism $(A_2,A_2) \sim (A_1,D_4)$ at the level of the trace formula.

\paragraph{Example 2}
Let us now consider the $(A_2,A_3)$ theory whose BPS quiver $ A_2 \sqprod A_3 $ is shown in the second line of Figure~\ref{fig: iso for quiver}. The charges $ \gamma_i $ can be divided into two sets given by
\begin{align}
    \G_+ = \{ \g_1 , \g_2 , \g_3 \} \, , \quad
    \G_- = \{ \g_4 , \g_5 , \g_6 \} \, .
\end{align}
The submatrix $ c_{i\bar{j}} $ of the Dirac pairing matrix can be decomposed as
\begin{align}
    c = 
    \left(
    \begin{array}{ccc}
        -1 & 1 & 0 \\
        1 & -1 & 1 \\
        0 & 1 & -1
    \end{array}
    \right) = ab^T \, , \quad
    b = {\bf 1}_{3\times 3} \, 
\end{align}
The variables $ x_i \equiv x_{\gamma_i} $ of the Weyl algebra can be rewritten as
\begin{gather}
    \begin{gathered}
        x_1 = -p_1+p_2 \, , \quad
        x_2 = p_1 - p_2 + p_3 \, , \quad
        x_3 = p_2 - p_3 \\
        x_4 = q_1 \, , \quad
        x_5 = q_2 \, , \quad
        x_6 = q_3 \, ,
    \end{gathered}
\end{gather}
where we set $ m_1=\cdots=m_6=0 $, since the kernels of the $ a $ and $ b $ matrices are trivial. This implies that $ (A_2, A_3) $ theory does not have a flavor symmetry. From the mutation described in section~\ref{sec: mutation}, one can find a finite chamber with 9 BPS particles. The trace formula is, then, given by
\begin{align}\label{eq: A2A3 trace}
    \begin{aligned}
        Z_{S_b^3}^{(A_2,A_3)} &= \Tr \big( \Phi_b(x_1) \Phi_b(x_2) \Phi_b(x_3) \Phi_b(x_1+x_4) \Phi_b(x_2+x_5) \Phi_b(x_3+x_6) \\
        &\qquad \quad \times \Phi_b(x_4) \Phi_b(x_5) \Phi_b(x_6) \times (x_i \to -x_i) \big) .
    \end{aligned}
\end{align}
This expression can be simplified to
\begin{align}
    \begin{aligned}\label{eq: (A2,A3) integral}
        Z_{S_b^3}^{(A_2,A_3)} &= \Phi_b(0)^{18} \Tr \Big( \Phi_b(q_3) \Phi_b(p_3+q_1) e^{\pi i (p_2-p_3+q_3)^2} e^{\pi i (p_1+q_3)^2} e^{\pi i q_2^2} e^{\pi i q_3^2} \\
        & \quad \times e^{\pi i (p_1-p_2+p_3)^2} e^{\pi i (p_1-p_2-q_1)^2} e^{\pi i q_2^2} e^{\pi i q_1^2} \Phi_b(q_1) e^{\pi i (p_1-p_2)^2} \Big) \, .
    \end{aligned}
\end{align}
By applying \eqref{eq:completeness}, one finds
\begin{align}\label{eq: (A2,A3) result}
    Z_{S_b^3}^{(A_2,A_3)} = i^{\frac{1}{2}}\Phi_b(0)^{18} \!\! \int \! du_1 du_2 du_3 \, e^{\pi i ( u_2^2+u_3^2 + 2u_1u_2 + 2u_1u_3 - 2u_2u_3)} \Phi_b(u_1) \Phi_b(u_2) \Phi_b(u_3) ,
\end{align}
which is exactly the same as that of the $(A_1,E_6)$ theory computed in \cite{Go:2025ixu},\footnote{One of the authors found a discrepancy of an overall $i^{1/2}$ factor in the previous work \cite{Go:2025ixu} which, however, does not affect the main results there.} thereby confirming the isomorphism $(A_2,A_3) \sim (A_1,E_6)$ at the level of the trace formula.

\paragraph{Example 3}

Consider the $(A_2,A_4)$ theory with a BPS quiver $A_2 \sqprod A_4$ as shown in the left side of the last line of Figure~\ref{fig: iso for quiver}. The basis of the charge lattice $ \gamma_i $ can be divided into two sets as
\begin{align}
    \G_+ = \{
    \g_1,\g_2,\g_3,\g_4
    \} \, , \quad
    \G_- = \{
    \g_5,\g_6,\g_7,\g_8
    \} \, .
\end{align}
From the Dirac pairing matrix, the variables $ x_i \equiv x_{\gamma_i} $ can be expressed in terms of the variables $ p_\alpha $ and $ q_\alpha $, which satisfy $ [p_\alpha, q_\beta] = (2\pi i)^{-1} \delta_{\alpha\beta} $, as
\begin{gather}
    \begin{gathered}
        x_1 = -p_1+p_2 \, , \quad
        x_2 = p_1-p_2+p_3 \, , \quad
        x_3 = p_2-p_3+p_4 \, , \quad
        x_4 = p_3-p_4 \, , \\
        x_5 = q_1 \, , \quad
        x_6 = q_2 \, , \quad
        x_7 = q_3 \, , \quad
        x_8 = q_4 \, .
    \end{gathered}
\end{gather}
The $ (A_2, A_4) $ theory has a finite chamber with 12 BPS particles and the trace formula is given by
\begin{align}
    \begin{aligned}\label{eq: (A2A4)-trace}
        Z_{S_b^3}^{(A_2,A_4)} &= \Tr \big( \Phi_b(x_1)\Phi_b(x_2)\Phi_b(x_3)\Phi_b(x_4) \Phi_b(x_1+x_5)\Phi_b(x_2+x_6) \Phi_b(x_3+x_7) \\
        &\qquad \quad \times \Phi_b(x_4+x_8) \Phi_b(x_5)\Phi_b(x_6)\Phi_b(x_7)\Phi_b(x_8) \times (x_i \to -x_i) \big) \, .
    \end{aligned}
\end{align}
Using the quantum dilogarithm identities and completeness relations, we find
\begin{align}
    Z_{S_b^{3}}^{(A_2,A_4)}
    &= i^{\frac{7}{2}} \Phi_b(0)^{26} \int \bigg[ \prod_{i=1}^5 du_i \bigg]
    e^{\pi i (u_1^2 - 2 u_1 u_3 + u_3^2 +2 u_1 u_4 + 2u_3 u_4 - 2u_1 u_5 + 2u_2 u_5 + 2u_5^2)}
    \prod_{i=1}^4 \Phi_b(u_i) .
\end{align}
This computation can be carried out using our \texttt{Mathematica} code, which is described in Appendix~\ref{app: code}.

On the other hand, let us consider the $(A_1,E_8)$ theory, whose BPS quiver is given by the $E_8$ Dynkin diagram depicted on the right side of the last line of Figure~\ref{fig: iso for quiver}. The basis charges $ \gamma_i $ can be divided into two sets $\G_\pm$ as
\begin{align}
    \G_+ = \{ \g_1,\g_3,\g_5,\g_7 \} \, , \quad
    \G_- = \{ \g_2,\g_4,\g_6,\g_8 \} \, .
\end{align}
The Dirac pairing matrix $ C_{ij} $ provides a parametrization for the Weyl algebra variables $ x_i \equiv x_{\gamma_i} $ in terms of the variables $ p_\alpha $ and $ q_\alpha $  as
\begin{gather}
    \begin{gathered}
        x_1 = p_1 \, , \quad
        x_3 = p_1+p_2+p_4 \, , \quad
        x_5 = p_2+p_3 \, , \quad
        x_7 = p_3 \, , \\
        x_2 = q_1 \, , \quad
        x_4 = q_2 \, , \quad
        x_6 = q_3 \, , \quad
        x_8 = q_4 \, ,
    \end{gathered}
\end{gather}
where $ [p_\alpha, q_\beta] = (2\pi i)^{-1} \delta_{\alpha\beta} $. The finite chamber of the $ E_8 $ BPS quiver contains 8 BPS particles, and the trace formula is given by
\begin{align}
    Z_{S_b^{3}}^{(A_1,E_8)} &= \Tr \big(
    \Phi_b(x_1)\Phi_b(x_3)\Phi_b(x_5)\Phi_b(x_7)
    \Phi_b(x_2)\Phi_b(x_4)\Phi_b(x_6)\Phi_b(x_8)
    \!\times\! (x_i \!\to\! -x_i) \big) .
\end{align}
Although the initial expression appears different from the trace formula of the $ (A_2, A_4) $ theory given in \eqref{eq: (A2A4)-trace}, they are in fact identical. This equivalence can be verified using the quantum dilogarithm identities \eqref{eq: Phi identity} and converting the expression into its integral form using \eqref{eq:completeness}. This verification can be also performed using our \texttt{Mathematica} code. This confirms the isomorphism $(A_2,A_4) \sim (A_1,E_8)$ at the level of the trace formula computation.

\paragraph{Example 4}

\begin{figure}
    \centering
    \begin{tabular}{C{30ex}cC{45ex}}
        \includegraphics{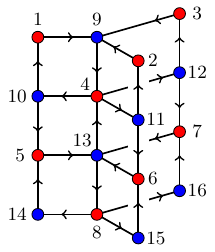} & $ \longleftrightarrow $ &
        \includegraphics{fig/A2E8.pdf} \\
        $ A_4 \sqprod D_4 $ & & $ A_2 \sqprod E_8 $
    \end{tabular}
    \caption{\label{fig: A4D4 and A2E8} BPS quivers of $(A_4,D_4)$ and $(A_2,E_8)$ theories which are expected to be isomorphic.}
\end{figure}

As the last example, consider the $(A_4,D_4)$ theory, whose BPS quiver is depicted in Figure~\ref{fig: A4D4 and A2E8}. The basis of the charge lattice $ \gamma_i $ can be divided into two sets as
\begin{align}
    \G_+ = \{ \g_1\,\cdots, \g_8 \} \, , \quad
    \G_- = \{ \g_9\,\cdots, \g_{16} \} \, .
\end{align}
We parametrize the variables $ x_i \equiv x_{\gamma_i} $ as
\begin{gather}
    \begin{gathered}
        x_1 = p_1-p_2 , \quad
        x_2 = p_1-p_3 , \quad
        x_3 = p_1-p_4 , \quad
        x_4 = p_2+p_3+p_4-p_1-p_5, \\
        x_5 = p_5-p_2-p_6, \quad
        x_6 = p_5-p_3-p_7, \quad
        x_7 = p_5-p_4-p_8,  \\
        x_8 = p_6+p_7+p_8-p_5, \quad
        x_{i+8}=q_i ,\quad (1 \leq i \leq 8) .
    \end{gathered}
\end{gather}
There is a finite chamber containing 40 BPS particles and the corresponding trace formula can be simplified to
\begin{align}\label{eq:A4D4-trace}
    Z_{S_b^3}^{(A_4,D_4)} = \Phi_b(0)^{80} 
    \int du_1 \cdots du_8\,
    e^{\pi i u^T L u}
    \bigg(\prod_{i=1}^8 \Phi_b(u_i)\bigg) \, ,
\end{align}
where $ u=(u_1, u_2, \cdots, u_8)^T $ is the vector of integration variables and the Chern-Simons level matrix $ L $ is given by
\begin{align}
    A = \left(
    \begin{array}{cccccccc}
        1 & 1 & 1 & 0 & 0 & 1 & 0 & -1 \\
        1 & 2 & 0 & -1 & 1 & -1 & 1 & -1 \\
        1 & 0 & 1 & 1 & 1 & 0 & 1 & 0 \\
        0 & -1 & 1 & 1 & 0 & 1 & 0 & 1 \\
        0 & 1 & 1 & 0 & 1 & -1 & 0 & 1 \\
        1 & -1 & 0 & 1 & -1 & 2 & 1 & -1 \\
        0 & 1 & 1 & 0 & 0 & 1 & 0 & 1 \\
        -1 & -1 & 0 & 1 & 1 & -1 & 1 & 2
    \end{array}
    \right)\,.
\end{align}

The $ (A_4, D_4) $ theory is expected to be isomorphic to the $ (A_2, E_8) $ theory whose BPS quiver $ A_2 \sqprod E_8 $ is shown in Figure~\ref{fig: A4D4 and A2E8}. The charge vectors can be divided into two sets as
\begin{align}
    \G_+ = \{\g_1,\cdots,\g_8\} \, , \quad
    \G_- = \{\g_9,\cdots,\g_{16}\} \, ,
\end{align}
and $ x_i \equiv x_{\gamma_i} $ can be reparametrized as
\begin{gather}
        x_1 = p_2-p_1  , \
        x_2 = p_1 - p_2 + p_3 , \
        x_3 = p_2 - p_3 + p_4 + p_5 , \
        x_4 = p_3 - p_4 , \
        x_5 = p_3 - p_5 + p_6 , \nonumber \\
        x_6 = p_5 - p_6 + p_7 , \
        x_7 = p_6 - p_7 + p_8 , \
        x_8 = p_7 - p_8 , \ 
        x_{i+8} = q_i , \ (1 \leq i \leq 8) \, .
\end{gather}
Using the BPS quiver $ A_2 \sqprod E_8 $, we find a finite chamber containing 24 BPS particles, and the trace formula exactly matches with \eqref{eq:A4D4-trace}. This confirms the isomorphism $ (A_4, D_4) \sim (A_2, E_8) $ at the level of the trace formula.

\section{\texttt{Mathematica} code for simplifying QDLs \label{app: code}}

In this appendix, we describe our \texttt{Mathematica} program for simplifying the trace formula, which can be found in \cite{code}. This simplification is useful for obtaining a simple 3d $\CN=2$ ACSM theory description of the twisted compactification of 4d $ \mathcal{N}=2 $ SCFTs. Moreover, since the structure of the trace formula and the identities of the Faddeev's quantum dilogarithm (QDL) are universal for the Schur index and the identities of the $ q $-exponential function \cite{Cordova:2015nma}, our program is also useful for computing the Schur index of the Argyres-Douglas theories, which is the vacuum character of the corresponding 2d VOAs.

The aim of the program is to rewrite the trace formula of a given theory using the smallest possible number of QDLs by using the identities \eqref{eq: Phi identity}. Thus, this task can be viewed as a minimization problem for the number of QDLs appearing in the expression. However, this task is highly nontrivial, as it is difficult to find a canonical rule for determining which QDL should be moved at each step. Local manipulations based on the pentagon relation \eqref{eq: Phi identity} and the shift property \eqref{eq: gaussian} of the Gaussian operators can easily lead to a local minimum, rather than the global minimum configuration. For instance, one may attempt an iterative approach as follows: test the movements of QDLs using \eqref{eq: Phi identity} and \eqref{eq: gaussian}, check whether the total number of QDLs decreases, and at each step choose the move that yields the simplest intermediate form. However, this procedure is often trapped in a local minimum, failing to reach the expression with a minimum number of QDLs.

\begin{figure}
    \centering
    \includegraphics[scale=0.6]{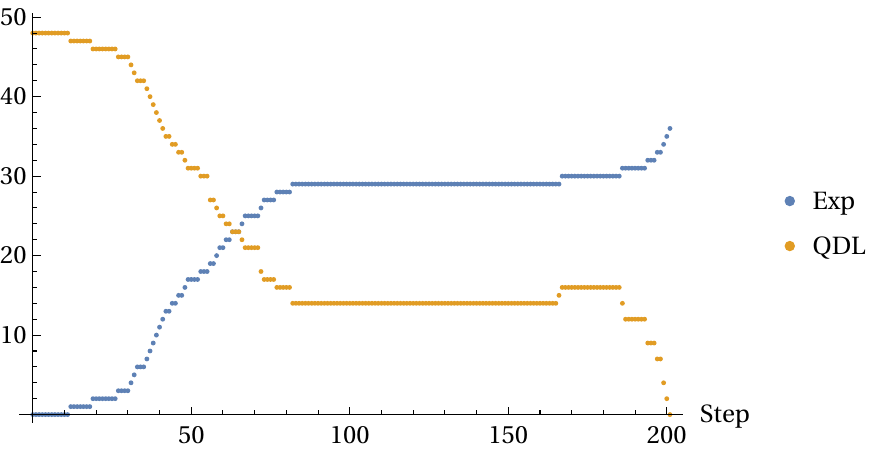}
    \caption{Example showing the number of QDLs (orange) and Gaussian operators (blue) at each step of the computation performed by the \texttt{Mathematica} code. The input expression is the trace formula of the $ (A_2, A_8) $ theory which has $ |\Gamma_{\mathrm{BPS}}| = 48 $.} \label{fig:code-eg}
\end{figure}

To overcome these difficulties, we adopt the following method. For a given expression, we first randomly choose a subset of the QDLs and move them using \eqref{eq: Phi identity} and \eqref{eq: gaussian}, until either the fusion \eqref{eq: Phi identity} occurs or no further movement is possible via the pentagon relation. Among all randomly chosen QDL movements, we select the most simplest configuration, in the sense that the number of QDLs is minimal. After repeating this step multiple times, we typically encounter a situation in which the number of QDLs no longer decreases. This situation is illustrated in Figure~\ref{fig:code-eg}. Here, we start with the trace formula of the $ (A_2,A_8) $ theory, which initially consists of 48 QDLs. After 82 steps, the number of QDLs is reduced to 14, while 29 Gaussian operators are generated. However, beyond this point, further QDL movements fail to reduce their number. To decrease the number of QDLs further, we introduce a random fluctuation into the computation: we randomly move QDLs without enforcing the minimization condition. Although this fluctuation may temporarily increase the number of QDLs, it helps the process escape from a local minimum in the QDL count. We then return to the first step and repeat the computation to further minimize the number of QDLs. In Figure~\ref{fig:code-eg}, the random fluctuations are introduced from the 165th step, during which the number of QDLs increases to 16 due to the fluctuation. After this fluctuation, the number of QDLs begins to decrease again and eventually reaches zero, while the number of Gaussian operators becomes 36 in this example.

The usage of the \texttt{Mathematica} code implementing above algorithm is follows. The input to the program is the ordered list $ \Gamma_{\mathrm{BPS}} $ written in terms of the variables $ \{p_\alpha, q_\beta, m_i\} $, where $ [p_\alpha, q_\beta] = (2\pi i)^{-1} \delta_{\alpha\beta} $ and $ m_i $ are $ c $-numbers. The main routine for simplifying the trace formula is
\begin{align}
    \texttt{QDilogSimp[$ \Gamma_{\mathrm{BPS}} $, Options]} \, .
\end{align}
Here, the input list $ \Gamma_{\mathrm{BPS}} $ must be written in terms of \texttt{P[i]}, \texttt{X[i]} and \texttt{M[i]}, which correspond to the variables $ p_i $, $ q_i $ and $ m_i $, respectively. Various options are available in the \texttt{QDilogSimp} function as summarized below.
\begin{itemize}
    \item \texttt{Repeat} (default: 1) -- Executes the function \texttt{Repeat} times and returns the simplest result.
    \item \texttt{Parallel} (default: \texttt{True}) -- If \texttt{Repeat} is greater than 1 and \texttt{Parallel} is set to \text{True}, \texttt{QDilogSimp} function is executed in parallel.
    \item \texttt{ShowProgress} (default: \texttt{True}) -- Monitors the progress of the computation and displays a graph of the number of QDLs and Gaussian operators at each step as Figure~\ref{fig:code-eg}.
    \item \texttt{RandomSimp} (default: \texttt{1/5}) -- During the computation, a subset of QDLs in the expression is randomly selected and moved. The parameter \texttt{RandomSimp} specifies the size of this subset: if it is set to zero, the subset is empty, whereas if it is set to one, all QDLs are selected.
    \item \texttt{Patience} (default: \texttt{2}) -- If the number of QDLs does not decrease after $ \texttt{Patience} \times (\texttt{Length of the list}) $ steps, random fluctuations are introduced.
    \item \texttt{RandomDepth} (default: \texttt{1/5}) -- Specifies the number of random fluctuations. The program randomly moves $ \texttt{RandomDepth} \times (\texttt{number of QDLs}) $ QDLs.
    \item \texttt{RandomNumber} (default: \texttt{5}) -- Random fluctuations may fail to escape a local minimum. In such cases, the program attempts additional random fluctuations. The parameter \texttt{RandomNumber} specifies the maximum number of repetitions of these random fluctuations.
    \item \texttt{MaxIteration} (default: \texttt{30}) -- If we define a cycle as the sequence consisting of (i) a random fluctuation and (ii) a subsequent reduction in the number of QDLs, then \texttt{MaxIteration} sets a hard limit on the number of repetitions of this cycle.
\end{itemize}
The output of the \texttt{QDilogSimp} function is an ordered list consisting of elements of the form \texttt{x} and \texttt{S[x]}, which correspond to $ \Phi_b(x) $ and $ e^{\pi i x^2} $, respectively. In addition, the function prints two integers indicating the number of Gaussian operators and QDLs. It should be noted that the result of \texttt{QDilogSimp} function implicitly includes an overall factor $ \Phi_b(0)^n $, where $ n $ is twice of the number of the Gaussian operators. The \texttt{QDilogSimp} function utilizes not only the QDL identities but also cyclicity of the trace. The program includes another function \texttt{QDilogSimpNoCyc}, which reduces the product of QDLs without using the cyclic property of the trace.

We have also implemented functions that convert the trace formula into its integral representation. The function \texttt{ConvertIntegral} transforms the trace into an integral form using the completeness relation \eqref{eq:completeness}. The input to \texttt{ConvertIntegral} is the output of the \texttt{QDilogSimp} function, and the output is of the form $ \{f(x_i, y_j, m_k), (x_i), (m_k), (y_j)\} $, which represents
\begin{align}\label{eq:ConvertIntegral}
    \int \bigg[\prod_i dx_i\bigg] \bigg[\prod_j dy_j \bigg] \, e^{\pi i f(x_i, y_j, m_k)} \prod_i \Phi_b(x_i) \, .
\end{align}
Here, $ y_i $ appears only in the exponential factor and can be integrated out using the Gaussian integral. The function \texttt{ConvertIntegral} also provides information about the integral: it prints the Jacobian factor arising from the basis change from $ \{\texttt{P[i]}, \texttt{X[j]}\} $ to $ \{x_i, y_j\} $, and displays the number of integration variables in the form `\texttt{a+b+c variables}'. Here, \texttt{a} denotes the number of $ x_i $ variables, \texttt{b+c} is the total number of $ y_j $ variables, and \texttt{c} represents the number of zero modes in the Gaussian integral. The Gaussian integral can be performed by executing \texttt{GaussianIntegral}. The input of this function is
\begin{align}
    \texttt{GaussianIntegral[$ F, L $]},
\end{align}
where $ F $ is the output of \texttt{ConvertIntegral} and $ L $ is an optional list of integers. If $ L $ is not specified, the function integrate over all $ y_j $ variables in \eqref{eq:ConvertIntegral}. If $ L $ is a non-empty integer list, the function integrates over all $ y_j $ except $ j\in L $.

\paragraph{Example 1}
Let us consider the $ (A_2, A_2) $ Argyres-Douglas theory whose trace formula is given in \eqref{eq: A2A2 integral}. We first define \texttt{PhiList}, which encodes the charges of the BPS particles as
\begin{align}
    &\text{\texttt{PhiList = \{M[1]+P[1], M[1]-P[1], M[1]+M[3]-P[1]+X[1], M[1]+M[3]+P[1]-X[1],}} \nonumber \\
    &\qquad \quad \text{\texttt{M[3]+X[1], M[3]-X[1], -M[1]-P[1], -M[1]+P[1], -M[1]-M[3]+P[1]-X[1],}} \nonumber \\
    &\qquad \quad \text{\texttt{-M[1]-M[3]-P[1]+X[1], -M[3]-X[1], -M[3]+X[1]\}}} \, .
\end{align}
This can be directly read from the trace formula \eqref{eq: A2A2 integral}. Here, \texttt{P[1]}, \texttt{X[1]} and \texttt{M[i]} denote $ p_1 $, $ q_1 $ and $ m_i $, respectively. The number of QDLs can be reduced using the \texttt{QDilogSimp} function:
\begin{align}
    \text{\texttt{PhiSimp = QDilogSimp[PhiList]}} \, .
\end{align}
The resulting expression \texttt{PhiSimp} is given by, for instance,
\begin{align}
    \begin{aligned}\label{eq:PhiSimp-eg1result}
        &\text{\texttt{\{S[M[1]+P[1]], S[M[1]-P[1]], S[M[1]+M[3]-P[1]+X[1]],}} \\
        &\quad \text{\texttt{S[M[1]+M[3]+P[1]-X[1]], S[M[3]+X[1]], S[M[3]-X[1]]\}}} \, ,
    \end{aligned}
\end{align}
where \texttt{S[x]} represents the Gaussian operator $ e^{\pi i x^2} $. Due to the randomness inherent in the computational method and the cyclicity of the trace, the result of \texttt{PhiSimp} may differ from above, however, all outcomes are equivalent. The result \eqref{eq:PhiSimp-eg1result} can be directly compared with \eqref{eq:(A2A2)-trace-2}. To convert the expression into its integral representation, we use
\begin{align}
    \texttt{PhiInt = ConvertIntegral[PhiSimp]} \, .
\end{align}
The function \texttt{ConvertIntegral} prints information about the resulting integral expression. If \texttt{PhiSimp} is given by \eqref{eq:PhiSimp-eg1result}, then \texttt{ConvertIntegral} prints `\texttt{0+6+0 variables}', which indicates that there are six integration variables $ y_1,\cdots,y_6 $ in the integral expression. This Gaussian integral can be performed using
\begin{align}
    \texttt{GaussianIntegral[PhiInt]} \, ,
\end{align}
which prints
\begin{align}
    \texttt{Variables:\{M[1],M[3]\}} \, , \quad
    \texttt{Overall factor:} -\frac{i}{2} \, , \quad
    K = \begin{pmatrix} 4 & 2 \\ 2 & 4 \end{pmatrix} \, .
\end{align}
The ouput shows that the result of the Gaussian integral is given by $ -\frac{i}{2} e^{4m_1^2 + 4m_1m_3 + 4m_2^2} $. This is the result of the integral \eqref{eq: (A2,A2) gaussian}. It is also possible to obtain the expression \eqref{eq: (A2,A2) gaussian} by executing
\begin{align}
    \texttt{GaussianIntegral[PhiInt, \{1,5\}]} \, ,
\end{align}
which performs the integral over $ y_2, y_3, y_4, y_6 $. The integrals over $ y_1 $ and $ y_5 $ remain in the final expression, corresponding to $ u_1 $ and $ u_2 $ in \eqref{eq: (A2,A2) gaussian}.

\paragraph{Example 2} Let us next consider the $ (A_2, A_3) $ theory. Based on the trace formula \eqref{eq: A2A3 trace}, the input to the program is
\begin{align}
    \begin{aligned}
        \texttt{PhiSimp = QDilogSimp[\{-P[1]+P[2], P[1]-P[2]+P[3], $ \cdots $, -X[2], -X[3]\},}\\
        \texttt{"Repeat"->10, "Parallel"->True]} .
    \end{aligned}
\end{align}
Due to the options \texttt{Repeat} and \texttt{Parallel}, this executes the \texttt{QDilogSimp} function 10 times in parallel and stores the simplest result in \texttt{PhiSimp}. The result is, for instance,
\begin{align}
    \begin{aligned}\label{eq:(A2,A3)-PhiSimp}
        &\texttt{\{X[3], P[3]+X[1], S[P[2]-P[3]+X[3]], S[P[1]+X[3]], S[X[2]],} \\
        &\quad \texttt{S[X[3]], S[-P[1]+P[2]-P[3]], S[P[1]-P[2]-X[1]], S[-X[2]],} \\
        &\quad \texttt{S[-X[1]], X[1], S[-P[1]+P[2]]\}},
    \end{aligned}
\end{align}
which corresponds to the result in \eqref{eq: (A2,A3) integral}. Because of the inherent randomness in the computation, one may obtain a different result; however, all such results represent the same trace formula. The equivalence of the results can be verified by converting the expressions into their integral representations and performing the Gaussian integrals. This can be done by
\begin{align}
    \texttt{GaussianIntegral[ConvertIntegral[PhiSimp]]} \, ,
\end{align}
which yields the effective CS level matrix $ K $ and the charge matrix $ Q $ of the chiral multiplet, resulting in \eqref{eq: (A2,A3) result}.

\bibliographystyle{JHEP}
\bibliography{ref}

\providecommand{\href}[2]{#2}\begingroup\raggedright\begin{thebibliography}{100}

\bibitem{Beem:2013sza}
C.~Beem, M.~Lemos, P.~Liendo, W.~Peelaers, L.~Rastelli and B.C.~van Rees, \emph{{Infinite Chiral Symmetry in Four Dimensions}}, \href{https://doi.org/10.1007/s00220-014-2272-x}{\emph{Commun. Math. Phys.} {\bfseries 336} (2015) 1359} [\href{https://arxiv.org/abs/1312.5344}{{\ttfamily 1312.5344}}].

\bibitem{Xie:2016evu}
D.~Xie, W.~Yan and S.-T.~Yau, \emph{{Chiral algebra of the Argyres-Douglas theory from M5 branes}}, \href{https://doi.org/10.1103/PhysRevD.103.065003}{\emph{Phys. Rev. D} {\bfseries 103} (2021) 065003} [\href{https://arxiv.org/abs/1604.02155}{{\ttfamily 1604.02155}}].

\bibitem{Song:2016yfd}
J.~Song, \emph{{Macdonald Index and Chiral Algebra}}, \href{https://doi.org/10.1007/JHEP08(2017)044}{\emph{JHEP} {\bfseries 08} (2017) 044} [\href{https://arxiv.org/abs/1612.08956}{{\ttfamily 1612.08956}}].

\bibitem{Song:2017oew}
J.~Song, D.~Xie and W.~Yan, \emph{{Vertex operator algebras of Argyres-Douglas theories from M5-branes}}, \href{https://doi.org/10.1007/JHEP12(2017)123}{\emph{JHEP} {\bfseries 12} (2017) 123} [\href{https://arxiv.org/abs/1706.01607}{{\ttfamily 1706.01607}}].

\bibitem{Beem:2017ooy}
C.~Beem and L.~Rastelli, \emph{{Vertex operator algebras, Higgs branches, and modular differential equations}}, \href{https://doi.org/10.1007/JHEP08(2018)114}{\emph{JHEP} {\bfseries 08} (2018) 114} [\href{https://arxiv.org/abs/1707.07679}{{\ttfamily 1707.07679}}].

\bibitem{Fluder:2017oxm}
M.~Fluder and J.~Song, \emph{{Four-dimensional Lens Space Index from Two-dimensional Chiral Algebra}}, \href{https://doi.org/10.1007/JHEP07(2018)073}{\emph{JHEP} {\bfseries 07} (2018) 073} [\href{https://arxiv.org/abs/1710.06029}{{\ttfamily 1710.06029}}].

\bibitem{Bonetti:2018fqz}
F.~Bonetti, C.~Meneghelli and L.~Rastelli, \emph{{VOAs labelled by complex reflection groups and 4d SCFTs}}, \href{https://doi.org/10.1007/JHEP05(2019)155}{\emph{JHEP} {\bfseries 05} (2019) 155} [\href{https://arxiv.org/abs/1810.03612}{{\ttfamily 1810.03612}}].

\bibitem{Creutzig:2018lbc}
T.~Creutzig, \emph{{Logarithmic W-algebras and Argyres-Douglas theories at higher rank}}, \href{https://doi.org/10.1007/JHEP11(2018)188}{\emph{JHEP} {\bfseries 11} (2018) 188} [\href{https://arxiv.org/abs/1809.01725}{{\ttfamily 1809.01725}}].

\bibitem{Oh:2019bgz}
J.~Oh and J.~Yagi, \emph{{Chiral algebras from $\Omega$-deformation}}, \href{https://doi.org/10.1007/JHEP08(2019)143}{\emph{JHEP} {\bfseries 08} (2019) 143} [\href{https://arxiv.org/abs/1903.11123}{{\ttfamily 1903.11123}}].

\bibitem{Jeong:2019pzg}
S.~Jeong, \emph{{SCFT/VOA correspondence via $\Omega$-deformation}}, \href{https://doi.org/10.1007/JHEP10(2019)171}{\emph{JHEP} {\bfseries 10} (2019) 171} [\href{https://arxiv.org/abs/1904.00927}{{\ttfamily 1904.00927}}].

\bibitem{Beem:2019snk}
C.~Beem, C.~Meneghelli, W.~Peelaers and L.~Rastelli, \emph{{VOAs and rank-two instanton SCFTs}}, \href{https://doi.org/10.1007/s00220-020-03746-9}{\emph{Commun. Math. Phys.} {\bfseries 377} (2020) 2553} [\href{https://arxiv.org/abs/1907.08629}{{\ttfamily 1907.08629}}].

\bibitem{Auger:2019gts}
J.~Auger, T.~Creutzig, S.~Kanade and M.~Rupert, \emph{{Braided Tensor Categories Related to ${\mathcal {B}}_{p}$ Vertex Algebras}}, \href{https://doi.org/10.1007/s00220-020-03747-8}{\emph{Commun. Math. Phys.} {\bfseries 378} (2020) 219} [\href{https://arxiv.org/abs/1906.07212}{{\ttfamily 1906.07212}}].

\bibitem{Xie:2019zlb}
D.~Xie and W.~Yan, \emph{{Schur sector of Argyres-Douglas theory and $W$-algebra}}, \href{https://doi.org/10.21468/SciPostPhys.10.3.080}{\emph{SciPost Phys.} {\bfseries 10} (2021) 080} [\href{https://arxiv.org/abs/1904.09094}{{\ttfamily 1904.09094}}].

\bibitem{Dedushenko:2019yiw}
M.~Dedushenko and M.~Fluder, \emph{{Chiral Algebra, Localization, Modularity, Surface defects, And All That}}, \href{https://doi.org/10.1063/5.0002661}{\emph{J. Math. Phys.} {\bfseries 61} (2020) 092302} [\href{https://arxiv.org/abs/1904.02704}{{\ttfamily 1904.02704}}].

\bibitem{Xie:2019vzr}
D.~Xie and W.~Yan, \emph{{4d $\mathcal{N}=2$ SCFTs and lisse W-algebras}}, \href{https://doi.org/10.1007/JHEP04(2021)271}{\emph{JHEP} {\bfseries 04} (2021) 271} [\href{https://arxiv.org/abs/1910.02281}{{\ttfamily 1910.02281}}].

\bibitem{Dedushenko:2019mzv}
M.~Dedushenko, \emph{{From VOAs to short star products in SCFT}}, \href{https://doi.org/10.1007/s00220-021-04066-2}{\emph{Commun. Math. Phys.} {\bfseries 384} (2021) 245} [\href{https://arxiv.org/abs/1911.05741}{{\ttfamily 1911.05741}}].

\bibitem{Adamovic:2020lvj}
D.~Adamovic, T.~Creutzig, N.~Genra and J.~Yang, \emph{{The Vertex Algebras $\mathcal {R}^{(p)}$ and $\mathcal {V}^{({p})}$}}, \href{https://doi.org/10.1007/s00220-021-03950-1}{\emph{Commun. Math. Phys.} {\bfseries 383} (2021) 1207} [\href{https://arxiv.org/abs/2001.08048}{{\ttfamily 2001.08048}}].

\bibitem{Dedushenko:2023cvd}
M.~Dedushenko, \emph{{On the 4d/3d/2d view of the SCFT/VOA correspondence}},  \href{https://arxiv.org/abs/2312.17747}{{\ttfamily 2312.17747}}.

\bibitem{Cordova:2015nma}
C.~Cordova and S.-H.~Shao, \emph{{Schur Indices, BPS Particles, and Argyres-Douglas Theories}}, \href{https://doi.org/10.1007/JHEP01(2016)040}{\emph{JHEP} {\bfseries 01} (2016) 040} [\href{https://arxiv.org/abs/1506.00265}{{\ttfamily 1506.00265}}].

\bibitem{Cecotti:2010fi}
S.~Cecotti, A.~Neitzke and C.~Vafa, \emph{{R-Twisting and 4d/2d Correspondences}},  \href{https://arxiv.org/abs/1006.3435}{{\ttfamily 1006.3435}}.

\bibitem{Cecotti:2015lab}
S.~Cecotti, J.~Song, C.~Vafa and W.~Yan, \emph{{Superconformal Index, BPS Monodromy and Chiral Algebras}}, \href{https://doi.org/10.1007/JHEP11(2017)013}{\emph{JHEP} {\bfseries 11} (2017) 013} [\href{https://arxiv.org/abs/1511.01516}{{\ttfamily 1511.01516}}].

\bibitem{Kim:2024dxu}
H.~Kim and J.~Song, \emph{{A Family of Vertex Operator Algebras from Argyres-Douglas Theory}},  \href{https://arxiv.org/abs/2412.20015}{{\ttfamily 2412.20015}}.

\bibitem{Go:2025ixu}
B.~Go, Q.~Jia, H.~Kim and S.~Kim, \emph{{From BPS spectra of Argyres-Douglas theories to families of 3d TFTs}}, \href{https://doi.org/10.1007/JHEP08(2025)012}{\emph{JHEP} {\bfseries 08} (2025) 012} [\href{https://arxiv.org/abs/2502.15133}{{\ttfamily 2502.15133}}].

\bibitem{Argyres:1995jj}
P.C.~Argyres and M.R.~Douglas, \emph{{New phenomena in SU(3) supersymmetric gauge theory}}, \href{https://doi.org/10.1016/0550-3213(95)00281-V}{\emph{Nucl. Phys. B} {\bfseries 448} (1995) 93} [\href{https://arxiv.org/abs/hep-th/9505062}{{\ttfamily hep-th/9505062}}].

\bibitem{Argyres:1995xn}
P.C.~Argyres, M.R.~Plesser, N.~Seiberg and E.~Witten, \emph{{New N=2 superconformal field theories in four-dimensions}}, \href{https://doi.org/10.1016/0550-3213(95)00671-0}{\emph{Nucl. Phys. B} {\bfseries 461} (1996) 71} [\href{https://arxiv.org/abs/hep-th/9511154}{{\ttfamily hep-th/9511154}}].

\bibitem{Eguchi:1996vu}
T.~Eguchi, K.~Hori, K.~Ito and S.-K.~Yang, \emph{{Study of N=2 superconformal field theories in four-dimensions}}, \href{https://doi.org/10.1016/0550-3213(96)00188-5}{\emph{Nucl. Phys. B} {\bfseries 471} (1996) 430} [\href{https://arxiv.org/abs/hep-th/9603002}{{\ttfamily hep-th/9603002}}].

\bibitem{Eguchi:1996ds}
T.~Eguchi and K.~Hori, \emph{{N=2 superconformal field theories in four-dimensions and A-D-E classification}},  in \emph{{Conference on the Mathematical Beauty of Physics (In Memory of C. Itzykson)}}, pp.~67--82, 7, 1996 [\href{https://arxiv.org/abs/hep-th/9607125}{{\ttfamily hep-th/9607125}}].

\bibitem{Cecotti:2011rv}
S.~Cecotti and C.~Vafa, \emph{{Classification of complete N=2 supersymmetric theories in 4 dimensions}},  \href{https://arxiv.org/abs/1103.5832}{{\ttfamily 1103.5832}}.

\bibitem{Keller:2010bq}
B.~Keller, \emph{{The Periodicity conjecture for pairs of Dynkin diagrams}},  \href{https://arxiv.org/abs/1001.1531}{{\ttfamily 1001.1531}}.

\bibitem{Xie:2012gd}
D.~Xie, \emph{{BPS spectrum, wall crossing and quantum dilogarithm identity}}, \href{https://doi.org/10.4310/ATMP.2016.v20.n3.a1}{\emph{Adv. Theor. Math. Phys.} {\bfseries 20} (2016) 405} [\href{https://arxiv.org/abs/1211.7071}{{\ttfamily 1211.7071}}].

\bibitem{code}
\url{https://github.com/minsung323/wcinv}.

\bibitem{Gaiotto:2024ioj}
D.~Gaiotto and H.~Kim, \emph{{3D TFTs from 4d $ \mathcal{N} $ = 2 BPS particles}}, \href{https://doi.org/10.1007/JHEP03(2025)173}{\emph{JHEP} {\bfseries 03} (2025) 173} [\href{https://arxiv.org/abs/2409.20393}{{\ttfamily 2409.20393}}].

\bibitem{Witten:1990bs}
E.~Witten, \emph{{Introduction to cohomological field theories}}, \href{https://doi.org/10.1142/S0217751X91001350}{\emph{Int. J. Mod. Phys. A} {\bfseries 6} (1991) 2775}.

\bibitem{Costello:2018fnz}
K.~Costello and D.~Gaiotto, \emph{{Vertex Operator Algebras and 3d $ \mathcal{N} $ = 4 gauge theories}}, \href{https://doi.org/10.1007/JHEP05(2019)018}{\emph{JHEP} {\bfseries 05} (2019) 018} [\href{https://arxiv.org/abs/1804.06460}{{\ttfamily 1804.06460}}].

\bibitem{Dedushenko:2018bpp}
M.~Dedushenko, S.~Gukov, H.~Nakajima, D.~Pei and K.~Ye, \emph{{3d TQFTs from Argyres{\textendash}Douglas theories}}, \href{https://doi.org/10.1088/1751-8121/abb481}{\emph{J. Phys. A} {\bfseries 53} (2020) 43LT01} [\href{https://arxiv.org/abs/1809.04638}{{\ttfamily 1809.04638}}].

\bibitem{Gang:2023rei}
D.~Gang, H.~Kim and S.~Stubbs, \emph{{Three-Dimensional Topological Field Theories and Nonunitary Minimal Models}}, \href{https://doi.org/10.1103/PhysRevLett.132.131601}{\emph{Phys. Rev. Lett.} {\bfseries 132} (2024) 131601} [\href{https://arxiv.org/abs/2310.09080}{{\ttfamily 2310.09080}}].

\bibitem{Ferrari:2023fez}
A.E.V.~Ferrari, N.~Garner and H.~Kim, \emph{{Boundary vertex algebras for 3d $\mathcal{N}=4$ rank-0 SCFTs}}, \href{https://doi.org/10.21468/SciPostPhys.17.2.057}{\emph{SciPost Phys.} {\bfseries 17} (2024) 057} [\href{https://arxiv.org/abs/2311.05087}{{\ttfamily 2311.05087}}].

\bibitem{ArabiArdehali:2024ysy}
A.~Arabi~Ardehali, M.~Dedushenko, D.~Gang and M.~Litvinov, \emph{{Bridging 4D QFTs and 2D VOAs via 3D high-temperature EFTs}},  \href{https://arxiv.org/abs/2409.18130}{{\ttfamily 2409.18130}}.

\bibitem{ArabiArdehali:2024vli}
A.~Arabi~Ardehali, D.~Gang, N.J.~Rajappa and M.~Sacchi, \emph{{3d SUSY enhancement and non-semisimple TQFTs from four dimensions}}, \href{https://doi.org/10.1007/JHEP09(2025)179}{\emph{JHEP} {\bfseries 09} (2025) 179} [\href{https://arxiv.org/abs/2411.00766}{{\ttfamily 2411.00766}}].

\bibitem{Maruyoshi:2016tqk}
K.~Maruyoshi and J.~Song, \emph{{Enhancement of Supersymmetry via Renormalization Group Flow and the Superconformal Index}}, \href{https://doi.org/10.1103/PhysRevLett.118.151602}{\emph{Phys. Rev. Lett.} {\bfseries 118} (2017) 151602} [\href{https://arxiv.org/abs/1606.05632}{{\ttfamily 1606.05632}}].

\bibitem{Maruyoshi:2016aim}
K.~Maruyoshi and J.~Song, \emph{{$ \mathcal{N}=1 $ deformations and RG flows of $ \mathcal{N}=2 $ SCFTs}}, \href{https://doi.org/10.1007/JHEP02(2017)075}{\emph{JHEP} {\bfseries 02} (2017) 075} [\href{https://arxiv.org/abs/1607.04281}{{\ttfamily 1607.04281}}].

\bibitem{Agarwal:2016pjo}
P.~Agarwal, K.~Maruyoshi and J.~Song, \emph{{$ \mathcal{N} $ =1 Deformations and RG flows of $ \mathcal{N} $ =2 SCFTs, part II: non-principal deformations}}, \href{https://doi.org/10.1007/JHEP12(2016)103}{\emph{JHEP} {\bfseries 12} (2016) 103} [\href{https://arxiv.org/abs/1610.05311}{{\ttfamily 1610.05311}}].

\bibitem{Kim:2025klh}
H.~Kim, H.~Kim and J.~Song, \emph{{Macdonald index from 3d TQFT}},  \href{https://arxiv.org/abs/2511.11186}{{\ttfamily 2511.11186}}.

\bibitem{Closset:2018ghr}
C.~Closset, H.~Kim and B.~Willett, \emph{{Seifert fibering operators in 3d $\mathcal{N}=2$ theories}}, \href{https://doi.org/10.1007/JHEP11(2018)004}{\emph{JHEP} {\bfseries 11} (2018) 004} [\href{https://arxiv.org/abs/1807.02328}{{\ttfamily 1807.02328}}].

\bibitem{Kapustin:2006hi}
A.~Kapustin, \emph{{Holomorphic reduction of N=2 gauge theories, Wilson-'t Hooft operators, and S-duality}},  \href{https://arxiv.org/abs/hep-th/0612119}{{\ttfamily hep-th/0612119}}.

\bibitem{Gang:2018huc}
D.~Gang and M.~Yamazaki, \emph{{Three-dimensional gauge theories with supersymmetry enhancement}}, \href{https://doi.org/10.1103/PhysRevD.98.121701}{\emph{Phys. Rev. D} {\bfseries 98} (2018) 121701} [\href{https://arxiv.org/abs/1806.07714}{{\ttfamily 1806.07714}}].

\bibitem{Gang:2021hrd}
D.~Gang, S.~Kim, K.~Lee, M.~Shim and M.~Yamazaki, \emph{{Non-unitary TQFTs from 3D $ \mathcal{N} $ = 4 rank 0 SCFTs}}, \href{https://doi.org/10.1007/JHEP08(2021)158}{\emph{JHEP} {\bfseries 08} (2021) 158} [\href{https://arxiv.org/abs/2103.09283}{{\ttfamily 2103.09283}}].

\bibitem{Alim:2011kw}
M.~Alim, S.~Cecotti, C.~Cordova, S.~Espahbodi, A.~Rastogi and C.~Vafa, \emph{{$\mathcal{N} = 2$ quantum field theories and their BPS quivers}}, \href{https://doi.org/10.4310/ATMP.2014.v18.n1.a2}{\emph{Adv. Theor. Math. Phys.} {\bfseries 18} (2014) 27} [\href{https://arxiv.org/abs/1112.3984}{{\ttfamily 1112.3984}}].

\bibitem{Gukov:1999ya}
S.~Gukov, C.~Vafa and E.~Witten, \emph{{CFT's from Calabi-Yau four folds}}, \href{https://doi.org/10.1016/S0550-3213(00)00373-4}{\emph{Nucl. Phys. B} {\bfseries 584} (2000) 69} [\href{https://arxiv.org/abs/hep-th/9906070}{{\ttfamily hep-th/9906070}}].

\bibitem{Shapere:1999xr}
A.D.~Shapere and C.~Vafa, \emph{{BPS structure of Argyres-Douglas superconformal theories}},  \href{https://arxiv.org/abs/hep-th/9910182}{{\ttfamily hep-th/9910182}}.

\bibitem{Shapere:2008zf}
A.D.~Shapere and Y.~Tachikawa, \emph{{Central charges of N=2 superconformal field theories in four dimensions}}, \href{https://doi.org/10.1088/1126-6708/2008/09/109}{\emph{JHEP} {\bfseries 09} (2008) 109} [\href{https://arxiv.org/abs/0804.1957}{{\ttfamily 0804.1957}}].

\bibitem{Cecotti:2013lda}
S.~Cecotti, M.~Del~Zotto and S.~Giacomelli, \emph{{More on the N=2 superconformal systems of type $D_p(G)$}}, \href{https://doi.org/10.1007/JHEP04(2013)153}{\emph{JHEP} {\bfseries 04} (2013) 153} [\href{https://arxiv.org/abs/1303.3149}{{\ttfamily 1303.3149}}].

\bibitem{Creutzig:2017qyf}
T.~Creutzig, \emph{{W-algebras for Argyres-Douglas theories}},  \href{https://arxiv.org/abs/1701.05926}{{\ttfamily 1701.05926}}.

\bibitem{Xie:2019yds}
D.~Xie and W.~Yan, \emph{{W algebras, cosets and VOAs for 4d $ \mathcal{N} $ = 2 SCFTs from M5 branes}}, \href{https://doi.org/10.1007/JHEP04(2021)076}{\emph{JHEP} {\bfseries 04} (2021) 076} [\href{https://arxiv.org/abs/1902.02838}{{\ttfamily 1902.02838}}].

\bibitem{Wang:2015mra}
Y.~Wang and D.~Xie, \emph{{Classification of Argyres-Douglas theories from M5 branes}}, \href{https://doi.org/10.1103/PhysRevD.94.065012}{\emph{Phys. Rev. D} {\bfseries 94} (2016) 065012} [\href{https://arxiv.org/abs/1509.00847}{{\ttfamily 1509.00847}}].

\bibitem{Drinfeld:1984qv}
V.G.~Drinfeld and V.V.~Sokolov, \emph{{Lie algebras and equations of Korteweg-de Vries type}}, \href{https://doi.org/10.1007/BF02105860}{\emph{J. Sov. Math.} {\bfseries 30} (1984) 1975}.

\bibitem{Feigin:1990pn}
B.~Feigin and E.~Frenkel, \emph{{Quantization of the Drinfeld-Sokolov reduction}}, \href{https://doi.org/10.1016/0370-2693(90)91310-8}{\emph{Phys. Lett. B} {\bfseries 246} (1990) 75}.

\bibitem{Kac:2003mjg}
V.G.~Kac, S.-s.~Roan and M.~Wakimoto, \emph{{Quantum Reduction for Affine Superalgebras}}, \href{https://doi.org/10.1007/s00220-003-0926-1}{\emph{Commun. Math. Phys.} {\bfseries 241} (2003) 307} [\href{https://arxiv.org/abs/math-ph/0302015}{{\ttfamily math-ph/0302015}}].

\bibitem{Arakawa:2018iyk}
T.~Arakawa, T.~Creutzig and A.R.~Linshaw, \emph{{W-algebras as coset vertex algebras}}, \href{https://doi.org/10.1007/s00222-019-00884-3}{\emph{Invent. Math.} {\bfseries 218} (2019) 145} [\href{https://arxiv.org/abs/1801.03822}{{\ttfamily 1801.03822}}].

\bibitem{Kac:1988qc}
V.G.~Kac and M.~Wakimoto, \emph{{Modular invariant representations of infinite dimensional Lie algebras and superalgebras}}, \href{https://doi.org/10.1073/pnas.85.14.4956}{\emph{Proc. Nat. Acad. Sci.} {\bfseries 85} (1988) 4956}.

\bibitem{Kac:1989}
V.G.~Kac and M.~Wakimoto, \emph{{Classification of modular invariant representations of affine algebras}}, \href{https://doi.org/10.1142/0869}{\emph{Adv. Ser. Math. Phys.} {\bfseries 7} (1989) 138}.

\bibitem{Gaberdiel:2001ny}
M.R.~Gaberdiel, \emph{{Fusion rules and logarithmic representations of a WZW model at fractional level}}, \href{https://doi.org/10.1016/S0550-3213(01)00490-4}{\emph{Nucl. Phys. B} {\bfseries 618} (2001) 407} [\href{https://arxiv.org/abs/hep-th/0105046}{{\ttfamily hep-th/0105046}}].

\bibitem{Zhu:1996gaq}
Y.~Zhu, \emph{{Modular invariance of characters of vertex operator algebras}}, \href{https://doi.org/10.1090/s0894-0347-96-00182-8}{\emph{J. Am. Math. Soc.} {\bfseries 9} (1996) 237}.

\bibitem{Buican:2016arp}
M.~Buican and T.~Nishinaka, \emph{{Conformal Manifolds in Four Dimensions and Chiral Algebras}}, \href{https://doi.org/10.1088/1751-8113/49/46/465401}{\emph{J. Phys. A} {\bfseries 49} (2016) 465401} [\href{https://arxiv.org/abs/1603.00887}{{\ttfamily 1603.00887}}].

\bibitem{Jiang:2024baj}
H.~Jiang, \emph{{Modularity in Argyres-Douglas theories with a = c}}, \href{https://doi.org/10.1007/JHEP06(2024)131}{\emph{JHEP} {\bfseries 06} (2024) 131} [\href{https://arxiv.org/abs/2403.05323}{{\ttfamily 2403.05323}}].

\bibitem{Pan:2025gzh}
Y.~Pan and P.~Yang, \emph{{Modularity, 4d mirror symmetry, and vertex operator algebra modules of 4D N=2 SCFTs with a=c}}, \href{https://doi.org/10.1103/38ph-h82n}{\emph{Phys. Rev. D} {\bfseries 112} (2025) 105011} [\href{https://arxiv.org/abs/2505.04706}{{\ttfamily 2505.04706}}].

\bibitem{Razamat:2014pta}
S.S.~Razamat and B.~Willett, \emph{{Down the rabbit hole with theories of class $ \mathcal{S} $}}, \href{https://doi.org/10.1007/JHEP10(2014)099}{\emph{JHEP} {\bfseries 10} (2014) 099} [\href{https://arxiv.org/abs/1403.6107}{{\ttfamily 1403.6107}}].

\bibitem{Cordova:2016uwk}
C.~Cordova, D.~Gaiotto and S.-H.~Shao, \emph{{Infrared Computations of Defect Schur Indices}}, \href{https://doi.org/10.1007/JHEP11(2016)106}{\emph{JHEP} {\bfseries 11} (2016) 106} [\href{https://arxiv.org/abs/1606.08429}{{\ttfamily 1606.08429}}].

\bibitem{Cordova:2017ohl}
C.~Cordova, D.~Gaiotto and S.-H.~Shao, \emph{{Surface Defect Indices and 2d-4d BPS States}}, \href{https://doi.org/10.1007/JHEP12(2017)078}{\emph{JHEP} {\bfseries 12} (2017) 078} [\href{https://arxiv.org/abs/1703.02525}{{\ttfamily 1703.02525}}].

\bibitem{Cordova:2017mhb}
C.~Cordova, D.~Gaiotto and S.-H.~Shao, \emph{{Surface Defects and Chiral Algebras}}, \href{https://doi.org/10.1007/JHEP05(2017)140}{\emph{JHEP} {\bfseries 05} (2017) 140} [\href{https://arxiv.org/abs/1704.01955}{{\ttfamily 1704.01955}}].

\bibitem{Cirafici:2017iju}
M.~Cirafici and M.~Del~Zotto, \emph{{Discrete integrable systems, supersymmetric quantum mechanics, and framed BPS states}}, \href{https://doi.org/10.1007/JHEP07(2022)005}{\emph{JHEP} {\bfseries 07} (2022) 005} [\href{https://arxiv.org/abs/1703.04786}{{\ttfamily 1703.04786}}].

\bibitem{Gang:2024loa}
D.~Gang, H.~Kim, B.~Park and S.~Stubbs, \emph{{Three Dimensional Topological Field Theories and Nahm Sum Formulas}},  \href{https://arxiv.org/abs/2411.06081}{{\ttfamily 2411.06081}}.

\bibitem{Cho:2020ljj}
G.Y.~Cho, D.~Gang and H.-C.~Kim, \emph{{M-theoretic Genesis of Topological Phases}}, \href{https://doi.org/10.1007/JHEP11(2020)115}{\emph{JHEP} {\bfseries 11} (2020) 115} [\href{https://arxiv.org/abs/2007.01532}{{\ttfamily 2007.01532}}].

\bibitem{Neitzke:2017cxz}
A.~Neitzke and F.~Yan, \emph{{Line defect Schur indices, Verlinde algebras and $U(1)_r$ fixed points}}, \href{https://doi.org/10.1007/JHEP11(2017)035}{\emph{JHEP} {\bfseries 11} (2017) 035} [\href{https://arxiv.org/abs/1708.05323}{{\ttfamily 1708.05323}}].

\bibitem{Kac:1990gs}
V.G.~Kac, \emph{{Infinite dimensional Lie algebras}}, {Cambridge University Press} (1990).

\bibitem{DiFrancesco:1997nk}
P.~Di~Francesco, P.~Mathieu and D.~Senechal, \emph{{Conformal Field Theory}}, Graduate Texts in Contemporary Physics, Springer-Verlag, New York (1997), \href{https://doi.org/10.1007/978-1-4612-2256-9}{10.1007/978-1-4612-2256-9}.

\bibitem{Feigin:1995xp}
B.~Feigin and F.~Malikov, \emph{{Modular functor and representation theory of SL(2) at a rational level}},  \href{https://arxiv.org/abs/q-alg/9511011}{{\ttfamily q-alg/9511011}}.

\bibitem{Arakawa:2016hkg}
T.~Arakawa and K.~Kawasetsu, \emph{{Quasi-lisse vertex algebras and modular linear differential equations}},  \href{https://arxiv.org/abs/1610.05865}{{\ttfamily 1610.05865}}.

\bibitem{Mathieu:1990dy}
P.~Mathieu and M.A.~Walton, \emph{{Fractional level Kac-Moody algebras and nonunitarity coset conformal theories}}, \href{https://doi.org/10.1143/PTP.102.229}{\emph{Prog. Theor. Phys. Suppl.} {\bfseries 102} (1990) 229}.

\bibitem{Mathieu:1991fz}
P.~Mathieu, D.~Senechal and M.~Walton, \emph{{Field identification in nonunitary diagonal cosets}}, \href{https://doi.org/10.1142/S0217751X92004002}{\emph{Int. J. Mod. Phys. A} {\bfseries 7S1B} (1992) 731} [\href{https://arxiv.org/abs/hep-th/9110003}{{\ttfamily hep-th/9110003}}].

\bibitem{Beltaos:2010ka}
E.~Beltaos and T.~Gannon, \emph{{The $W_{N}$ minimal model classification}}, \href{https://doi.org/10.1007/s00220-012-1473-4}{\emph{Commun. Math. Phys.} {\bfseries 312} (2012) 337} [\href{https://arxiv.org/abs/1004.1205}{{\ttfamily 1004.1205}}].

\bibitem{Nishinaka:2025nbe}
T.~Nishinaka and H.~Sasaki, \emph{{On bosonic vertex algebras associated with 3D reductions of Argyres-Douglas theories}},  \href{https://arxiv.org/abs/2508.15315}{{\ttfamily 2508.15315}}.

\bibitem{Gaiotto:2008ak}
D.~Gaiotto and E.~Witten, \emph{{S-Duality of Boundary Conditions In N=4 Super Yang-Mills Theory}}, \href{https://doi.org/10.4310/ATMP.2009.v13.n3.a5}{\emph{Adv. Theor. Math. Phys.} {\bfseries 13} (2009) 721} [\href{https://arxiv.org/abs/0807.3720}{{\ttfamily 0807.3720}}].

\bibitem{Gang:2018wek}
D.~Gang and K.~Yonekura, \emph{{Symmetry enhancement and closing of knots in 3d/3d correspondence}}, \href{https://doi.org/10.1007/JHEP07(2018)145}{\emph{JHEP} {\bfseries 07} (2018) 145} [\href{https://arxiv.org/abs/1803.04009}{{\ttfamily 1803.04009}}].

\bibitem{Assel:2022row}
B.~Assel, Y.~Tachikawa and A.~Tomasiello, \emph{{On $ \mathcal{N} $ = 4 supersymmetry enhancements in three dimensions}}, \href{https://doi.org/10.1007/JHEP03(2023)170}{\emph{JHEP} {\bfseries 03} (2023) 170} [\href{https://arxiv.org/abs/2209.13984}{{\ttfamily 2209.13984}}].

\bibitem{Garozzo:2019ejm}
I.~Garozzo, G.~Lo~Monaco, N.~Mekareeya and M.~Sacchi, \emph{{Supersymmetric Indices of 3d $S$-fold SCFTs}}, \href{https://doi.org/10.1007/JHEP08(2019)008}{\emph{JHEP} {\bfseries 08} (2019) 008} [\href{https://arxiv.org/abs/1905.07183}{{\ttfamily 1905.07183}}].

\bibitem{Jeong:2025xid}
K.~Jeong and S.~Lee, \emph{{QFT Realization of Non-Unitary $\mathfrak{sl}(2,\mathbb{C})$ WRT Invariants and Their Galois Conjugations}},  \href{https://arxiv.org/abs/2511.16380}{{\ttfamily 2511.16380}}.

\bibitem{Gang:2024tlp}
D.~Gang, H.~Kang and S.~Kim, \emph{{Non-hyperbolic 3-manifolds and 3D field theories for 2D Virasoro minimal models}},  \href{https://arxiv.org/abs/2405.16377}{{\ttfamily 2405.16377}}.

\bibitem{Gang:2022kpe}
D.~Gang and D.~Kim, \emph{{Generalized non-unitary Haagerup-Izumi modular data from 3D S-fold SCFTs}}, \href{https://doi.org/10.1007/JHEP03(2023)185}{\emph{JHEP} {\bfseries 03} (2023) 185} [\href{https://arxiv.org/abs/2211.13561}{{\ttfamily 2211.13561}}].

\bibitem{Gang:2023ggt}
D.~Gang, D.~Kim and S.~Lee, \emph{{A non-unitary bulk-boundary correspondence: Non-unitary Haagerup RCFTs from S-fold SCFTs}}, \href{https://doi.org/10.21468/SciPostPhys.17.2.064}{\emph{SciPost Phys.} {\bfseries 17} (2024) 064} [\href{https://arxiv.org/abs/2310.14877}{{\ttfamily 2310.14877}}].

\bibitem{haagerup1994principal}
U.~Haagerup, \emph{Principal graphs of subfactors in the index range $4<[m: N]< 3+\sqrt{2}$}, {\emph{Subfactors (Kyuzeso, 1993)} (1994) 1}.

\bibitem{Asaeda_1999}
M.~Asaeda and U.~Haagerup, \emph{Exotic subfactors of finite depth with jones indices $(5+\sqrt{13})/2$ and $(5+\sqrt{17})/2$}, \href{https://doi.org/10.1007/s002200050574}{\emph{Communications in Mathematical Physics} {\bfseries 202} (1999) 1–63}.

\bibitem{Izumi:2000qa}
M.~Izumi, \emph{{The structure of sectors associated with Longo-Rehren inclusions. I: General theory}}, \href{https://doi.org/10.1007/s002200000234}{\emph{Commun. Math. Phys.} {\bfseries 213} (2000) 127}.

\bibitem{Terashima:2011qi}
Y.~Terashima and M.~Yamazaki, \emph{{SL(2,R) Chern-Simons, Liouville, and Gauge Theory on Duality Walls}}, \href{https://doi.org/10.1007/JHEP08(2011)135}{\emph{JHEP} {\bfseries 08} (2011) 135} [\href{https://arxiv.org/abs/1103.5748}{{\ttfamily 1103.5748}}].

\bibitem{Dimofte:2011ju}
T.~Dimofte, D.~Gaiotto and S.~Gukov, \emph{{Gauge Theories Labelled by Three-Manifolds}}, \href{https://doi.org/10.1007/s00220-013-1863-2}{\emph{Commun. Math. Phys.} {\bfseries 325} (2014) 367} [\href{https://arxiv.org/abs/1108.4389}{{\ttfamily 1108.4389}}].

\bibitem{Gang:2018gyt}
D.~Gang, \emph{{Quantum Approach to Dehn Surgery Problem}},  \href{https://arxiv.org/abs/1803.11143}{{\ttfamily 1803.11143}}.

\bibitem{Choi:2022dju}
S.~Choi, D.~Gang and H.-C.~Kim, \emph{{Infrared phases of 3D class R theories}}, \href{https://doi.org/10.1007/JHEP11(2022)151}{\emph{JHEP} {\bfseries 11} (2022) 151} [\href{https://arxiv.org/abs/2206.11982}{{\ttfamily 2206.11982}}].

\bibitem{Baek:2025uev}
S.~Baek and H.~Kang, \emph{{Non-hyperbolic 3-manifolds and bulk field theories for supersymmetric/$W_N$ minimal models}},  \href{https://arxiv.org/abs/2511.04524}{{\ttfamily 2511.04524}}.

\bibitem{Creutzig:2024ljv}
T.~Creutzig, N.~Garner and H.~Kim, \emph{{Mirror Symmetry and Level-rank Duality for 3d $\mathcal{N} = 4$ Rank 0 SCFTs}},  \href{https://arxiv.org/abs/2406.00138}{{\ttfamily 2406.00138}}.

\bibitem{Kucharski:2025lcr}
P.~Kucharski, P.~Longhi, D.~Noshchenko, S.~Park and P.~Sulkowski, \emph{{Quivers and BPS states in 3d and 4d}},  \href{https://arxiv.org/abs/2508.09729}{{\ttfamily 2508.09729}}.

\bibitem{Baek:2024tuo}
S.~Baek and D.~Gang, \emph{{3D bulk field theories for 2D non-unitary $ \mathcal{N} $ = 1 supersymmetric minimal models}}, \href{https://doi.org/10.1007/JHEP01(2025)027}{\emph{JHEP} {\bfseries 01} (2025) 027} [\href{https://arxiv.org/abs/2405.05746}{{\ttfamily 2405.05746}}].

\bibitem{Nekrasov:2009uh}
N.A.~Nekrasov and S.L.~Shatashvili, \emph{{Supersymmetric vacua and Bethe ansatz}}, \href{https://doi.org/10.1016/j.nuclphysbps.2009.07.047}{\emph{Nucl. Phys. B Proc. Suppl.} {\bfseries 192-193} (2009) 91} [\href{https://arxiv.org/abs/0901.4744}{{\ttfamily 0901.4744}}].

\bibitem{Nekrasov:2009ui}
N.A.~Nekrasov and S.L.~Shatashvili, \emph{{Quantum integrability and supersymmetric vacua}}, \href{https://doi.org/10.1143/PTPS.177.105}{\emph{Prog. Theor. Phys. Suppl.} {\bfseries 177} (2009) 105} [\href{https://arxiv.org/abs/0901.4748}{{\ttfamily 0901.4748}}].

\bibitem{Closset:2019hyt}
C.~Closset and H.~Kim, \emph{{Three-dimensional {\ensuremath{\mathscr{N}}} = 2 supersymmetric gauge theories and partition functions on Seifert manifolds: A review}}, \href{https://doi.org/10.1142/S0217751X19300114}{\emph{Int. J. Mod. Phys. A} {\bfseries 34} (2019) 1930011} [\href{https://arxiv.org/abs/1908.08875}{{\ttfamily 1908.08875}}].

\bibitem{Closset:2023vos}
C.~Closset and O.~Khlaif, \emph{{Twisted indices, Bethe ideals and 3d $ \mathcal{N} $ = 2 infrared dualities}}, \href{https://doi.org/10.1007/JHEP05(2023)148}{\emph{JHEP} {\bfseries 05} (2023) 148} [\href{https://arxiv.org/abs/2301.10753}{{\ttfamily 2301.10753}}].

\bibitem{Closset:2023jiq}
C.~Closset and O.~Khlaif, \emph{{On the Witten index of 3d $\mathcal{N}=2$ unitary SQCD with general CS levels}}, \href{https://doi.org/10.21468/SciPostPhys.15.3.085}{\emph{SciPost Phys.} {\bfseries 15} (2023) 085} [\href{https://arxiv.org/abs/2305.00534}{{\ttfamily 2305.00534}}].

\bibitem{Closset:2023bdr}
C.~Closset and O.~Khlaif, \emph{{Grothendieck lines in 3d $ \mathcal{N} $ = 2 SQCD and the quantum K-theory of the Grassmannian}}, \href{https://doi.org/10.1007/JHEP12(2023)082}{\emph{JHEP} {\bfseries 12} (2023) 082} [\href{https://arxiv.org/abs/2309.06980}{{\ttfamily 2309.06980}}].

\bibitem{Closset:2023izb}
C.~Closset and O.~Khlaif, \emph{{New results on 3d {\ensuremath{\mathscr{N}}}=2 SQCD and its 3d GLSM interpretation}}, \href{https://doi.org/10.1142/S0217751X24460114}{\emph{Int. J. Mod. Phys. A} {\bfseries 39} (2024) 2446011} [\href{https://arxiv.org/abs/2312.05076}{{\ttfamily 2312.05076}}].

\bibitem{Closset:2024sle}
C.~Closset, E.~Furrer and O.~Khlaif, \emph{{One-form symmetries and the 3d $\mathcal{N}=2$ $A$-model: Topologically twisted indices and CS theories}}, \href{https://doi.org/10.21468/SciPostPhys.18.2.066}{\emph{SciPost Phys.} {\bfseries 18} (2025) 066} [\href{https://arxiv.org/abs/2405.18141}{{\ttfamily 2405.18141}}].

\bibitem{Closset:2025lqt}
C.~Closset, E.~Furrer, A.~Keyes and O.~Khlaif, \emph{{The 3d $A$-model and generalised symmetries, Part I: bosonic Chern-Simons theories}},  \href{https://arxiv.org/abs/2501.11665}{{\ttfamily 2501.11665}}.

\bibitem{Gukov:2015sna}
S.~Gukov and D.~Pei, \emph{{Equivariant Verlinde formula from fivebranes and vortices}}, \href{https://doi.org/10.1007/s00220-017-2931-9}{\emph{Commun. Math. Phys.} {\bfseries 355} (2017) 1} [\href{https://arxiv.org/abs/1501.01310}{{\ttfamily 1501.01310}}].

\bibitem{Benini:2015noa}
F.~Benini and A.~Zaffaroni, \emph{{A topologically twisted index for three-dimensional supersymmetric theories}}, \href{https://doi.org/10.1007/JHEP07(2015)127}{\emph{JHEP} {\bfseries 07} (2015) 127} [\href{https://arxiv.org/abs/1504.03698}{{\ttfamily 1504.03698}}].

\bibitem{Benini:2016hjo}
F.~Benini and A.~Zaffaroni, \emph{{Supersymmetric partition functions on Riemann surfaces}}, {\emph{Proc. Symp. Pure Math.} {\bfseries 96} (2017) 13} [\href{https://arxiv.org/abs/1605.06120}{{\ttfamily 1605.06120}}].

\bibitem{Closset:2016arn}
C.~Closset and H.~Kim, \emph{{Comments on twisted indices in 3d supersymmetric gauge theories}}, \href{https://doi.org/10.1007/JHEP08(2016)059}{\emph{JHEP} {\bfseries 08} (2016) 059} [\href{https://arxiv.org/abs/1605.06531}{{\ttfamily 1605.06531}}].

\bibitem{Faddeev:1993rs}
L.D.~Faddeev and R.M.~Kashaev, \emph{{Quantum Dilogarithm}}, \href{https://doi.org/10.1142/S0217732394000447}{\emph{Mod. Phys. Lett. A} {\bfseries 9} (1994) 427} [\href{https://arxiv.org/abs/hep-th/9310070}{{\ttfamily hep-th/9310070}}].

\end{thebibliography}\endgroup
\end{document}